%% file: main.tex
\documentclass[11pt,letterpaper]{article}
\usepackage[margin=1in]{geometry}
\usepackage{graphicx}
\usepackage{amsmath}
\usepackage{amssymb}
\usepackage{amsthm}
\usepackage{algorithm}
\usepackage{algpseudocode}
\usepackage{xcolor}
\usepackage{tikz}
\usetikzlibrary{arrows.meta,shapes.geometric}
\definecolor{citationgreen}{RGB}{0,100,0}
\usepackage[colorlinks=true,citecolor=citationgreen,linkcolor=red]{hyperref}
\usepackage[nameinlink]{cleveref}

\newtheorem{lemma}{Lemma}
\newtheorem{corollary}{Corollary}
\newtheorem{observation}{Observation}
\newtheorem{definition}{Definition}
\newtheorem{theorem}{Theorem}

\newcommand{\dep}{\operatorname{dep}}
\newcommand{\dist}{\operatorname{dist}}

\title{Constant-Time Dynamic Edge Coloring,\\via Dynamic Splitting and LLL}
\author{%
  Mohsen Ghaffari\\
  {\small MIT}\\
  {\small\texttt{ghaffari@mit.edu}}
  \and
  Sijin Peng\\
  {\small MIT}\\
  {\small\texttt{sijinp@mit.edu}}%
}
\date{}

\newcommand{\poly}{\mathrm{poly}}

\newcommand{\cnt}{\mathrm{cnt}}
\newcommand{\dir}{\mathrm{dir}}
\newcommand{\eval}{\mathrm{eval}}

\newcommand{\routepath}{\mathrm{path}}
\newcommand{\sample}{\mathrm{sample}}
\newcommand{\update}{\mathrm{update}}
\newcommand{\vbl}{\mathrm{vbl}}
\newcommand{\Search}{\mathsf{Search}}
\newcommand{\polylog}{\mathrm{polylog}}
\renewcommand{\P}{\mathrm{P}}
\newcommand{\LC}{\mathrm{LC}}
\newcommand{\RC}{\mathrm{RC}}

\newcommand{\bad}{\mathrm{bad}}

\newcommand{\old}{\mathrm{old}}
\newcommand{\new}{\mathrm{new}}

\begin{document}
\hypersetup{pageanchor=false}

\maketitle
\thispagestyle{empty}

\begin{abstract}
We give a dynamic algorithm that maintains a near-optimal edge coloring in constant expected time per update. For every fixed $\epsilon>0$, the algorithm maintains a $(1+\epsilon)\Delta$ edge coloring under edge insertions and deletions in $\poly(1/\epsilon)$ worst-case expected update time against an oblivious adversary, provided $\Delta\ge\Delta_\epsilon=\widetilde\Theta(1/\epsilon^2)$. The previous constant-time algorithm of Bhattacharya, Costa, Panski, and Solomon (BCPS) required $\Delta\ge(\log n/\epsilon^4)^{\Theta(\log(1/\epsilon)/\epsilon)}$, where $n$ is the number of vertices. We remove this dependence on $n$, answering their open question.

Our approach differs substantially from that of BCPS, which relied on coloring in locally tree-like subgraphs. Instead, we draw on distributed algorithms that organize partial coloring through successive applications of the Lov\'asz Local Lemma (LLL). We develop a dynamic LLL solver with worst-case expected update bounds and a dynamic splitting algorithm that partitions edges into low-degree subgraphs. A key technical contribution is a bottom-up method for iterative splitting, which avoids the logarithmic overhead of maintaining successive levels. It achieves constant expected update time by adaptively choosing which levels to inspect and repair, rather than inspecting every level after each update. We believe these dynamic splitting and LLL tools can be of interest beyond edge coloring.
\end{abstract}


\clearpage
\thispagestyle{empty}
\setcounter{tocdepth}{1}
\vspace{2\baselineskip}
{\small
\hypersetup{linkcolor=blue}
\renewcommand{\contentsname}{{\normalsize Contents}}
\tableofcontents
}

\clearpage
\pagenumbering{arabic}
\hypersetup{pageanchor=true}
\input{introduction-draft.tex}

\input{technical-overview.tex}
\input{dynamic-LLL}
\input{edge-coloring-poly-delta.tex}
\input{edge-coloring-log-delta.tex}
\input{edge-coloring-const.tex}
\input{witness.tex}

\section{AI Statement}

The authors designed the algorithms and analyses presented in this paper. We have used AIs (GPT~6.1 Sol, GPT 6.0 Astra, and GPT 5.6~Sol) for reviewing our write-ups and help with the writing and copyediting. In addition, they helped us simplify \Cref{def:witness} in \Cref{sec:witness}. We have verified all parts and assume responsibility for the content.

\bibliographystyle{alpha}
\bibliography{ref}

\end{document}

%% file: introduction-draft.tex
\section{Introduction}
\label{sec:introduction-draft}

This paper presents a constant-time dynamic algorithm for near-optimal edge
coloring. We next review the context and the most directly related work, and
then state our results and technical contributions. This is then followed by a
discussion of other related work.

\subsection{Context and our contribution}
\label{sec:introduction-context}

Vizing's theorem shows that every simple graph of maximum degree $\Delta$ admits
a $(\Delta+1)$ edge coloring. However, maintaining such a coloring efficiently
in dynamic graphs, as edges are inserted or deleted, remains open: There is no
known dynamic algorithm for it with subpolynomial update time, except in the
special case of low-degree graphs. We
discuss these and other related results, including static algorithms, in
\Cref{sec:introduction-related-work}. In contrast, the problem becomes more
tractable when we relax the number of colors slightly, to $(1+\epsilon)\Delta$
colors. The state of the art is the algorithm of
Bhattacharya, Costa, Panski, and Solomon (BCPS)~\cite{bhattacharya2023nibbling}, which 
maintains a $(1+\epsilon)\Delta$ edge coloring in $\poly(1/\epsilon)$ worst-case
expected update time against an oblivious adversary, provided that
\[
    \Delta\ge
    (\log n/\epsilon^4)^{\Theta(\log(1/\epsilon)/\epsilon)}.
\]
Their algorithm thus gives constant update time for fixed $\epsilon$, but
requires a large maximum degree as a function of the number of vertices $n$.
They explicitly left improving this degree requirement while retaining constant
update time as a ``\emph{challenging open
question}''~\cite[Section~1.4]{bhattacharya2023nibbling}.

As a side comment, we note that constant-time dynamic algorithms remain rare, 
limited to only a handful of problems, all of which have much simpler structures than
(near-optimal) edge coloring, at least in the static setting.\footnote{Examples
include maximal matching and its dual, $2$-approximate vertex
cover~\cite{solomon2016constant};
$(\Delta+1)$-vertex
coloring~\cite{bhattacharya2022constantcoloring,henzinger2020constantcoloring,
ghaffari2027dynamiccoloring};
and constant-stretch spanners~\cite{baswana2012spanners}.}

\paragraph{Our main result.}
We present a constant-time dynamic algorithm for $(1+\epsilon)\Delta$ edge
coloring that removes the $n$-dependence in the degree requirement and assumes
only $\Delta\ge\Delta_\epsilon$, where
$\Delta_\epsilon=\widetilde O(1/\epsilon^2)$ is a constant depending only on
$\epsilon$.

\begin{theorem}\label{thm:introduction-main}
For every $0<\epsilon<1$, there is a threshold $\Delta_\epsilon=\widetilde
O(1/\epsilon^2)$ and a randomized dynamic algorithm with the following
guarantee. Given a fixed degree bound $\Delta\ge\Delta_\epsilon$, the algorithm
maintains a proper edge coloring with at most $(1+\epsilon)\Delta$ colors under
edge insertions and deletions in a simple graph of maximum degree at most
$\Delta$. Against an oblivious adversary, each update takes $\poly(1/\epsilon)$
worst-case expected time.
\end{theorem}

This answers the open question posed by BCPS. We emphasize that the theorem
bounds the expected cost of each update, rather than an amortized average over
the whole update sequence. The oblivious adversary in the statement fixes the
update sequence knowing our algorithm, but without knowing its random choices.
The $\widetilde O$ notation hides factors polylogarithmic in $1/\epsilon$.

\paragraph{Our Method, and Technical Contributions.}
At a high level, our algorithm builds on R\"odl's nibble
approach~\cite{rodl1985packing}, which BCPS also followed and which is one of
the major approaches to near-optimal edge coloring. Beyond this high-level
similarity, however, we depart completely from their method. The nibble colors
the graph gradually, preserving regularity conditions on the remaining degrees
and available palettes after each partial-coloring round. The local randomized
choices make this
organization attractive for dynamic maintenance, but dependencies between the
choices make the probabilistic analysis delicate.
BCPS~\cite{bhattacharya2023nibbling} address these dependencies by
partitioning the edges into locally tree-like subgraphs, exploiting the
probabilistic independence between the branches of the local neighborhoods
during the iterations of coloring each subgraph.

Instead, we take a different route by bringing in viewpoints from distributed
graph algorithms. We abstract the guarantees desired from each iteration of
partial coloring as a Lov\'asz Local Lemma (LLL) problem, following Elkin,
Pettie, and Su~\cite{elkin2015edgecoloring} and Chang, He, Li, Pettie, and
Uitto~\cite{chang2017complexity}. We develop a dynamic LLL
solver to maintain these instances without requiring locally tree-like
neighborhoods. This first gives an update time polynomial in the degree. We
then remove that dependence through a dynamic splitting algorithm, which
effectively cuts down the relevant degree of the edge-coloring instances to
$\poly(1/\epsilon)$. This ingredient itself is inspired by algorithmic LLL
techniques in the distributed literature, namely graph
shattering~\cite{halldorsson2022fast}, and making it run in constant update time
is the most technical part of our contribution in this paper. To summarize, we
can organize our contributions into the following three categories (and we are
hopeful that ingredients (I) and (III) will be of broader interest, beyond
edge coloring):

\begin{enumerate}
    \renewcommand{\labelenumi}{(\Roman{enumi})}
    \item \textbf{Dynamic LLL.}
    We describe a dynamic LLL algorithm that maintains a satisfying assignment
    in $\poly(\Delta)$ worst-case expected time per update in the LLL (a
    clause--variable update), assuming the polynomial LLL criterion $pd^C<1$
    for a sufficiently large absolute constant $C$.
    Here $p$ bounds the bad-event probabilities, $d$ bounds the dependency
    degree, and $\Delta$ bounds the local instance parameters and the costs
    of sampling and evaluating events.
    This criterion is readily satisfied in all our applications.\footnote{Concentration
    bounds, which in our applications give failure probabilities that decay
    exponentially in suitable functions of the degree parameters, make the
    bad-event probabilities smaller than any fixed inverse polynomial in
    those parameters. Indeed, we could work with even stronger LLL
    criteria---e.g., requiring inverse-quasipolynomial failure probabilities
    or even inverse exponentials with small polynomial
    exponents---by trading off some parameters, while keeping both the
    degree threshold and the update time polynomial in $1/\epsilon$. In
    addition, this polynomial criterion makes the expected repair work and recourse
    small, which is essential when we compose several LLL iterations.}
    A clause--variable update adds or removes a variable
    from a clause. We process batches of these changes, allowing several
    variables or clauses
    to be modified simultaneously. We require the LLL criterion to hold before
    and after each batch, so that we always have a meaningful problem, without
    requiring every intermediate instance to satisfy the criterion. Our
    algorithm for this part is simple and builds on the observation that the
    distributed resampling algorithm of Chung, Pettie, and
    Su~\cite{chung2014distributed} is stable under dynamic updates when we
    preserve its randomness: the changes to its execution are small in
    expectation. This directly exploits the locality of the algorithm and its
    witness-tree analysis: the probability of resampling at later rounds
    decreases fast enough to offset the growth of the neighborhood that an
    update can affect. We view this as a general tool that may be useful beyond
    edge coloring.

    \item \textbf{Low-degree edge coloring.}
    We obtain a dynamic $(1+\epsilon)\Delta$ edge-coloring algorithm with
    $\poly(\Delta)$ worst-case expected update time for
    $\Delta\ge\Delta_\epsilon$. In particular, we obtain $\poly(1/\epsilon)$
    update time when $\Delta=\poly(1/\epsilon)$. We achieve this by applying our
    dynamic LLL solver iteratively to the nibble-based static partial-coloring
    algorithm of Chang et al.~\cite{chang2017complexity}, with a more careful
    bound on how many changes one iteration passes to the next. Here, we note
    that a direct black-box application would lose a polynomial factor in
    $\Delta$ in each iteration and lead to a
    $\Delta^{O(\log(1/\epsilon))}$ update time. We avoid this and achieve
    $\poly(\Delta)$ update time by coupling color proposals before and after a
    palette change, so that changing one color in a large palette rarely
    changes the proposal. We use this result as the low-degree coloring
    ingredient for our final algorithm.

    \item \textbf{Bottom-up splitting.}
    Our third contribution is a scheme that partitions the graph into at most
    $(1+O(\epsilon))\Delta/D$ edge-disjoint subgraphs of maximum degree at most
    $D=\poly(1/\epsilon)$, in constant expected update time and recourse. This
    allows us to solve the edge-coloring problem in each subgraph independently
    with a separate palette, obtaining $\poly(1/\epsilon)$ worst-case expected
    update time for the original graph. Maintaining this partition in constant
    expected update time is the technical heart of the paper.
    We first develop a warm-up top-down iterative splitting algorithm, which
    gives $O(\log\Delta)+\poly(1/\epsilon)$ update time for edge coloring. We
    then develop a bottom-up variant that represents the hierarchy implicitly
    and discovers which levels need attention. The algorithm and analysis here
    use LLL-inspired notions such as locally freezing variables during a
    partial assignment when certain constraints become at risk of violation,
    and graph shattering to show that the unresolved regions break into small
    connected components. We analyze the dependencies between levels through
    a joint witness argument. This lets us inspect and repair only the relevant
    parts of the hierarchy, with expected cost independent of the number of
    partitioning levels, allowing us to achieve constant expected update time
    for its maintenance. We believe that this bottom-up splitting technique can
    also be useful beyond edge coloring.
\end{enumerate}

In \Cref{sec:technical-overview}, we give a closer overview of these ingredients
and how we combine them to obtain the dynamic edge-coloring result of
\Cref{thm:introduction-main}.

\paragraph{Two comments on the $\epsilon$-dependencies of our constants.}
\textbf{(A)} The $\poly(1/\epsilon)$ dependence in our update time is
currently bottlenecked by the $\poly(D)$ update time of our dynamic
edge-coloring algorithm on graphs of maximum degree at most
$D=O(\epsilon^{-5})$, a constant for fixed $\epsilon$. This is the degree
bound reached by our present reduction (\Cref{lem:edge-coloring-small-deg}).
The splitting procedures incur only $O(1)$ expected work and generate
$O(1)$ expected updates to these low-degree instances per input update.
Thus, improving the worst-case expected update time for maintaining
$(1+\epsilon)D$ edge colorings in this degree range would also improve our
update-time bound for graphs of unbounded degree.
\textbf{(B)} The degree requirement $\Delta\ge\Delta_\epsilon$, where
$\Delta_\epsilon=\widetilde O(\epsilon^{-2})$, is inherited from the
nibble-based coloring approach.
Intuitively, sampling counts with expectation $d$ and variance $\Theta(d)$
have fluctuations of order $\sqrt d$, or relative fluctuations of order
$1/\sqrt d$. Keeping these fluctuations within an $\epsilon$ fraction of
the expectation therefore requires $d$ to be at least on the order of
$1/\epsilon^2$, with additional logarithmic factors to obtain the
concentration needed by the LLL. This sampling fluctuation is the source
of the $1/\epsilon^2$ bottleneck in our current degree threshold.

\subsection{Other Related Work}
\label{sec:introduction-related-work}

\paragraph{Randomized dynamic edge coloring.}
Duan, He, and Zhang~\cite{duan2019dynamic} gave a randomized dynamic
$(1+\epsilon)\Delta$ edge-coloring algorithm with $\poly(\log n,1/\epsilon)$
amortized update time when $\Delta=\Omega(\log^2 n/\epsilon^2)$. Bhattacharya,
Grandoni, and Wajc~\cite{bhattacharya2021online} obtained $\poly(1/\epsilon)$
expected recourse (recolorings) per update in a sufficiently large-degree
regime, without a corresponding constant update-time guarantee.
Christiansen~\cite{christiansen2023power} removed the additional large-degree
requirement and obtained $\poly(\log n,1/\epsilon)$ worst-case update time with
high probability. Setting $\epsilon=1/\Delta$ gives a dynamic $(\Delta+1)$
edge-coloring algorithm with $\poly(\Delta\log n)$ update time, which is
polylogarithmic in $n$ for graphs of degree at most $\polylog n$.
BCPS~\cite{bhattacharya2023nibbling} then achieved constant
worst-case expected update time for fixed $\epsilon$ under the degree
requirement discussed above. We compare most directly with their update
guarantee, while also reaching the degrees for which Christiansen's
polylogarithmic-time algorithm was previously the relevant alternative.

Haeupler, Mitrovi\'{c}, Ramachandran, Sheu, and
Tarjan~\cite{haeupler2026dynamic} gave a dynamic $(1+\epsilon)\Delta$
edge-coloring algorithm against an adaptive adversary, with amortized update time
$\bigl(\Delta\log n\bigr)^{O(\log(1/\epsilon))}$
over polynomial-length update sequences, for
$\epsilon=\omega(\log^{2.5}\Delta/\sqrt\Delta)$. They also
build on the distributed partial-coloring algorithm of Chang et
al.~\cite{chang2017complexity}. Their guarantee supports an adaptive adversary,
who knows the algorithm's outputs and random choices so far and can choose
subsequent updates accordingly. We focus on an oblivious adversary and obtain
\emph{constant} worst-case expected bounds independent of $n$ and $\Delta$.

\paragraph{Deterministic dynamic edge coloring.}
Christiansen~\cite{christiansen2026deterministic} gave a deterministic algorithm
maintaining a $(1+\epsilon)\Delta$ edge coloring in $2^{\widetilde
O_{\log(1/\epsilon)}(\sqrt{\log n})}$ amortized update time. This provides a
subpolynomial bound for fixed $\epsilon$ without randomization. Our focus is the
randomized setting, where we seek constant update time.

\paragraph{Dynamic LLL.}
Haeupler et al.~\cite{haeupler2026dynamic} developed a general framework for
maintaining solutions under constraint insertions and deletions against adaptive
adversaries. In standard variable-based LLL settings with fixed slack, their
framework gives $\widetilde O(\poly(k,d))$ amortized update time, with an
additional initialization cost, where $k$ bounds the number of variables per
clause and $d$ bounds the dependency degree. They obtain this guarantee by
bounding the total number of resampling steps over the update sequence, and
derive several applications, including their edge-coloring result. We work
under a polynomial LLL criterion, which is readily satisfied in all our
applications and in many other algorithmic applications of the LLL. We support
batches of clause--variable changes against an oblivious adversary. We obtain
worst-case expected update bounds and small expected recourse, which lets us
compose the solver with successive partial-coloring instances.

\paragraph{Static edge coloring.}
Karloff and Shmoys~\cite{karloff1987efficient} gave a near-linear-time
algorithm using $\Delta+\widetilde O(\sqrt\Delta)$ colors. More recent work has
developed several approaches beyond the nibble. Elkin and
Khuzman~\cite{elkin2024deterministic} gave a deterministic $(1+\epsilon)\Delta$
edge-coloring algorithm in $O(m\log n/\epsilon)$ time.
Assadi~\cite{assadi2025faster} used fair matchings and a reduction to bipartite
edge coloring to obtain $\Delta+O(\log n)$ colors in $O(m\log\Delta)$ expected
time; he also obtained $(1+\epsilon)\Delta$ colors in $O(m\log(1/\epsilon))$
expected time for $\epsilon=\omega(\log n/\Delta)$. Bernshteyn and
Dhawan~\cite{bernshteyn2025linear} used randomized multi-step Vizing chains to
obtain a linear-time algorithm with high probability for every fixed
$\epsilon>0$ and $\Delta\ge1/\epsilon$.

For the exact Vizing bound, Assadi et al.~\cite{assadi2025vizing} gave a
randomized $(\Delta+1)$ edge-coloring algorithm in $O(m\log n)$ time with high
probability, using batched recoloring with Vizing fans and alternating paths.
These static results do not directly bound the cost of repairing a coloring
after an individual update. Vizing-chain methods also underlie dynamic
algorithms, including those of Duan, He, and Zhang~\cite{duan2019dynamic} and
Christiansen~\cite{christiansen2023power}; our route to constant update time
instead develops dynamic LLL and splitting tools.

\paragraph{Distributed edge coloring.}
Several ingredients of our algorithms are directly inspired by distributed
approaches to near-optimal edge coloring. Elkin, Pettie, and
Su~\cite{elkin2015edgecoloring} obtained randomized
$(1+\epsilon)\Delta$-edge-coloring algorithms for every fixed $\epsilon>0$
and sufficiently large $\Delta$ depending only on $\epsilon$, using
distributed LLL algorithms to enforce the desired guarantees in successive
nibble iterations (lower bounds on available palette sizes and upper bounds
on remaining degrees and on the number of uncolored edges incident to a
vertex whose palettes contain a given color).
Chang, He, Li, Pettie, and Uitto~\cite{chang2017complexity} refined this
approach to use $O(\log(1/\epsilon))$ LLL instances, permitting palettes as
small as $\Delta+\widetilde O(\sqrt{\Delta})$. Our dynamic edge-coloring
algorithm for constant-degree graphs essentially shows how to maintain their
partial-coloring procedure dynamically as edges are inserted or deleted.
Subsequently, Halld\'orsson, Maus,
and Nolin~\cite{halldorsson2022fast} further improved the distributed
complexity for fixed $\epsilon$ by using shattering to solve splitting instances and
reduce degrees before applying LLL-based edge coloring. Their splitting
approach directly inspires our degree-reduction machinery, though our
constant-time dynamic splitting requires substantial new technical ingredients
and analysis beyond this high-level connection.

%% file: technical-overview.tex

\section{Technical Overview}
\label{sec:technical-overview}

The overall merit of our approach is to maintain the execution transcripts of static randomized algorithms, using persistent randomness for each variable. After an input update, we repair the transcript so that it agrees with a static execution on the new input using the same randomness. This approach also appears in the prior work of Bhattacharya, Costa, Panski, and Solomon~\cite{bhattacharya2023nibbling}.

Static distributed algorithms are natural candidates for this approach: their locality limits how an input change can propagate through the transcript. The main challenge is to identify the parts of the transcript that need repair without inspecting every part that might have changed.

\subsection{Dynamic Lov\'asz Local Lemma}

Our first ingredient, described in \Cref{sec:dynamic-lll}, is a dynamic algorithm to maintain a satisfying assignment of an LLL instance. The instance goes through batches of clause--variable updates, in which each update inserts a variable into a clause or removes it from a clause. When the LLL instances have large slack, the expected recourse of our algorithm will be extremely small, while the expected runtime will only depend on how efficiently one can maintain the state of clauses under the initial assignment after each update.  

We observe that the distributed LLL algorithm of Chung, Pettie, and Su~\cite{chung2014distributed}, a parallel variant of the Moser--Tardos resampling algorithm~\cite{moser2010constructive}, is a nice candidate to maintain dynamically. It starts with a random assignment and consists of several resampling iterations until the assignment satisfies every clause. In each iteration, it finds a particular independent set of bad events that occur under the current assignment and resamples all variables involved in those events.

We leverage the following two properties of the static algorithm during the analysis of our dynamic algorithm.
\begin{itemize}
    \item First, changes propagate locally from one iteration to the next. More precisely, say the information for a clause at stage $r$ records whether it is falsified by the assignment at stage $r$ and whether any variable is resampled in that stage. We are able to demonstrate that, any clause whose information at stage $r$ changes must lie within distance $\Theta(r)$ of a clause being modified in the dependency graph. Thus, for a single update, at most $d^{\Theta(r)}$ clauses can be affected at stage $r$, where $d$ is the degree of the dependency graph. This locality also relates consecutive iterations, and we can use changes at earlier stages to guide efficient repair at later stages.
    \item Second, thanks to the witness-tree analysis given in \cite{chung2014distributed}, which is originally used to give the round complexity of the static algorithm, we are able to show a rapidly decreasing probability that any particular clause needs attention at a late stage during the repair. Specifically, a clause needs attention in iteration $r$ with probability roughly $(pd^2)^{\Theta(r)}$, where $p$ bounds the maximum probability of a clause being falsified. This argument is independent of the locality property above, so we can apply it to all $d^{\Theta(r)}$ clauses in the local neighborhood.
\end{itemize}

From the properties, we are able to show that the expected repair work is very small when the slack is large, which is indeed the case for our edge-coloring applications. The main bottleneck becomes certifying that no repair is needed. Careful maintenance can remove the overhead of maintaining the dependency graph and other auxiliary information, leaving one dominant cost: checking whether a clause is falsified under the initial random assignment after a clause update. A naive way to implement this check takes time polynomial in the LLL parameter, which is why the algorithm in \Cref{sec:edge-coloring-poly-delta} is useful only in the small-degree regime. We later combine it with the degree-splitting procedures in \Cref{sec:edge-coloring-log-delta}, where bad events can be maintained much more efficiently.

\subsection{$(1+\epsilon)\Delta$ Edge Coloring: Small Degree Regime}

In \Cref{sec:edge-coloring-poly-delta}, we obtain a dynamic $(1+\epsilon)\Delta$ edge coloring with $\poly(\Delta)$ expected update time when $\Delta = \tilde \Omega(\epsilon^{-2})$. This algorithm will be used when a graph has maximum degree $\poly(1/\epsilon)$.

\paragraph{The Static Algorithm and its Black-Box Maintenance} The dynamic algorithm maintains the transcript of the static edge-coloring algorithm by Chang et al.~\cite{chang2017complexity}. The static algorithm consists of two phases, and the first phase contains $O(\log(1/\epsilon))$ partial-coloring iterations. In each iteration, edges that have not received a color independently propose a color from their palettes, and an edge gets colored if no adjacent edge proposes the same color. An LLL instance for each iteration ensures the invariants needed for the next iteration. The second phase will do a trivial cleanup, and it is easy to maintain in the dynamic setting in $O(\Delta)$ time per update.

In \Cref{sec:edge-coloring-poly-delta-first}, we describe our first attempt, using the dynamic LLL algorithm described above to maintain the satisfying assignment to the LLL instances in the first phase. In these instances, to check whether a bad event happens, one need to collect a constant-hop neighborhood and perform a computation with $\mathrm{poly}(\Delta)$ time. As a result, each LLL instance takes $\mathrm{poly}(\Delta)$ time per update. 

We also need to consider how a single update propagates in these iterations. In the worst case, a single edge insertion or deletion in iteration $i$ creates constant many edge insertions or deletions for iteration $i+1$ along with $O(\Delta)$ palette changes. Although each palette update inserts or removes only one color from the palette of an edge, it changes the possible values and distribution of the corresponding variable. Treating every such change as a variable replacement introduces a polynomial blowup in the number of updates at every iteration, resulting $\Delta^{O(\log 1/\epsilon)}$ update time. 

\paragraph{The Improvement} In \Cref{sec:edge-coloring-poly-delta-second}, we describe how we improve the $\Delta^{O(\log 1/\epsilon)}$ algorithm above and get $\mathrm{poly}(\Delta)$ update time. The key observation for the improvement is that every edge has a palette of size $\Omega(\Delta)$, while a typical palette update changes only one color. Intuitively speaking, each palette change should only affect the LLL solution with probability $O(1/\Delta)$.

To make this intuition formal, we still represent a palette update by replacing the corresponding variable, but we couple their first randomness in the LLL algorithm. This allows us to argue that when the palette of an edge $e$ changes from $\Psi(e)$ to $\Psi'(e)$, with probability $1 - O(|\Psi(e) \oplus \Psi'(e)| / \Delta)$, where $\oplus$ denotes symmetric difference, the palette update on $e$ does not affect the transcript of the dynamic LLL algorithm.

Consequently, for every update in iteration $i$, among $O(\Delta)$ palette changes in iteration $i+1$ generated from this update, only constant of them in expectation has an effect to the next iteration. As a result, the expected number of meaningful updates grows by only a constant factor between consecutive iterations, giving a total expected number of updates polynomial in $1/\epsilon$ and total update time $\mathrm{poly}(\Delta)$.

\subsection{$(1+\epsilon)\Delta$ Edge Coloring: Iterative Degree Splitting}

In \Cref{sec:edge-coloring-log-delta}, we describe dynamic algorithms to maintain $(1+\epsilon)\Delta$ edge coloring when $\Delta \ge \Delta_\epsilon$ in time $O(\log \Delta) + \poly(1/\epsilon)$. The main idea is to partition the input graph into several subgraphs of degree $\mathrm{poly}(1/\epsilon)$, and then apply the  small-degree algorithm in \Cref{sec:edge-coloring-poly-delta} on each of them with distinct palette.

\paragraph{Dynamic Degree Splitting}

In the algorithm, we wish to maintain an edge bipartition for an graph $G$ of maximum degree $\Delta$ such that the degree discrepancy for every vertex is small. In the static setting, we can achieve every small discrepancy in linear time by computing Euler tours, but this is not a good choice for the dynamic setting.

Instead, we relax the discrepancy requirement to be at most $\Delta/\mathrm{poly}(\log \Delta)$. A concentration bound implies that a random bipartition satisfies the requirement for a single vertex with probability $\exp(-\tilde \Omega(\Delta))$. Based on this, we can formulate an LLL instance with large slack to find a good random bipartition, and use the dynamic LLL algorithm to maintain a satisfying assignment. 

A good property for this LLL instance, compared to that in the low-degree case, is that one can update in constant time whether a bad event occurs under the initial random assignment for each edge insertion or deletion. Our dynamic LLL algorithm can therefore maintain such bipartition in constant update time.

We apply this bipartition recursively for $R = O(\log \Delta)$ times until the maximum degree of every resulting subgraph reaches $\exp(\poly(1/\epsilon))$, and a final partition reduces the degree further to $\poly(1/\epsilon)$. Each time, the degree upper bound is slightly larger than an exact halving. We choose the factor carefully so that these losses accumulate to only a $1+O(\epsilon)$ factor over the entire recursion. We then apply the algorithm in \Cref{sec:edge-coloring-poly-delta} to every final subgraph. The total palette will then be $(1+O(\epsilon))\Delta$.

The large LLL slack ensures that the recourse at any level is extremely small. As a result, the typical execution for an update is that, we apply it to some LLL instance for every level, and each of them reports no change in its maintained LLL solution. The bottleneck of the algorithm is to visit all $O(\log \Delta)$ levels to certify that no repair is needed and apply the update to the corresponding final low-degree instance, giving total expected update time $O(\log \Delta)+\poly(1/\epsilon)$.

\paragraph{An Alternative Approach Based on Graph Shattering}

In order to get constant-time algorithm, we introduce an alternative approach in~\cite{halldorsson2022fast} for degree bipartition, based on graph shattering. 

The algorithm has two phases. In the pre-shattering phase, edges propose independent random directions. A local procedure detects problematic regions and marks some edges as bad. All unmarked edges keep their original random directions. In the post-shattering phase, each connected component of bad edges is bipartitioned deterministically with small discrepancy.

An important property for this algorithm is that only the local pre-shattering phase decides which edges must deviate from their initial random directions. The post-shattering phase may inspect an entire bad component, but it does not cause new edges outside that component to become bad. This will provide useful locality to build our constant-time algorithm, while the general LLL algorithm in \Cref{sec:dynamic-lll} does not have this property.

\subsection{$(1+\epsilon)\Delta$ Edge Coloring: Bottom-Up Degree Splitting}

The work bottleneck for the algorithm above is visiting all $O(\log \Delta)$ levels and certifying that their very rare bad events do not occur, rather than fixing bad events. A simple lazy implementation does not avoid this cost: In the top-down algorithm, determining which subgraph at a given level contains a newly inserted edge requires knowing the decisions made at all earlier levels. Thus, the algorithm appears to have to process the levels in order before it even knows where the edge will end up.

To obtain constant update time, we observe that only the final partition is passed to the small-degree coloring algorithm. Correctness requires the degree bound to hold only in this final partition. Therefore, we do not need to repair every bad event at every intermediate level; we only need to fix the failures that can affect the degree of a final subgraph.

At a high level, based on the graph-shattering algorithm in \Cref{sec:edge-coloring-log-delta}, the algorithm in \Cref{sec:edge-coloring-const} samples all randomness for an edge at once. This specify in advance the final subgraph that the edge would reach if no repair were necessary. We can therefore apply an edge update directly to that final subgraph in constant time. We then perform a bottom-up local search for the bad events responsible for a possible degree overflow there. The search keeps only enough bad events to recover the final degree guarantee and ignores intermediate failures that cannot affect it. We then repair exactly the part of the execution exposed by this search based on a modification of the algorithm in \Cref{sec:edge-coloring-log-delta}.

To avoid visiting every level, we store the execution explicitly only near detected failures. Everywhere else, the execution agrees with the initially sampled directions and can be represented implicitly. Updating one final subgraph therefore updates the implicit information at all intermediate levels without visiting them one by one.

\paragraph{Multi-Level Shattering Analysis} The amount of repair work done by the dynamic algorithm is related to sizes of the ``bad components'' affected by the update. The usual shattering analysis says that every bad component has logarithmic size with high probability. Because our LLL instances have large slack, with the same proof, the expected bad-component size around any fixed vertex is extremely small. In the top-down algorithm, each level has independent randomness and we can conclude that the total work and recourse is small.

The same argument does not apply to the bottom-up algorithm. Its local search reveals information about the random choices at several levels, so the components encountered at different levels can no longer be treated as independent. We therefore analyze all levels jointly in \Cref{sec:edge-coloring-const,sec:witness}.

We begin by identifying the \emph{critical failures} at which the ideal random execution first goes wrong. A critical failure is extremely unlikely: the execution at the preceding level is still well behaved, leaving enough slack for a strong concentration bound at the current level.

We draw a ball of radius $\mathrm{poly}(\log \Delta_i)$ for a critical failure of level $i$ inside the subgraph where the failure occurs. $\Delta_i$ denotes the degree upper bound for level $i$. The ball will roughly contain $\exp(\poly(\log \Delta_i))$ vertices. Because the pre-shattering procedure is local, whenever an edge deviates from its initial random direction, we can trace this back to find a critical failure whose ball contains the edge. Viewing the iterative bipartition as a execution tree, the collection of balls in the ancestors of a node contain everything that may need to be updated or recomputed at that node.

To measure how these balls interact, we form an auxiliary graph $\hat G$ of the balls and show in \Cref{sec:edge-coloring-const} that the cost of the local search for essential bad events and the final repair is small if we can bound the component size in $\hat G$.

In \Cref{sec:witness}, we bound the expected size of this component through joint witness. Each witness, contains the shortest path in $\hat G$ between two balls and a low-degree local structure for each ball in the path to certify that these vertices and edges are indeed in $\hat G$. Counting the expected component size then reduces to counting the expected number of witness.

The remaining difficulty is that a witness can span several levels. We handle this by exposing the randomness one level at a time, from top to bottom. At each step, we bound the conditional probability that a partially verified witness survives the next level. Multiplying these bounds gives the probability bound for a complete witness. Finally, we sum this probability over all possible witness structures. The probability decreases faster than the number of structures grows, proving that the expected component size, and hence the expected update work, is small.

Finally, our technique largely relies on the fact that the witness is a constant-degree structure, and we expect that this can be generalized and only be generalized to bound any constant moment of the connected component size distribution.

%% file: dynamic-LLL.tex
\section{Dynamic Lov\'asz Local Lemma}
\label{sec:dynamic-lll}

In this section, we give a dynamic algorithm that maintains a satisfying assignment for a Lov\'asz Local Lemma instance under clause modifications. 

\subsection{Model and Main Result}

There is a set $V$ of mutually independent random variables and a fixed set $\mathcal{C}$ of $m$ clauses, each with a unique identifier in $[m]$. Every variable $x\in V$ has a domain $\Omega_x$ and a distribution $\mathcal{D}_x$ over $\Omega_x$. Write $\vbl(C)$ for the variables on which $C$ depends and $\mathcal{C}(x)=\{C:x\in\vbl(C)\}$. Initially, $V=\varnothing$ and every clause is empty.

For $C\in\mathcal{C}$ and $U\subseteq V$, a function $f_{C,U}:\prod_{x\in U}\Omega_x\to\{0,1\}$ specifies the truth table of $C$ when $\vbl(C)=U$; in our applications, this family of truth tables is represented implicitly. By convention, a clause with no variables is satisfied, so the initial instance is satisfiable. For a fixed instance, let $\Pr[C]$ denote the probability that $C$ is \emph{falsified} under the product distribution of its variables.

The \emph{dependency graph} $H$ of the instance has one vertex for each clause and an edge between two clauses when they share a variable. For a graph $H$, a vertex $v$, and $r \ge 1$, we define $\Gamma_H^r(v)$ to contain every vertex, including $v$, within distance at most $r$ from $v$, and define $\Gamma_H(v)$ to be $\Gamma_H^1(v)$. We use the following parameters as uniform upper bounds throughout the update sequence:
\begin{center}
\begin{tabular}{ll}
\hline
Parameter & Upper bound on \\
\hline
$k$ & Maximum number of clauses that a variable is in \\
$s$ & Maximum number of variables that a clause depends on \\
$d$ & $\max_{C\in\mathcal{C}}|\Gamma_H(C)|$ \\
$T_{\sample}$ & Expected time to sample from any $\mathcal{D}_x$ \\
$T_{\eval}$ & Expected time to evaluate a clause on a given assignment \\
\hline
\end{tabular}
\end{center}
The clauses containing a fixed variable form a clique in $H$, so we may take $k\le d$. We take $k,s\ge1$ and $d\ge2$, as otherwise the instance is trivial.

We assume two additional properties. First, each clause has a data structure that maintains its value under insertions and deletions of variables in $T_{\update}$ expected time per operation. For variable insertion, its value is also given. Reevaluating the clause after each operation gives a trivial implementation, so we may take $T_{\update}\le T_{\eval}$.
Second, there is a uniform parameter $q\in(0,1/2)$ such that, for every instance, some function $x:\mathcal{C}\to[0,1)$ satisfies $\max_C x(C)/(1-x(C))=O(1)$ and
\begin{equation}\label{eq:dynamic-lll-slack}
    \Pr[A]\le\frac{q}{4d^2}
        \left(x(A)\prod_{B\in\Gamma_H^2(A)}(1-x(B))\right)
\end{equation}
for every $A\in\mathcal{C}$. In the typical symmetric LLL setting, we will pick $x$ to be the constant function $x(A)=1/d^2$, and $\Pr[A]\le q/(4ed^4)$ for every event $A$ suffices. The restriction $q<1/2$ gives $\sum_{r\ge1}q^r=O(q)$ and $\sum_{r\ge1}rq^r=O(q)$.

\begin{theorem}\label{thm:dynamic-lll}
Under these assumptions, there is a dynamic algorithm that maintains a satisfying assignment against an oblivious adversary. It supports the following updates:
\begin{itemize}
    \item Introduce an isolated variable $x$ in $O(T_{\sample})$ expected time.
    \item Remove an isolated variable $x$ in $O(1)$ time.
    \item Apply a batch $\mathcal{B}=\{(C_i,x_i):i\in[|\mathcal{B}|]\}$ of distinct clause updates. Each update $(C_i,x_i)$ adds $x_i \in V$ to $\vbl(C_i)$ when $x_i \notin \vbl(C_i)$ and removes $x_i$ otherwise. The expected update time is
    \[
        O\!\left(|\mathcal{B}|\left(
            T_{\update}
            +k^2s^2dq\bigl(T_{\eval}+T_{\sample}\bigr)
        \right)\right),
    \]
    and the expected recourse, defined as the number of value changes in the final output, is 
    \[
        O(|\mathcal{B}|s^2q).
    \]
\end{itemize}
\end{theorem}

In our applications, $1/q$ will be larger than any polynomial in the other parameters. The expected time per update is then dominated by $T_{\update}$, and the expected recourse is correspondingly small.

\subsection{The Static Algorithm}

We first review the static algorithm of Chung, Pettie, and Su~\cite{chung2014distributed} and introduce notations. Independently for every variable $x$, sample a \emph{randomness table} $(R(x,i))_{i\ge0}$ of independent samples from $\mathcal{D}_x$. The initial assignment is $\sigma_0(x)=R(x,0)$. Resampling a variable advances its value to the next entry of its table.

The static algorithm consists of stages. At stage $r\ge1$, the algorithm evaluates every clause under $\sigma_{r-1}$ and lets $b_{C,r}$ denote whether $C \in \mathcal{C}$ is falsified. After that, we take an independent set of falsified clauses in $H$ by including every ID local minimum. Specifically, define
\begin{equation}\label{eq:LLL-dynamic-def-cnt}
    \cnt_{C,r}
    =b_{C,r}\sum_{C'\in\Gamma_H(C)}
    [\mathrm{ID}(C')<\mathrm{ID}(C),\ b_{C',r}=1]
\end{equation}
and set $I_r=\{C:b_{C,r}=1,\ \cnt_{C,r}=0\}$. The set $I_r$ is independent in $H$, so each variable belongs to at most one clause in $I_r$. Resampling all variables in these clauses produces $\sigma_r$. The execution stops when every clause is satisfied, and we extend the execution by infinitely many empty stages so that two transcripts remain aligned even if they terminate at different stages.

The following table summarizes the transcript of the static algorithm and the additional quantities that will be maintained by the dynamic algorithm.
\begin{table}[htbp]
\centering
\begin{tabular}{ll}
\hline
Quantity & Meaning \\
\hline
$b_{C,r}$ & Whether $C$ is falsified by $\sigma_{r-1}$ \\
$\cnt_{C,r}$ & The counter in \Cref{eq:LLL-dynamic-def-cnt} \\
$I_r=\{C:b_{C,r}=1, \cnt_{C,r}=0\}$ & Clauses selected for resampling at stage $r$ \\ \hline
$b_{x,r}=[I_r\cap\mathcal{C}(x)\ne\varnothing]$ & Whether $x$ is resampled at stage $r$ \\
$s_x=\sum_{r\ge1}b_{x,r}$ & Total number of resamplings of $x$ \\
$r_x=\max(\{0\}\cup\{r:b_{x,r}=1\})$ & Last stage at which $x$ is resampled \\
$\ell_C=\max(\{0\}\cup\{r_x:x\in\vbl(C)\})$ & Last stage affecting a variable of $C$ \\
\hline
\end{tabular}
\caption{Data in an LLL transcript}
\label{tab:static-transcript}
\end{table}

It is worth noting that the bits $b_{x,r}$ for $r \ge 1$ implicitly define the assignment $\sigma_i(x) = R(x,\sum_{j=1}^i b_{x,j})$ for $i \ge 0$. Similarly, the final output assignment $\sigma_\infty$ assigns $x$ the value $R(x,s_x)$.

In the following, we review the probabilistic statement of Chung, Pettie, and Su~\cite{chung2014distributed} based on witness trees. For an execution, its \emph{resampling log} $L$ concatenates $I_1,I_2,\ldots$, ordering each set by increasing clause ID.
\begin{definition}[$2$-witness tree]
    For a resampling event $L_i$ in an execution, the \emph{$2$-witness tree} for $L_i$ is a rooted tree $T$ with a labeling $f$ generated in the following way: Initially, $T$ consists of a single root with label $L_i$. Iterate $j$ downward from $i-1$ to $1$. If $L_j\in\Gamma_H^2(f(x))$ for some $x\in V(T)$, create a node with label $L_j$ and make it a child of any deepest such node $x$.
\end{definition}

\begin{lemma}[\cite{chung2014distributed}, Lemma 3.6]\label{lem:resampling-round-to-witness-tree-size}
If $C\in I_r$, then the execution contains a $2$-witness tree with root label $C$ and size at least $r$.
\end{lemma}
\begin{lemma}[\cite{chung2014distributed}, Appendix B]\label{lem:witness-tree-prob}
For every clause $C$ and $r\ge1$, the probability that the execution contains a $2$-witness tree with root label $C$ and at least $r$ nodes is at most $(q/4d^2)^r x(C) / (1-x(C))$.
\end{lemma}

From \Cref{lem:resampling-round-to-witness-tree-size,lem:witness-tree-prob} and a union bound over the root labels, the probability that $\sigma_t$ does not satisfy every clause is at most $(q/4d^2)^{t+1}\sum_{C\in\mathcal{C}}x(C) / (1-x(C)).$ Thus the algorithm terminates almost surely, and within $O(\log(m+1))$ stages with high probability.

In the following we establish some useful consequences of the previous lemmas.

\begin{lemma}\label{lem:dynamic-lll-r-value-tail}
For every variable $x$ and $r\ge1$,
$\Pr[b_{x,r}=1]\le\Pr[r_x\ge r]=O(k(q/4d^2)^r)$.
\end{lemma}
\begin{proof}
If $r_x\ge r$, some clause $C\in\mathcal{C}(x)$ is resampled at a stage at least $r$. The corresponding witness tree has at least $r$ nodes by \Cref{lem:resampling-round-to-witness-tree-size}. Applying \Cref{lem:witness-tree-prob} and taking a union bound over at most $k$ choices of $C$ gives the desired bound. The first inequality follows from that $b_{x,r}=1$ implies $r_x\ge r$.
\end{proof}
\begin{corollary}\label{cor:dynamic-lll-falsifiability-value-tail}
For every clause $C$ and $r\ge1$,
$\Pr[b_{C,r}=1]\le\Pr[\ell_C\ge r]=O(ks(q/4d^2)^r)$.
\end{corollary}
\begin{proof}
The event $\ell_C\ge r$ requires $r_x\ge r$ for some $x\in\vbl(C)$, so its probability follows from \Cref{lem:dynamic-lll-r-value-tail} and a union bound over at most $s$ variables. If $b_{C,r}=1$, the clause is falsified by $\sigma_{r-1}$ but satisfied by the final assignment. Some variable in $C$ must therefore be resampled at a stage at least $r$, implying $\ell_C\ge r$.
\end{proof}

\subsection{Maintaining the Transcript}

Insertions and deletions of isolated variables will be handled trivially, so let us fix a batch $\mathcal{B}$ of clause updates. Let $\mathcal{T}^\old$ and $\mathcal{T}^\new$ be the static transcripts of the instances before and after the batch, with the same randomness for each variable. With a plausible initialization for the empty instance, it therefore suffices for the dynamic algorithm to transform the old transcript into the new one after each batch.

For a transcript item $X$, write $X^\alpha$ for its value in $\mathcal{T}^\alpha$, where $\alpha\in\{\old,\new\}$. If $X^\old$ equals $X^\new$, we will simply use $X$ to denote its value in both $\mathcal{T}^\old$ and $\mathcal{T}^\new$.

Define $H^\cup=H^\old\cup H^\new$ and $\mathcal{M}=\{C_i:i\in[|\mathcal{B}|]\}$. Since all information in \Cref{tab:static-transcript} depends on $b_{C,r}$ and $\cnt_{C,r}$, we will pay attention to \emph{dirty} pairs $(C,r)$ such that $(b_{C,r}^\old, \cnt_{C,r}^\old) \ne (b_{C,r}^\new, \cnt_{C,r}^\new)$. In the following, we establish some properties of dirty pairs.

\begin{observation}\label{obs:dynamic-lll-prob-dirty}
    For any $C \in \mathcal{C}$ and $r \ge 1$, $\Pr[(C,r)\text{\ is\ dirty}] = O(ks(q/4d^2)^r)$.
\end{observation}
\begin{proof}
    $(C,r)$ is dirty only when $b_{C,r}^\old = 1$ or $b_{C,r}^\new = 1$ since $\cnt_{C,r} = 0$ whenever $b_{C,r}=0$, and the statement holds by applying \Cref{cor:dynamic-lll-falsifiability-value-tail} separately to $\mathcal{T}^\old$ and $\mathcal{T}^\new$. 

    Notice that, although the randomness in two transcripts are coupled, the randomness within each transcript is mutually independent and follows the prescribed distributions. The oblivious adversary fixes both instances independently of the randomness, so we can apply \Cref{lem:dynamic-lll-r-value-tail,cor:dynamic-lll-falsifiability-value-tail} to each of them.
\end{proof}

\begin{lemma}\label{lem:dynamic-lll-dirty-pair-locality}
Every dirty pair $(C,r)$ satisfies $\operatorname{dist}_{H^\cup}(C,\mathcal{M})\le2r-1$.
\end{lemma}
\begin{proof}
We induct on $r$. It suffices to consider dirty pairs $(C,r)$ with $\mathrm{dist}_{H^\cup}(C, \mathcal{M}) \ge 2$. This means that, $\Gamma_{H}(C)$ and $\mathrm{vbl}(C')$ for every clause $C' \in \Gamma_H(C)$ do not change under the update. In this case, $\sigma_0^\old=\sigma_0^\new$, so $(C,1)$ cannot be dirty, and this gives the base case.

For $r\ge2$, a dirty pair requires some $C'\in\Gamma_{H^\cup}(C)$ with $b_{C',r}^\old\ne b_{C',r}^\new$. Since $\mathrm{vbl}(C')$ is unchanged, there must exist some $x\in\vbl(C')$ with different values under $\sigma_{r-1}^\old$ and $\sigma_{r-1}^\new$. As $\sigma_{r-1}(x)$ is defined to be $R(x,\sum_{j=1}^{r-1} b_{x,j})$ in both transcripts and the randomness for two transcripts are the same, $b_{x,j}^\old\ne b_{x,j}^\new$ for some $1\le j<r$.

If $\mathcal{C}^\old(x)\ne\mathcal{C}^\new(x)$, a clause $C''$ in their symmetric difference belongs to $\mathcal{M}$ and is within one hop of $C'$. Thus $C$ is within two hops of $\mathcal{M}$. Otherwise, some $C''\in\mathcal{C}(x)$ enters or leaves $I_j$, so $(C'',j)$ is dirty. By the induction hypothesis, $C''$ is within $2j-1$ hops of $\mathcal{M}$. Since $\operatorname{dist}_{H^\cup}(C,C')\le1$ and $\operatorname{dist}_{H^\cup}(C',C'')\le1$, the distance between $C$ and $\mathcal{M}$ is at most $2j+1\le2r-1$.
\end{proof}

\begin{corollary}\label{cor:dynamic-lll-num-dirty-pair}
For each $r\ge1$, the expected number of dirty pairs $(C,r)$ is $O(sq^r|\mathcal{B}|)$.
\end{corollary}
\begin{proof}
The graph $H^\cup$ has maximum degree at most $2d$. By \Cref{lem:dynamic-lll-dirty-pair-locality}, there are at most $|\mathcal{B}|(2d)^{2r-1}$ possible clauses $C$ with $(C,r)$ dirty. Each of these pairs is dirty with probability $O(ks(q/4d^2)^r)$ according to \Cref{obs:dynamic-lll-prob-dirty}. As a result, the expected number of dirty pairs is $O((ks/d)q^r|\mathcal{B}|)=O(sq^r|\mathcal{B}|)$ because $k\le d$.
\end{proof}

\begin{corollary}\label{cor:dynamic-lll-recourse}
Let $Z_r=\{x:b_{x,r}^\old\ne b_{x,r}^\new\}$. For every $r\ge1$, $\mathbb{E}[|Z_r|]=O(s^2q^r|\mathcal{B}|)$. Consequently, the expected recourse of the dynamic algorithm is $O(s^2q|\mathcal{B}|)$.
\end{corollary}
\begin{proof}
There are two possibilities where $b_{x,r}$ changes after the update:

\begin{itemize}
    \item Either a clause $C$ enters or leaves $I_r$, meaning that $(C,r)$ is dirty. Each dirty pair $(C,r)$ changes $b_{x,r}$ for at most $2s$ variables $x$. By \Cref{cor:dynamic-lll-num-dirty-pair}, this case accounts for $O(s^2q^r|\mathcal{B}|)$ variables in $Z_r$.
    \item Or, $x$ enters or leaves $\vbl(C)$ for some $C \in I_r^\old \cap I_r^\new$, meaning that $C$ belongs to $\mathcal{M}\cap I_r^\old$. For each $C\in\mathcal{M}$, $b_{C,r}^\old=1$ happens with probability $O(ks(q/4d^2)^r)=O(sq^r)$ by \Cref{cor:dynamic-lll-falsifiability-value-tail} and $k\le d$. Each such clause again affects at most $2s$ variables, so this case accounts for $O(s^2q^r|\mathcal{B}|)$ variables in $Z_r$.
\end{itemize}

This shows that $\mathbb{E}[|Z_r|] = O(s^2q^r|\mathcal{B}|)$. The algorithm shifts its output from $\sigma_\infty^\old(x)$ to $\sigma_\infty^\new(x)$. Every variable $x$ with $\sigma_\infty^\old(x) \ne \sigma_\infty^\new(x)$ belongs to $\cup_{r \ge 1}Z_r$, and summing $\mathbb{E}[|Z_r|]$ over $r\ge1$ proves the final claim.
\end{proof}

Our algorithm will find dirty pairs and construct $\mathcal{T}^\new$ stage by stage. At the start of stage $r\ge1$, the algorithm will hold a \emph{candidate set} $\mathcal{C}_r$ containing every clause $C$ for which $(C,r)$ is dirty, and also know $(b_{C,r}^\new,\cnt_{C,r}^\new)$ for every clause $C$. The initialization implemented below will construct $\mathcal{C}_1$ and compute the information for stage one. It also makes available
\[
    \ell_C^\star=
    \max\!\left(\{0\}\cup
    \{r_x^\old:x\in\vbl^\old(C)\cup\vbl^\new(C)\}\right).
\]
Stage $r$ consists of the following steps:
\begin{enumerate}
    \item For each $C\in\mathcal{C}_r$, update membership in $I_r$ using $(b_{C,r}^\new,\cnt_{C,r}^\new)$. Every $C \in I_r^\old\oplus I_r^\new$ forms a dirty pair $(C,r)$, so this recovers $I_r^\new$.
    \item For every clause $C$ in the union of $I_r^\old\oplus I_r^\new$ and $\mathcal{M}\cap I_r^\old\cap I_r^\new$, update $b_{x,r}$ for $x \in \vbl^\old(C) \cup \vbl^\new(C)$ as appropriate. This will recover $b_{x,r}^\new$ and identify $Z_r$, since $b_{x,r}^\new \ne b_{x,r}^\old$ either due to a clause entering or leaving $I_r$, or due to $x$ entering or leaving $\vbl(C)$ for some $C \in I_r^\old \cap I_r^\new$.
    \item Update $s_x$, $r_x$ and $\sigma_\infty(x)$ for every variable $x\in Z_r$.
    \item Construct $\bar{\mathcal{C}}_{r+1}$ from all clauses satisfying at least one of the following conditions:
    \begin{itemize}
        \item[(1)] $\vbl^\new(C)\cap Z_r\ne\varnothing$;
        \item[(2)] $C\in\mathcal{C}_r\cup\mathcal{M}$ and $\ell_C^\star\ge r$;
        \item[(3)] $C\in\mathcal{C}_r$ and $\cnt_{C,r}^\new>0$.
    \end{itemize}
    \item Set $\mathcal{C}_{r+1} = \Gamma_{H^\cup}(\bar{\mathcal{C}}_{r+1})$. We compute $(b_{C,r+1}^\new,\cnt_{C,r+1}^\new)$ for every clause $C \in \mathcal{C}_{r+1}$, and continue to the next stage if $\mathcal{C}_{r+1} \ne \varnothing$.
\end{enumerate}
Notice that stage one will be executed even when $\mathcal{C}_1$ is empty. After this process ends, we output the final assignment $\sigma_\infty^\new$ and compute $\ell_C^\new$ according to its definition in \cref{tab:static-transcript}.

In the following, we argue the correctness of the algorithm.

\begin{lemma}\label{lem:dynamic-lll-correctness}
Before stage $r\ge1$ in the algorithm, the following invariants hold:
\begin{enumerate}
    \item The algorithm has recovered $I_j^\new$ and $b_{x,j}^\new$ for $1\le j<r$ and also $\sigma_j^\new$ for $0 \le j < r$;
    \item The set $\mathcal{C}_r$ contains every clause $C$ for which $(C,r)$ is dirty;
    \item The algorithm has recovered $(b_{C,j}^\new, \cnt_{C,j}^\new)$ for every clause $C$ and $1 \le j \le r$.
\end{enumerate}
\end{lemma}
\begin{proof}
We prove the invariants by induction. At stage one, invariant 1 holds trivially, while invariants 2 and 3 hold due to the initialization discussed below.

Suppose the invariants hold before stage $r$. Steps 1 and 2 in stage $r$ recover $I_r^\new$ and $b_{x,r}^\new$ for $x \in V$. Since $\sigma_r(x) = R(x,\sum_{j=1}^r b_{x,j})$, this also shows that the algorithm constructs $\sigma_r^\new$, and invariant 1 holds.

To show invariant 2, define $\mathcal{A}_{r+1}$ to include every clause $C$ satisfying either $b_{C,r+1}^\old \ne b_{C,r+1}^\new$, or $b_{C,r+1}^\new = 1$ and $C \in \mathcal{M}$. We claim that $\mathcal{A}_{r+1} \subseteq \bar{\mathcal C}_{r+1}$. Fix $C \in \mathcal{A}_{r+1}$ and consider two cases:
\begin{itemize}
    \item $\ell_C^\star\ge r$. If $C\in\mathcal{M}$, condition~(2) includes it. Otherwise, its variable set does not change, but $b_{C,r+1}^\old \ne b_{C,r+1}^\new$. This means some $x \in \vbl(C)$ satisfies $\sigma_r^\old(x) \ne \sigma_r^\new(x)$, and $b_{x,j}^\old \ne b_{x,j}^\new$ for some $j\le r$. Condition~(1) puts $C$ in $\bar{\mathcal{C}}_{j+1}\subseteq\mathcal{C}_{j+1}$. Condition (2) then adds $C$ to $\bar{\mathcal C}_{k}$ and $\mathcal{C}_k$ for every $k \in [j+1,r + 1]$.
    \item $\ell_C^\star < r$. $C$ is satisfied in both $\sigma_{r-1}^\old$ and $\sigma_r^\old$, and $b_{x,r}^\old = 0$ for every $x \in \vbl^\new(C)$. As a result, $C \in \mathcal{A}_{r+1}$ implies $b_{C,r+1}^\new=1$.
    
    If $b_{x,r}^\new = 1$ for some $x \in \vbl^\new(C)$, condition~(1) includes $C$. Otherwise, $C$ has the same assignment under $\sigma_r^\new$ and $\sigma_{r-1}^\new$, hence $b_{C,r}^\new=1$ and $(C,r)$ is dirty. Invariant~2 in the induction hypothesis gives $C\in\mathcal{C}_r$. As $C$ is falsified by $\sigma_{r-1}^\new$ but does not belong to $I_r^\new$, $\cnt_{C,r}^\new > 0$, and condition~(3) includes it in $\bar{\mathcal C}_{r+1}$.
\end{itemize}

Finally, there are two ways for $(C,r+1)$ to be dirty:

\begin{itemize}
    \item $\Gamma_{H^\cup}(C)$ contain a clause $C'$ with $b_{C',r+1}^\old \ne b_{C',r+1}^\new$, where $C' \in \mathcal{A}_{r+1} \subseteq \bar {\mathcal{C}}_{r+1}$ and $C \in \mathcal{C}_{r+1}$.
    \item $b_{C,r+1}^\old = b_{C,r+1}^\new = 1$, but there is a clause $C'$ with $(C,C') \in H^\old \oplus H^\new$ and $b_{C',r+1}^\old = b_{C',r+1}^\new = 1$. In this case, either $C$ or $C'$ belongs to $\mathcal{M}$, and it belongs to $\mathcal{A}_{r+1}$, so $C \in \mathcal{C}_{r+1}$.
\end{itemize}

This proves invariant 2 for stage $r+1$, and invariant 3 follows immediately from invariant 2 and step 5.
\end{proof}

\begin{corollary}\label{cor:dynamic-lll-correctness-after-stopping}
    When the algorithm terminates, the invariants in \cref{lem:dynamic-lll-correctness} hold for every $r \ge 1$ and the algorithm reconstructs $\mathcal{T}^\new$.
\end{corollary}
\begin{proof}
    From \cref{lem:dynamic-lll-correctness}, it suffices to consider the stopping condition. Suppose $\mathcal{C}_{r+1}=\varnothing$ for some $r \ge 1$. Then $\bar{\mathcal{C}}_{r+1}=\varnothing$, and condition~(2) implies $\ell_C^\star<r$ for every $C\in\mathcal{M}$. There are no dirty pairs at stage $r+1$, so $I_{r+1}^\old = I_{r+1}^\new$. Additionally, there is no clause $C\in\mathcal{M}$ with $\ell_C^\star \ge r+1$, so steps~1--3 of stage $r+1$ do nothing, and $b_{x,r+1}^\old = b_{x,r+1}^\new$. This also shows that conditions~(1)--(3) in step 4 of stage $r+1$ produce no clause in $\bar{\mathcal C}_{r+2}$, so $\mathcal{C}_{r+2} = \varnothing$ and all resampling bits and clause falsification bits in subsequent iterations coincide between $\mathcal{T}^\old$ and $\mathcal{T}^\new$.

In sum, all invariants hold for every $r \ge 1$ at the end. Since we update $s_x$, $r_x$, and $\sigma_{\infty}$ along the way according to their definitions and finally compute $\ell_C^\new$, the algorithm reconstructs $\mathcal{T}^\new$.
\end{proof}

\subsection{Implementation Details and Analysis}

\paragraph{Transcript storage.}
For each clause $C$, we store $\ell_C$ and sequences $b_{C,r}$, $\cnt_{C,r}$, and $[C\in I_r]$; for each variable $x$, we store its randomness table, $s_x$, $r_x$, and the sequence $b_{x,r}$. Each infinite-length sequence is represented by a linked list of its nonzero entries in increasing order of stage. Since everything is zero for an empty stage, the length of each linked list is finite almost surely. Accessing or locating index $r$ in each array now costs $O(r)$ rather than constant.

We use hash tables of size $s$ and $k$, respectively, to store $\vbl(C)$ and $\mathcal{C}(x)$, with chaining~\cite{knuth2005art} to resolve collisions. Each clause $C$ will lazily generate a random key $k_C \in [k]$, and each variable lazily generates $k_x \in [s]$. The expected number of collisions is $O(1)$ for each insertion or deletion, so each update takes $O(1)$ time. The buckets for $\mathcal{C}(x)$ are generated lazily when $x$ is inserted to avoid $O(k)$ time for variable insertion. Iterating through $\vbl(C)$ and $\mathcal{C}(x)$ takes $O(s)$ and $O(k)$ time, respectively, and we can iterate through $\Gamma_{H}(C)$ in $O(ks)$ time since $d \le ks$.

\paragraph{Initialization.}
During initialization, we first apply updates in $\mathcal{B}$ to get $\vbl^\new(C)$ and $\mathcal{C}^\new(x)$ for clauses $C$ and variables $x$ in $O(|\mathcal{B}|)$ expected time. We can also easily compute $\ell_C^\star$ for each clause $C$ along the way in $O(|\mathcal{B}|)$ time.

We maintain a data structure $A_C$ for each clause $C$ to evaluate it on $\sigma_0$. Since $\sigma_0^\old = \sigma_0^\new$, we only need to apply updates in $\mathcal{B}$ to $A_C$ in $O(|\mathcal{B}|T_\update)$ expected time to get $b_{C,1}^\new$ for $C \in \mathcal{M}$; clause $C \notin \mathcal{M}$ satisfies $b_{C,1}^\old = b_{C,1}^\new$.

There are several possibilities a clause $C$ satisfies $\cnt_{C,1}^\new \ne \cnt_{C,1}^\old$:

\begin{itemize}
    \item $b_{C,1}^\old \ne b_{C,1}^\new$, where $C \in \mathcal{M}$.
    \item $b_{C,1}^\old = b_{C,1}^\new = 1$ but $b_{C',1}^\old \ne b_{C',1}^\new$ for some $C' \in \Gamma_{H^\cup}(C)$, meaning that $C' \in \mathcal{M}$.
    \item $b_{C,1}^\old = b_{C,1}^\new = 1$, $b_{C',1}^\old = b_{C',1}^\new = 1$, but $(C,C')$ belongs to only one of $H^\old$ and $H^\new$, meaning that either $C$ or $C'$ belongs to $\mathcal{M}$.
\end{itemize}

In sum, if $\cnt_{C,1}^\new \ne \cnt_{C,1}^\old$, then there is $C' \in \Gamma_{H^\cup}(C)$ such that $b_{C',1}^\old$ or $b_{C',1}^\new$ is one and $C' \in \mathcal{M}$. We go through every $C' \in \mathcal{M}$ such that $b_{C',1}^\old$ or $b_{C',1}^\new$ is one, and then compute $\cnt_{C,1}^\new$ for every $C \in \Gamma_{H^\cup}(C')$.

By \Cref{eq:dynamic-lll-slack}, each of the events $b_{C',1}^\old = 1$ and $b_{C',1}^\new = 1$ has probability $O(q/d^2)$. Hence, in expectation, $O(q|\mathcal{M}|)$ clauses need to update their counters, and the expected work to perform these updates is $O(ksq|\mathcal{M}|)$. Finally, include in $\mathcal{C}_1$ every clause whose $(b,\cnt)$ pair changed. We have $\mathbb{E}[|\mathcal{C}_1|] = O(q|\mathcal{M}|)$.

For use in later stages, during initialization, we also construct $\mathcal{M}_r=\{C\in\mathcal{M}:\ell_C^\star\ge r\}$ for every $r \ge 1$ in $O(|\mathcal{M}|+\sum_{r\ge1}|\mathcal{M}_r|)$ time. Applying the same idea as in \Cref{cor:dynamic-lll-falsifiability-value-tail}, we have
\begin{equation}\label{eq:dynamic-lll-modified-history-tail}
    \mathbb{E}[|\mathcal{M}_r|]
    =O\!\left(ks\left(\frac{q}{4d^2}\right)^r |\mathcal{B}|\right)
    =O(sq^r |\mathcal{B}|).
\end{equation}
The total expected cost to construct $\mathcal{M}_r$ for $r \ge 1$ is $O(|\mathcal{B}|(1+sq))$, and the expected total initialization cost is therefore $O(|\mathcal{B}|(T_{\update}+ksq))$.

\paragraph{Implementing a stage.} Next, we discuss implementations for stage $r$.
\begin{itemize}
    \item Step~1 takes $O(r|\mathcal{C}_r|)$ time to access $r$-th index in an array for every $C \in \mathcal{C}_r$.
    \item To implement step~2, it suffices to inspect every clause $C \in \mathcal{C}_r\cup\mathcal{M}_r$ and update $b_{x,r}$ as appropriate for $x \in \vbl^\old(C) \cup \vbl^\new(C)$, since $\mathcal{M}\cap I_r^\old\subseteq\mathcal{M}_r$. Each clause affects $2s$ variables, and each variable contributes $O(r)$ work to access and update $b_{x,r}$, so the total work is $O(sr(|\mathcal{C}_r|+|\mathcal{M}_r|))$.
    \item Step 2 yields the variable set $Z_r = \{x: b_{x,r}^\old \ne b_{x,r}^\new\}$. In step 3, every variable $x \in Z_r$ updates $s_x$ and $r_x$ in $O(1)$ time, and updating $\sigma_\infty(x)$ takes $O(T_\sample)$ time since it may generate a new entry in its randomness table. So the total cost for step 3 is $O(T_\sample|Z_r|)$.
    \item In step~4, condition (1) requires every $x \in Z_r$ to add every clause in $\mathcal{C}^\new(x)$ to $\bar{\mathcal{C}}_{r+1}$, while conditions (2) and (3) require the algorithm to inspect every clause in $\mathcal{C}_r \cup \mathcal{M}_r$, so constructing $\bar{\mathcal C}_{r+1}$ costs $O(k|Z_r|+r|\mathcal{C}_r|+|\mathcal{M}_r|)$. 
    \item Constructing $\mathcal{C}_{r+1}$ in step~5 costs $O(ks|\bar{\mathcal{C}}_{r+1}|)$. For each clause in $\mathcal{C}_{r+1}$, recovering its assignment in $\sigma_r^\new$ incurs $O(rs)$ work, and evaluation takes $T_{\eval}$ expected time. The counters are computed in $O(ksr)$ time afterwards, so step~5 costs $O((T_{\eval}+ksr)|\mathcal{C}_{r+1}|)$ in expectation since $\bar{\mathcal C}_{r+1} \subseteq \mathcal{C}_{r+1}$.
\end{itemize}

Charge the work in step~5 of stage $r$ to stage $r+1$. By \Cref{cor:dynamic-lll-recourse,eq:dynamic-lll-modified-history-tail}, the expected work charged to stage $r$ is
\begin{equation}\label{eq:dynamic-lll-expected-work-per-stage}
    O\!\left(
        |\mathcal{B}|s^2q^r(k+r+T_{\sample})
        +(ksr+T_{\eval})\mathbb{E}[|\mathcal{C}_r|]
    \right).
\end{equation}
From the initialization, we know that $\mathbb{E}[|\mathcal{C}_1|] = O(q|\mathcal{B}|)$, and we will bound $\mathbb{E}[|\mathcal{C}_r|]$ for $r \ge 2$ in the following.
\begin{lemma}\label{lem:dynamic-lll-repair-locality}
For every $r\ge1$ and $C\in\bar{\mathcal{C}}_{r+1}$, $\operatorname{dist}_{H^\cup}(C,\mathcal{M})\le2r$ and $\max(\ell_C^\star,\ell_C^\new)\ge r.$
\end{lemma}
\begin{proof}
The distance claim is trivial if $C \in \mathcal{M}$, so we consider the case where $C \notin \mathcal{M}$, meaning that $\vbl^\old(C) = \vbl^\new(C)$. Consider the three candidate conditions that include $C$ in $\bar {\mathcal C}_{r+1}$.

Under condition~(1), some $x \in \vbl^\new(C)$ satisfies $b_{x,r}^\old \ne b_{x,r}^\new$. One possibility is that a clause $C' \in \mathcal{C}^\old(x) \cup \mathcal{C}^\new(x)$ enters or leaves $I_r$, in which case \Cref{lem:dynamic-lll-dirty-pair-locality} shows $\dist_{H^\cup}(C',\mathcal{M}) \le 2r-1$ and $\dist_{H^\cup}(C,C') \le 1$. The other case is that some clause $C'$ belongs to $\mathcal{C}^\old(x) \oplus \mathcal{C}^\new(x)$, where $C' \in \mathcal{M}$ and $\dist_{H^\cup}(C,C') \le 1$. In either case, the distance between $C$ and $\mathcal{M}$ is at most $2r$.

For conditions~(2) and~(3), either $C\in\mathcal{M}$, where the distance is clearly zero, or $C\in\mathcal{C}_r$. At $r=1$, initialization gives $\mathcal{C}_1\subseteq\Gamma_{H^\cup}(\mathcal{M})$. For $r\ge2$, a direct induction gives that every clause in $\mathcal{C}_r=\Gamma_{H^\cup}(\bar{\mathcal{C}}_r)$ is within $2(r-1)+1$ hops of $\mathcal{M}$. Both cases imply the claimed distance bound.

For the second assertion, condition~(1) supplies $x \in \vbl^\new(C)$ with $b_{x,r}^\old \ne b_{x,r}^\new$. When $b_{x,r}^\old = 1$, $\ell_C^\star\ge r$, and when $b_{x,r}^\new = 1$, $\ell_C^\new\ge r$. Condition~(2) states $\ell_C^\star\ge r$ directly. Under condition~(3), $C$ is falsified by $\sigma_{r-1}^\new$, so some variable must be resampled at stage $r$ or later before the clause becomes satisfied. Hence $\ell_C^\new\ge r$.
\end{proof}

For a fixed clause $C$, applying the same idea as in \Cref{cor:dynamic-lll-falsifiability-value-tail} and a union bound over $O(s)$ variables gives that the probability that $\max(\ell_C^\star,\ell_C^\new) \ge r$ is $O(ks(q/4d^2)^r)$. The oblivious update sequence fixes the $2r$-hop neighborhood of $\mathcal{M}$ independently of the randomness tables, and the neighborhood contains $O((4d^2)^r|\mathcal{B}|)$ clauses, so \Cref{lem:dynamic-lll-repair-locality} implies
\begin{gather*}
    \mathbb{E}[|\bar{\mathcal{C}}_{r+1}|]=O(ksq^r|\mathcal{B}|),\mathbb{E}[|\mathcal{C}_{r+1}|]=O(ksdq^r|\mathcal{B}|).
\end{gather*}
We conclude that the total expected work for stages $r \ge 2$ is
\begin{equation}
    \label{eq:dynamic-lll-total-repair-work}
    |\mathcal{B}| \sum_{r \ge 1} O\left( s^2q^{r+1}(k+r+T_\sample) + (ksr+T_\eval) ksdq^r \right) = O(k^2s^2dq(T_\eval + T_\sample)|\mathcal{B}|).
\end{equation}
The expected work for stage one is $O(|\mathcal{B}|s^2q(k+T_\sample+T_\eval))$ since $\mathbb{E}[|\mathcal{C}_1|] = O(q|\mathcal{B}|)$, and it is dominated by the sum above. \Cref{eq:dynamic-lll-total-repair-work} therefore gives the total expected work for these stages.

\paragraph{Final cleanup.}
Finally, we compute $\ell_C^\new$ only for clauses in $\bigcup_{r\ge1}\mathcal{C}_r$. To justify this, consider a clause outside it. Due to condition~(1), $b_{x,r}^\old = b_{x,r}^\new$ for every $x \in \vbl^\new(C)$ and $r \ge 1$. If $\vbl^\old(C) = \vbl^\new(C)$, then $\ell_C^\old = \ell_C^\new$. Otherwise, it belongs to $\mathcal{M}$. If $\ell_C^\star\ge1$, condition~(2) at stage one would put it in $\bar{\mathcal{C}}_2\subseteq\mathcal{C}_2$, a contradiction. Thus $\ell_C^\star=0$, and this implies $\ell_C^\old = \ell_C^\new=0$. Each computation scans at most $s$ variables, so the expected cleanup cost is at most
\[
    O\!\left(s\sum_{r\ge1}\mathbb{E}[|\mathcal{C}_r|]\right)
    =O(ks^2dq|\mathcal{B}|).
\]

We finally wrap up the discussion and prove the main theorem of this section.

\begin{proof}[Proof of \Cref{thm:dynamic-lll}]
For a batch $\mathcal{B}$ of clause updates, correctness follows from \Cref{lem:dynamic-lll-correctness}. The expected recourse is $O(s^2q|\mathcal{B}|)$ by \Cref{cor:dynamic-lll-recourse}. Initialization takes $O(|\mathcal{B}|(T_{\update}+ksq))$ expected time, repair takes $O(|\mathcal{B}|k^2s^2dq(T_{\eval}+T_{\sample}))$ expected time, and cleanup takes $O(|\mathcal{B}|ks^2dq)$ expected time. Their sum gives the stated work bound.

To introduce an isolated variable, initialize its transcript data to be all zero and generate $R(x,0)$ as its value in the output assignment. This takes $O(T_{\sample})$ expected time. To remove an isolated variable, discard its record and remove it from the output, taking constant time.
\end{proof}

%% file: edge-coloring-poly-delta.tex
\section{Dynamic $(1+\epsilon)\Delta$ edge coloring: $\poly(\Delta)$ update time}
\label{sec:edge-coloring-poly-delta}

In this section, we give a dynamic algorithm that maintains a $(1+\epsilon)\Delta$ edge coloring in $\poly(\Delta)$ expected update time. We first illustrate a black-box application of \Cref{thm:dynamic-lll} in the iterative edge-coloring framework of~\cite{chang2017complexity}, leading to a $\Delta^{O(\log(1/\epsilon))}$ time algorithm, and then improve it to $\poly(\Delta)$ by a more careful application of \Cref{thm:dynamic-lll}.


Notice that the static algorithm in \cite{chang2017complexity} requires $\epsilon = \omega(\log^{2.5}\Delta / \sqrt{\Delta})$, and this will be the parameter range in which all the following edge-coloring algorithms operate. For a fixed $\epsilon$, define $\Delta_\epsilon = \tilde \Theta(\epsilon^{-2})$ to be the smallest maximum degree for which the static algorithm in \cite{chang2017complexity} works.

\subsection{The Static Algorithm and a Black-Box Dynamic Algorithm}

\label{sec:edge-coloring-poly-delta-first}

We begin by reviewing the two phases of the static algorithm of~\cite{chang2017complexity}. Its first phase consists of $r=O(\log(1/\epsilon))$ iterations of a partial-coloring procedure. It uses a palette $\mathcal C_1=[(1+\epsilon_{\mathrm{pal}})\Delta]$ for some $\epsilon_{\mathrm{pal}} \in (0,\epsilon)$, and leaves an uncolored graph of maximum degree $\Delta'=(\epsilon-\epsilon_{\mathrm{pal}})\Delta/5 = \Theta(\epsilon \Delta)$. The second phase colors this remaining graph with the colors outside $\mathcal C_1$.

At the beginning of iteration $i$, let $G_i$ be the graph of uncolored edges and let $\Psi_i(e)\subseteq\mathcal C_1$ be the palette of $e\in E(G_i)$. Write $\deg_i(v)$ for the degree of $v$ in $G_i$, and write $\deg_{i,c}(v)$ for the number of edges of $G_i$ incident to $v$ whose palettes contain $c$. The algorithm maintains, for suitable parameter sequences $\{d_i\},\{t_i\},\{p_i\}$, the invariants
\begin{gather*}
    \max_v\deg_i(v)\le d_i, \\
    \max_{v,c}\deg_{i,c}(v)\le t_i, \\
    \min_{e\in E(G_i)}|\Psi_i(e)|\ge p_i.
\end{gather*}
We do not need the exact definitions of these sequences. We only use that $\max\{d_i,t_i,p_i\}=O(\Delta)$, $t_i/p_{i+1} \le C$, and $p_i/p_{i+1} \le C$ for every $i\in[r]$ and some constant $C$ not depending on $i$ or $\epsilon$. These bounds follow from \cite[Lemma~4]{chang2017complexity}.

The first step of iteration $i$ is to form an augmented instance $(G_i',\Psi_i')$ to ensure uniform progress across the instance in the following way:
\begin{itemize}
    \item First, each edge in $G_i$ drops arbitrary colors to achieve uniform palette size $p_i$.
    \item After that, define $\mathcal{C}_0$ to be a set of $p_i$ placeholder colors disjoint from $\mathcal{C}_1$. We attach an \emph{imaginary subtree} of depth three to every vertex $v\in V(G_i)$. The palette of each imaginary edge either contains all of $\mathcal{C}_0$ or contains all but one color of $\mathcal{C}_0$ together with one color from $\mathcal{C}_1$. The goal is that at every vertex within a two-hop neighborhood of the real vertices, the color degree of every color in $\mathcal{C}_1$, including imaginary edges, is exactly $t_i$.
\end{itemize}
Notice that the degree of a vertex may exceed $d_i$ or the color degree of a color in $\mathcal{C}_0$ at a vertex may exceed $t_i$ in the augmented instance, but we will only keep real edges in $G_i$ and colors in $\mathcal{C}_1$ after the iteration, so this causes no issue. We assume that there are enough imaginary edges around each internal vertex with palette $\mathcal{C}_0$.

The procedure $\mathsf{OneShotColoring}$ is then applied to the augmented instance. It independently selects a uniformly random color $c_i(e)\in\Psi_i'(e)$ for every edge $e\in E(G_i')$. An edge is successfully colored if none of its adjacent edges in $G_i'$ selects the same color. The unsuccessful real edges form $G_{i+1}$, and colors used by successful adjacent edges are removed from their palettes.

The LLL instance for iteration $i$ contains clauses asserting the degree, color-degree, and palette invariants for vertices and edges in $G_{i+1}$. As observed in~\cite[Section~10.1]{haeupler2026dynamic}, every variable occurs in $O(\Delta^3)$ clauses, and the maximum failure probability is $\exp(-\epsilon^2\Delta/\log^{4+o(1)}\Delta) = \exp(-\omega(\log \Delta))$ when $\Delta \ge \Delta_\epsilon$. Hence all the parameters in \Cref{thm:dynamic-lll} are $\poly(\Delta)$, except $q = \Delta^{-\omega(1)}$.

\begin{lemma}\label{lem:edge-coloring-for-small-degree-weak}
There is a dynamic algorithm that maintains a $(1+\epsilon)\Delta$ edge coloring of a graph of maximum degree $\Delta \ge \Delta_\epsilon$ under edge insertions and deletions with expected update time $\Delta^{O(\log(1/\epsilon))}$.
\end{lemma}
\begin{proof}
For every iteration $i$, maintain an instance $\mathcal L_i$ using \Cref{thm:dynamic-lll}, and process $\mathcal L_1,\ldots,\mathcal L_r$ in order. Let $\mathcal B_i$ be the updates to $(G_i,\Psi_i)$ generated by the preceding iteration, where $\mathcal B_1$ consists of the input update.

Since $\mathcal L_i$ is built on the augmented instance $(G_i',\Psi_i')$, we first construct the corresponding batch $\mathcal B_i'$ of updates to $G_i'$. In addition to inserting or deleting the real edge itself, an insertion or deletion in $G_i$ changes the palettes of $O(p_i)$ imaginary edges to maintain the uniform color degree; enough imaginary edges with palette $\mathcal{C}_0$ are available for this purpose. Implementing every such palette change by deleting and reinserting the edge gives $|\mathcal B_i'|=O(p_i|\mathcal B_i|)$.

Now we apply the updates in $\mathcal{B}_i'$ to $\mathcal{L}_i$. We first introduce the new variables for insertions in $\mathcal B_i'$, apply one batch of updates to the variable sets of the $O(\Delta^3|\mathcal B_i'|)$ affected clauses, and then remove the variables that have become isolated. By \Cref{thm:dynamic-lll}, these operations take $|\mathcal B_i'|\poly(\Delta)$ expected work. The variable insertions and deletions incur $O(|\mathcal B_i'|)$ recourse, while the clause updates incur $|\mathcal B_i'|\Delta^{-\omega(1)}$ additional expected recourse. Thus the total expected recourse is $O(|\mathcal B_i'|)$.

We next generate $\mathcal B_{i+1}$ from the recourse of $\mathcal L_i$. If the selected color of an edge $e$ changes, only $e$ and adjacent edges selecting its old or new color can change their success status. At most one such adjacent edge at each endpoint can be successful for either color, so at most five edges change their success status and at most five real edges enter or leave $G_{i+1}$. A changed successful color in $\mathcal C_1$, on either a real or an imaginary edge, affects the palettes of $O(t_i)$ adjacent real edges. Thus one change in the selected color generates $O(t_i)$ updates to $(G_{i+1},\Psi_{i+1})$. Consequently, $\mathbb E[|\mathcal B_{i+1}|]=O(t_i\mathbb E[|\mathcal B_i'|])=O(t_ip_i\mathbb E[|\mathcal B_i|])$. Iterating through $r=O(\log(1/\epsilon))$ levels and using $\max\{t_i,p_i\}=O(\Delta)$ shows that the expected work and recourse of the first phase are $\Delta^{O(\log(1/\epsilon))}$.

In the second phase, we color the remaining graph, whose maximum degree is $\Delta'=(\epsilon-\epsilon_{\mathrm{pal}})\Delta/5$, with the $5\Delta'$ colors outside $\mathcal C_1$. We maintain this coloring greedily: delete the color of an edge that leaves the graph, and assign any color unused at either endpoint to an edge that enters the graph. This takes $O(\Delta)$ time per change, so the total expected update time remains $\Delta^{O(\log(1/\epsilon))}$.
\end{proof}

\subsection{A $\poly(\Delta)$ algorithm}
\label{sec:edge-coloring-poly-delta-second}

The loss in \Cref{lem:edge-coloring-for-small-degree-weak} comes from implementing a palette change as a variable replacement. In the following theorem, we show that only $O(1/p_i+\poly(\Delta)q)$ fraction of palette changes affect the selected colors in $\mathcal{L}_i$, if we carefully couple the randomness of the variables. With this, we can show that the number of updates grows by only a constant factor per iteration.

\begin{theorem}\label{thm:edge-coloring-poly-delta}
There is a dynamic algorithm that maintains a $(1+\epsilon)\Delta$ edge coloring of a graph of maximum degree $\Delta \ge \Delta_\epsilon$ under edge insertions and deletions with expected update time $\poly(\Delta)$.
\end{theorem}
\begin{proof}
    We first specify how the entries in the randomness table are sampled. For every variable $x_e\in\mathcal L_i$ and $j\ge0$, lazily generate a uniformly random permutation $\pi_{x_e,j}$ of $\mathcal C_1\cup\mathcal C_0$, and let $R(x_e,j)$ be the first color of $\Psi_i'(e)$ in this permutation. This is a uniformly random element of $\Psi_i'(e)$, and $T_\sample=|\mathcal C_1\cup\mathcal C_0|=O(\Delta)$.

    As in \Cref{lem:edge-coloring-for-small-degree-weak}, process $\mathcal L_1,\ldots,\mathcal L_r$ in order, and let $\mathcal B_i$ and $\mathcal B_i'$ be the updates to $(G_i,\Psi_i)$ and $(G_i',\Psi_i')$, respectively. Let $S_i$ be the edge insertions and deletions in $\mathcal B_i$, and let $T_i$ be the palette changes in $\mathcal{B}_i$. A palette change in $T_i$ is a pair $(e,c)$ with $e\notin S_i$ that adds or removes $c$ from $\Psi_i(e)$. Define $S_i'$ and $T_i'$ analogously. Initially, $|S_1|=1$ and $T_1=\varnothing$.

    From \Cref{lem:edge-coloring-for-small-degree-weak}, each edge in $S_i$ generates one update in $S_i'$ and $O(p_i)$ updates in $T_i'$. For an insertion of a color into $\Psi_i(e)$, or a deletion $(e,c)\in T_i$ with $c\notin\Psi_i'(e)$, no change to the augmented instance is needed. Otherwise, we need to replace $c$ in $\Psi_i'(e)$ by an arbitrary color of $\Psi_i(e)\setminus\Psi_i'(e)$ and update a constant number of edges in the imaginary subtrees to maintain uniform color degree. Hence $|S_i'|=|S_i|$ and $|T_i'|=O(p_i|S_i|+|T_i|)$.

    We apply $\mathcal{B}_i'$ to $\mathcal{L}_i$ in the same way as \Cref{lem:edge-coloring-for-small-degree-weak}. For each $e$ with palette changes, we replace its old variable $x_e^\old$ with a new variable $x_e^\new$, implemented by an edge deletion and an edge insertion. However, we will couple the zeroth randomness-table entries of $x_e^\old$ and $x_e^\new$ by setting $\pi_{x_e^\new,0}=\pi_{x_e^\old,0}$; all entries with index $j\ge1$ are generated independently.

    Conditional on the entire input history from earlier iterations, the update sequence and palettes for this iteration are fixed independently of its randomness. We first argue that the coupling preserves the bounds in \Cref{thm:dynamic-lll}. Indeed, we only couple pairs of variables in which one variable belongs exclusively to clauses in $\mathcal{T}^\old$ and the other belongs exclusively to clauses in $\mathcal{T}^\new$. Thus, within each of $\mathcal{T}^\old$ and $\mathcal{T}^\new$, the randomness tables of all variables remain mutually independent with the prescribed distributions. Therefore, \Cref{lem:witness-tree-prob,lem:dynamic-lll-r-value-tail} apply separately to $\mathcal T^\old$ and $\mathcal T^\new$. The analysis of \Cref{thm:dynamic-lll} uses these two tail bounds only through union bounds and does not require the two transcripts to be independent. Thus the update takes $|\mathcal B_i'|\poly(\Delta)$ expected work, and the batch of clause updates generates $O(|\mathcal B_i'|\poly(\Delta)q)$ expected recourse.

    Now we bound the changes propagated to the next iteration. Let $D_i$ count the number of edges in $G_i'$ that either change their selected colors or are inserted or deleted. Every update in $S_i'$ contributes one to $D_i$, and the clause updates in $\mathcal{L}_i$ contributes $O(|\mathcal B_i'|\poly(\Delta)q)$ edges to $D_i$ in expectation.

    Now consider an edge $e$ whose palette changes from $\Psi'^\old(e)$ to $\Psi'^\new(e)$. Define $c_i^\old(e)$ and $c_i^\new(e)$ to be its color before and after the update. If they differ, then either $\max\{r_{x_e^\old}^\old,r_{x_e^\new}^\new\}>0$ or $R(x_e^\old,0)\ne R(x_e^\new,0)$. The first event has probability $O(\poly(\Delta)q)$ by \Cref{lem:dynamic-lll-r-value-tail}. For the second event, as $\pi_{x_e^\old,0} = \pi_{x_e^\new,0}$, the permutation must place an element of $\Psi'^\old(e)\oplus\Psi'^\new(e)$ first among $\Psi'^\old(e)\cup\Psi'^\new(e)$. As a result,
    \[
    \Pr[R(x_e^\old,0)\ne R(x_e^\new,0)] = \frac{|\Psi'^\old(e)\oplus\Psi'^\new(e)|}{|\Psi'^\old(e)\cup\Psi'^\new(e)|}=O\left(\frac{|\Psi'^\old(e)\oplus\Psi'^\new(e)|}{p_i}\right),
    \]
    and each palette change in $T_i'$ contributes $O(1/p_i+\poly(\Delta)q)=O(1/p_i)$ edges to $D_i$ in expectation. In sum, we have
    \[
        \mathbb E[D_i]=O\left(\mathbb E\left[|S_i'|+\frac{|T_i'|}{p_i}+\poly(\Delta)q|\mathcal B_i'|\right]\right) = O\left(\mathbb E\left[|S_i|+\frac{|T_i|}{p_i}\right]\right).
    \]

    As in the analysis of \Cref{lem:edge-coloring-for-small-degree-weak}, each edge counted by $D_i$ changes the status of $O(1)$ real edges and generates $O(t_i)$ palette updates in $G_{i+1}$. At the same time, changes in $T_i$ could propagate to the next iteration. Consequently,
    \[
        \mathbb E[|S_{i+1}|]=O(\mathbb E[D_i]),\mathbb E[|T_{i+1}|]=O(t_i\mathbb E[D_i]+\mathbb E[|T_i|]).
    \]
    Define the quantity $Z_i=\mathbb E[|S_i|+|T_i|/p_i]$. Since $t_i/p_{i+1} \le C$ and $p_i/p_{i+1} \le C$, the preceding bounds give $Z_{i+1} \le C' Z_i$ for some constant $C'$, and $Z_i\le\exp(O(r))=(1/\epsilon)^{O(1)}=\poly(\Delta)$ for every $i\le r+1$. Since $\mathbb E[|\mathcal B_i'|]=O(p_iZ_i)$, the process for all iterations takes $\poly(\Delta)$ expected work in total. Finally, the same greedy method handles the second phase with an additional $O(\Delta)$ factor. The total expected work remains $\poly(\Delta)$.
\end{proof}

%% file: edge-coloring-log-delta.tex
\section{Dynamic $(1+\epsilon)\Delta$ edge coloring: $O(\log \Delta) + \poly(1/\epsilon)$ update time}
\label{sec:edge-coloring-log-delta}

In this section, we build a dynamic algorithm for $(1+\epsilon)\Delta$ edge coloring with update time $O(\log\Delta)+\poly(1/\epsilon)$. We first illustrate the iterative degree splitting approach using the algorithm in \cref{thm:dynamic-lll}, and then introduce an alternative algorithm based on graph shattering.

\subsection{Iterative Degree Splitting}

The main idea is to apply the following lemma recursively to reduce the degree of the input instance.

\begin{lemma}\label{lem:edge-partitioning-one-level}
Let $\Delta$ bound the maximum degree of a dynamic graph $G$, and let $F\in\{1,\ldots,\lfloor\Delta\rfloor\}$ and $0<\delta\le1$ satisfy $\delta^2\Delta/F=\omega(\log\Delta)$. There is a dynamic algorithm that partitions $G$ into $F$ subgraphs, each of maximum degree at most $(1+\delta)\Delta/F$. Its expected update time is $O(1)$, and its expected recourse, defined by the sum of edge membership changes across all subgraphs, is $1+\exp(-\Omega(\delta^2\Delta/F))$.
\end{lemma}
\begin{proof}
Consider the random process in which each edge $e$ independently chooses $x_e$ uniformly from $[F]$ and joins the $x_e$-th subgraph. For each vertex $v$, let $B_v$ be the event that its degree exceeds $(1+\delta)\Delta/F$ in some subgraph. A Chernoff bound followed by a union bound over $F$ subgraphs gives $\Pr[B_v]=\exp(-\Omega(\delta^2\Delta/F))$. The event $B_v$ depends only on the variables for edges incident to $v$, so the dependency graph has degree at most $\Delta$. Since $\delta^2\Delta/F = \omega(\log\Delta)$, this is an LLL instance.

We maintain a satisfying assignment of the instance with \Cref{thm:dynamic-lll}. Its parameters are $k=2$, $s=\Delta$, $d=\Delta+1$, $q=4ed^4\exp(-\Omega(\delta^2\Delta/F)) = \exp(-\omega(\log\Delta))$, $T_\eval=O(\Delta)$, and $T_\sample=O(1)$. In this instance, $T_\update=O(1)$: for every vertex, we store its degree in each subgraph and the number of subgraphs that exceed the threshold in $\sigma_0$, and updating these quantities after a single insertion or deletion takes constant time.

Finally, an edge update introduces or removes one variable, which participates in two bad events. By \Cref{thm:dynamic-lll}, the expected work is $O(1)+q\poly(\Delta)=O(1)$, and the expected recourse is $1+q\poly(\Delta)=1+\exp(-\Omega(\delta^2\Delta/F))$.
\end{proof}

We first use this procedure once to handle a bounded range of degrees.

\begin{lemma}\label{lem:edge-coloring-small-deg}
There is an algorithm that maintains a $(1+O(\epsilon))\Delta$ edge coloring for a dynamic graph of maximum degree $\Delta \in [\Delta_\epsilon,\exp(\Delta_\epsilon)]$ under edge insertions and deletions with expected update time $\poly(1/\epsilon)$.
\end{lemma}
\begin{proof}
Set $D=C\epsilon^{-5}$ for a sufficiently large constant $C$. If $\Delta\le3D$, the claim follows directly from \Cref{thm:edge-coloring-poly-delta}. Suppose instead that $\Delta\in(3D,\exp(\Delta_\epsilon)]$. We first apply \Cref{lem:edge-partitioning-one-level} with $\delta=\epsilon$ and $F=\lfloor\Delta/D\rfloor$. Every resulting subgraph has maximum degree at most $(1+\epsilon)\Delta/F\in[D,3D]$, and the lemma applies because $\delta^2\Delta/F\ge C\epsilon^{-3}=\omega(\log\Delta)$. We then apply independent instances of \Cref{thm:edge-coloring-poly-delta} to these subgraphs with distinct palettes, so the total palette size is $(1+\epsilon)^2\Delta + F = (1+O(\epsilon))\Delta$.

The partition algorithm has $O(1)$ expected work and recourse. The update sequence for each \Cref{thm:edge-coloring-poly-delta} instance is oblivious to the randomness, so each instance has $\poly(1/\epsilon)$ expected work and recourse. Every reassigned edge produces one deletion and one insertion among the instances, so the total expected update time across all instances is $\poly(1/\epsilon)$.
\end{proof}

It remains to consider $\Delta > \exp(\Delta_\epsilon)$. We define $U_0=\Delta$, and for every $d\ge0$
\[
    U_{d+1}=\frac{U_d}{2} \left(1+\frac{10}{\log^2 U_d}\right).
\]
We additionally define $q_d = 1/\log^2 U_d$; thus $U_{d+1} = U_d(1+10q_d)/2$. The sequence $U_d$ gives the degree bounds across the levels of our recursive execution.

Let $R$ be the smallest index for which $U_R\le\exp(\Delta_\epsilon)$. Then $R = O(\log \Delta)$ and $U_R>\exp(\Delta_\epsilon)/2$. We may assume that $\Delta_\epsilon$ is large enough so that $10q_d \le 1/2$ for every $d \le R$, and thus $U_{d+1} \le 3U_d/4$ for $d \in [0,R-1]$. It follows that
\[
    U_R=\frac{\Delta}{2^R}\prod_{d=0}^{R-1}(1+10q_d)
    =\frac{\Delta}{2^R} \exp\left(O\left(\sum_{d = 0}^{R-1} q_d\right)\right)
    =\frac{\Delta}{2^R}\exp(O(1/\Delta_\epsilon)) =
    \frac{\Delta}{2^R}  (1+O(\epsilon)).
\]
To see why $\sum_{d=0}^{R-1}q_d = O(1/\Delta_\epsilon)$, observe that $\log U_{R-1} > \Delta_\epsilon$ and $\log U_d > \log U_{d+1} + \log(4/3)$ for every $d \in [0,R-2]$. Thus,
\[
    \sum_{d=0}^{R-1} q_d \le \sum_{d \ge 0} \frac{1}{(\Delta_\epsilon + d \log(4/3))^2} = O\left(\int_{\Delta_\epsilon}^\infty \mathrm{d} x /x^2\right) = O(1/\Delta_\epsilon).
\]

The argument above similarly gives $U_d = (1+O(\epsilon))\Delta/2^d$ for every $d \in [0,R]$ and thus $U_d / U_{d'} = (1\pm O(\epsilon)) 2^{d'-d}$ for $d,d' \in [0,R]$.

\begin{lemma}\label{lem:edge-coloring-log-delta}
There is an algorithm that maintains a $(1+O(\epsilon))\Delta$ edge coloring of a graph of maximum degree $\Delta \ge \Delta_\epsilon$ under edge insertions and deletions with expected update time $\poly(1/\epsilon) + O(\log \Delta)$.
\end{lemma}
\begin{proof}
By \Cref{lem:edge-coloring-small-deg}, it suffices to consider $\Delta>\exp(\Delta_\epsilon)$. Our algorithm consists of $R$ levels, with independent randomness at every level. At level $d \in [0,R]$, the algorithm maintains a partition $\mathcal P_d$ into $2^d$ subgraphs of maximum degree at most $U_d$. Starting with $\mathcal P_0=\{G\}$, we apply \Cref{lem:edge-partitioning-one-level} with $F=2$ and $\delta=10q_{d-1}$ to every member of $\mathcal P_{d-1}$, and add the resulting bipartitions to $\mathcal{P}_d$ for level $d$. We have $(10q_{d-1})^2U_{d-1}/F = \omega(\log U_{d-1})$, so the lemma can be applied. Finally, for every graph in $\mathcal{P}_R$, we apply \Cref{lem:edge-coloring-small-deg} to obtain a $(1+\epsilon)U_R$ edge coloring. Since $U_R = (1+O(\epsilon))\Delta/2^R$, the final coloring uses $(1+O(\epsilon))\Delta$ colors.

At level $d$, an update causes expected recourse $1+\exp(-\Omega(U_{d-1}q_{d-1}^2))=1+\exp(-U_{d-1}^{1-o(1)})$. Every reassigned edge produces one deletion and one insertion at the next level, so each update in the $d$-th level passes in expectation $1+\exp(-U_{d-1}^{1-o(1)})$ updates to the next level. As a result, a single edge update to $G$, equivalent to an update to $\mathcal{P}_0$, generates in expectation
\begin{align*}
    \prod_{d'=0}^{d-1}(1+\exp(-\Omega(U_{d'}q_{d'}^2))) \le \exp\!\left(\sum_{d'=0}^{d-1}\exp(-U_{d'}^{1-o(1)})\right)=O(1)
\end{align*}
updates to $\mathcal{P}_d$ for every $d\in[R]$. Summing over $R = O(\log \Delta)$ levels shows that maintaining $\mathcal{P}_R$ takes $O(\log \Delta)$ expected work and generates $O(1)$ recourse to $\mathcal{P}_R$. Finally, each update to $\mathcal{P}_R$ is passed to the corresponding instance in \Cref{lem:edge-coloring-small-deg}, so the total expected cost is $\poly(1/\epsilon)$.
\end{proof}

\subsection{Alternative Approach Based on Graph Shattering}

\label{sec:edge-coloring-log-delta-shattering}

Here, we review an alternative approach to compute the iterative degree bipartition in the static setting given in \cite{halldorsson2022fast}. This will be the backbone of our final constant-time dynamic edge coloring algorithm.

The recursion has the same structure as in \Cref{lem:edge-coloring-log-delta} and is represented by a full binary tree $T$ with levels $0,\ldots,R$. We use \emph{nodes} for elements of $T$ and reserve \emph{vertices} for elements of $V(G)$. The output at depth $d$ is a partition $\mathcal P_d$ of the input graph into $2^d$ subgraphs.

Write $r_T$ for the root of $T$, $\dep(a)$ for the depth of $a$ in $T$, $\mathcal L_T$ for the set of leaves, and $\mathcal L_T(a)$ for the set of leaves inside the subtree of a node $a$. A node $a$ is \emph{internal} if it is not a leaf. For an internal node, write $\LC(a)$ and $\RC(a)$ for its left and right children; for $a\ne r_T$, write $\P(a)$ for its parent. Two nodes are \emph{comparable} if one is an ancestor of the other. Here and below, a node is considered an ancestor of itself.

For $0\le d<R$, set $m_d=\lceil1/q_d\rceil$. Independently for every edge $e$ and depth $d<R$, sample a uniform direction $\dir_{e,d}\in\{\LC,\RC\}$ and a uniform category $\mathrm{cat}_{e,d}\in[m_d]$; all category and direction choices are mutually independent.

\paragraph{Algorithm Description} In the algorithm, each node $a$ in $T$ maintains an output edge set $E_a$. Initially, $E_{r_T}=E(G)$ and all other sets are empty. The algorithm calls $\mathcal A(r_T)$, where $\mathcal A(a)$ performs the following steps.
\begin{enumerate}
    \item Let $d=\dep(a)$. If $d=R$, stop the process.
    \item Partition $E_a$ into $E_{a,1},\ldots,E_{a,m_d}$ by placing each edge into its category. Initialize $E_{a,i}^\bad = \varnothing$ for $i \in [m_d]$, and set $E_a^\bad = \cup_{i \in [m_d]} E_{a,i}^\bad$.
    \item Process the categories in the order $i=1,\ldots,m_d$.
    \begin{itemize}
        \item[(a)] Let $H_{a,i}$ consist of edges in $E_{a,i}$ that have not entered $E_a^\bad$. Each such edge $e$ proposes $\dir_{e,d} \in \{\LC, \RC\}$.
        \item[(b)] Perform the following tests for every vertex $x$ in parallel:
        \begin{itemize}
            \item the \emph{degree test} fails if the number of incident edges in $H_{a,i}$ exceeds $2U_d/m_d$;
            \item the \emph{discrepancy test} fails if the numbers of incident edges in $H_{a,i}$ directed $\LC$ and $\RC$ differ by more than $U_dq_d/m_d$.
        \end{itemize}
        If either test fails, call $(x,a,i)$ a \emph{failure}. For each failure $(x,a,i)$, add all its incident edges in $H_{a,i}$ to $E_{a,i}^\bad$. For every neighbor $y$ of $x$ in $H_{a,i}$ and $j > i$, also \emph{freeze} all edges in $E_{a,j}$ incident to $y$ by adding them to $E_{a,i}^\bad$.
        \item[(c)] Send every surviving edge of $H_{a,i}$ to $E_{\LC(a)}$ or $E_{\RC(a)}$ according to its proposed direction.
    \end{itemize}
    \item For every connected component of $(V(G),E_a^\bad)$, deterministically compute an edge bipartition with degree discrepancy at most two at every vertex. Send its two parts to $E_{\LC(a)}$ and $E_{\RC(a)}$. This can be done deterministically by computing an Euler tour of the component after pairing odd-degree vertices.
    \item Call $\mathcal A(\LC(a))$ and $\mathcal A(\RC(a))$.
\end{enumerate}
We call steps 2 and 3 the \emph{pre-shattering process} and step 4 the \emph{post-shattering process}.

The following table summarizes the input and data that the algorithm keeps track of. For quantities index by a node $a$, write $d = \dep(a)$.

\begin{table}[htbp]
\centering
\begin{tabular}{ll}
\hline
Quantity & Meaning \\
\hline
$\dir_{e,d}$ & The sampled direction of edge $e$ at depth $d$ \\
$\mathrm{cat}_{e,d}$ & The sampled category of edge $e$ at depth $d$ \\
\hline
$E_a$ & The edges assigned to node $a$ \\
$E_{a,i}$ & The set of category-$i$ edges at $a$ \\
$H_{a,i}$ & The edges of $E_{a,i}$ that are not already bad when category $i$ begins \\
$E_{a,i}^\bad$ & The edges added to the bad set while processing category $i$, including frozen edges \\
$E_a^\bad$ & $\bigcup_{i=1}^{m_d}E_{a,i}^\bad$ \\
\hline
\end{tabular}
\caption{Input and data in a static edge-splitting transcript}
\label{tab:edge-coloring-static-transcript-top-down}
\end{table}

\paragraph{Algorithm Correctness} In the following, we prove the correctness of the algorithm. Let $\mathcal{P}_d = \{E_a \mid \dep(a) = d\}$.

\begin{observation}\label{obs:edge-coloring-log-delta-top-down-partition}
    For every $0 \le d \le R$, $\mathcal{P}_d$ is a partition of $E(G)$.
\end{observation}
\begin{proof}
    We prove by induction on $d$. The base case $d=0$ is trivial. Suppose that the statement holds for $d$ and consider any node $a$ of depth $d$. After the pre-shattering process of $\mathcal{A}(a)$, every edge in $E_a$ belongs to exactly one of $E_{\LC(a)}$, $E_{\RC(a)}$ and $E_a^\bad$. The post-shattering process further partitions $E_a^\bad$ into two parts, and merge them with $E_{\LC(a)}$ and $E_{\RC(a)}$. This implies that $E_{\LC(a)}$ and $E_{\RC(a)}$ partitions $E_a$, and thus $\mathcal{P}_{d+1}$ partitions $E(G)$ from the induction hypothesis. 
\end{proof}

\begin{lemma}\label{lem:edge-coloring-log-delta-top-down-degree}
    For every $0 \le d \le R$, the maximum degree of any graph in $\mathcal{P}_d$ is at most $U_d$.
\end{lemma}
\begin{proof}
    We prove by induction on $d$. The base case $d=0$ is trivial as the input graph is of maximum degree $\Delta = U_0$.
    
    Suppose that the statement holds for $d$. Consider any node $a$ of depth $d$ and any vertex $w$. $\mathcal{A}(a)$ produces an edge bipartition $(E_{\LC(a)}, E_{\RC(a)})$ of $E_a$ according to \Cref{obs:edge-coloring-log-delta-top-down-partition}, and to show that the degree of $w$ in both edge sets does not exceed $U_{d+1}$, it suffices to show that the degree discrepancy is at most $2U_{d+1}-U_d = 10U_dq_d$. The post-shattering process always produces at most two discrepancy for each vertex, so we pay attention to pre-shattering process next.
    
    During $\mathcal{A}(a)$, Let $i^\star$ be the first category in which $w$ or a neighbor of $w$ in $H_{a,i^\star}$ fails, and set $i^\star = m + 1$ if it does not exist. Consider the contribution to the discrepancy for each category:
    \begin{itemize}
        \item For $i < i^\star$, every edge incident to $w$ in $H_{a,i}$ is not moved to $E_{a,i}^\bad$ and the discrepancy test bounds their discrepancy by $U_dq_d/m_d$.
        \item For $i^\star$, either $w$ fails and contributes nothing, or its degree test passes and the surviving edges contribute at most $2U_d/m_d\le2U_dq_d$ to the disrepancy.
        \item For $i > i^\star$, since neighboring failure in category $i^\star$ freezes every edge incident to $w$ in later categories, category $i$ contributes zero to the disrepancy.
    \end{itemize}
    
    In sum, the pre-shattering discrepancy is therefore at most $3U_dq_d$, and the total discrepancy is at most $2+3U_dq_d \le 10U_dq_d$.
\end{proof}

%% file: edge-coloring-const.tex
\section{Dynamic $(1+\epsilon)\Delta$ edge coloring: $\poly(1/\epsilon)$ update time}
\label{sec:edge-coloring-const}

In this section, we develop the algorithm that removes the dependence on $\Delta$ from the update time based on the static algorithm in \Cref{sec:edge-coloring-log-delta-shattering}.

\subsection{A Modified Static Algorithm}

We follow all definitions and notations in \Cref{sec:edge-coloring-log-delta-shattering}. For each edge $e$, we notice that the random directions $\mathrm{dir}_{e,0},\dots,\mathrm{dir}_{e,R-1}$ defines a \emph{routing path} $\routepath_{e,0},\ldots,\routepath_{e,R}$ for $e$, where $\routepath_{e,0}=r_T$ and
\begin{equation}\label{eq:def-a}
    \routepath_{e,d+1}=
    \begin{cases}
        \LC(\routepath_{e,d}), & \dir_{e,d}=\LC,\\
        \RC(\routepath_{e,d}), & \dir_{e,d}=\RC.
    \end{cases}
\end{equation}

\begin{definition}[$\bar E_a$]
    For a node $a$ of depth $d$, define $\bar E_a=\{e\in E(G):\routepath_{e,d}=a\}$ to be the set of edges that would reach $a$ if they followed their routing paths.
\end{definition}

By a slight abuse of notation, in the following, we also use an edge set such as $\bar E_a$ to denote the graph with vertex set $V(G)$ and edge set $\bar E_a$.

The edge sets $\bar E_a$ describe the ideal execution of the algorithm, in which every edge follows its sampled directions. However, the algorithm may need to reroute some edges when the random choices are unfavorable. The algorithm described in \Cref{sec:edge-coloring-log-delta-shattering} tests every possible vertex in the pre-shattering process to compute $E_a^\bad$, the set of edges being rerouted. In our modified algorithm,the rerouting is instead guided by a set $\mathcal F$ of \emph{seed failures}, defined as follows.

\begin{definition}[Seed failure]
A pair $(u,a)$ with $d=\dep(a)$ is a \emph{seed failure} if $\deg_{\bar E_a}(u)>U_d$, or if $a$ is internal and some category $i\in[m_d]$ satisfies at least one of the following conditions in $\bar E_a$:
\begin{itemize}
    \item more than $2U_d/m_d$ category-$i$ edges are incident to $u$;
    \item the numbers of incident category-$i$ edges with direction $\LC$ and $\RC$ differ by more than $U_dq_d/m_d$.
\end{itemize}
\end{definition}

Notice that the tests in the seed failure match those in step 3(b) of the algorithm in \Cref{sec:edge-coloring-log-delta-shattering}, but use different underlying edge sets. Seed failures identify the starting points where the algorithm may need to deviate from the proposed edge set $\bar E_a$.

For $0\le d\le R$, define $L_d=\sum_{i=d}^{R-1}2m_i=O(\log^3 U_d)$, where the empty sum gives $L_R=0$. The next definition characterizes the ``effect'' of a seed failure.

\begin{definition}[Cluster]\label{def:edge-coloring-cluster}
The \emph{cluster} of $(w,b)\in V(G)\times T$ consists of all pairs $(x,c)$ such that $c$ is a descendant of $b$ and $\bar E_b$ contains a path from $w$ to $x$ of length at most $L_{\dep(b)}$ on which every vertex except the terminal vertex $x$ has degree at most $U_{\dep(b)}$. Paths of length zero are allowed.
\end{definition}
Since we require every vertex except the terminal vertex to have degree at most $U_{\dep(b)}$, one can show that a cluster contains at most $\exp(O(\log^4 U_{\dep(b)}))$ pairs. Roughly speaking, in our algorithm below, each seed failure is responsible for degree overflow within its cluster during the execution. We now state the completeness condition required of $\mathcal{F}$ by the static algorithm.

\begin{definition}[Completeness]
A set $\mathcal F\subseteq V(G)\times T$ is \emph{complete} for an instance if it satisfies all the following.
\begin{itemize}
    \item[(I)] Every pair in $\mathcal F$ is a seed failure.
    \item[(II)] Every seed failure in the cluster of a pair in $\mathcal F$ also belongs to $\mathcal F$.
    \item[(III)] If $(w,b)\in\mathcal F$, $b\ne r_T$, and $\deg_{\bar E_b}(w)>U_{\dep(b)}$, then $(w,\P(b))\in\mathcal F$.
    \item[(IV)] If $c$ is a leaf and $\deg_{\bar E_c}(x)>U_R$, then $(x,\P(c))\in\mathcal F$.
\end{itemize}
\end{definition}

\paragraph{Algorithm Description}

Fix a complete set $\mathcal F$. Each node $a$ maintains an output edge set $E_a$ and a set $V_a^\star$ of \emph{active} vertices. Initially, $E_{r_T}=E(G)$ and all other sets are empty. The algorithm calls $\bar {\mathcal A}(r_T)$, where $\bar {\mathcal A}(a)$ performs the following steps. Steps different from the algorithm $\mathcal{A}(a)$ in \Cref{sec:edge-coloring-log-delta-shattering} are marked red.
\begin{enumerate}
    \item Let $d=\dep(a)$. If $d=R$, stop the process.
    \item \textcolor{red}{Define $\bar V_{a,0}$ to be $V_a^\star\cup\{x:(x,a)\in\mathcal F\}$.}
    \item Partition $E_a$ into $E_{a,1},\ldots,E_{a,m_d}$ by placing each edge into its category. Initialize $E_{a,i}^\bad = \varnothing$ for $i \in [m_d]$, and set $E_a^\bad = \cup_{i \in [m_d]} E_{a,i}^\bad$.
    \item Process the categories in the order $i=1,\ldots,m_d$.
    \begin{itemize}
        \item[(a)] Let $H_{a,i}$ consist of edges in $E_{a,i}$ that have not entered $E_a^\bad$. Each such edge $e$ proposes $\dir_{e,d} \in \{\LC, \RC\}$.
        \item[(b)] \textcolor{red}{Initialize $\bar V_{a,i}$ to be $\bar V_{a,i-1}$, and perform the following tests for every vertex $x \in \bar V_{a,i-1}$ in parallel:}
        \begin{itemize}
            \item the \emph{degree test} fails if the number of incident edges in $H_{a,i}$ exceeds $2U_d/m_d$;
            \item the \emph{discrepancy test} fails if the numbers of incident edges in $H_{a,i}$ directed $\LC$ and $\RC$ differ by more than $U_dq_d/m_d$.
        \end{itemize}
        If either test fails, call $(x,a,i)$ a \emph{failure}. For each failure $(x,a,i)$, add all its incident edges in $H_{a,i}$ to $E_{a,i}^\bad$. For every neighbor $y$ of $x$ in $H_{a,i}$ and $j > i$, also \emph{freeze} all edges in $E_{a,j}$ incident to $y$ by adding them to $E_{a,i}^\bad$. \textcolor{red}{Whenever an edge enters $E_{a,i}^\bad$, add both endpoints to $\bar V_{a,i}$.}
        \item[(c)] Send every surviving edge of $H_{a,i}$ to $E_{\LC(a)}$ or $E_{\RC(a)}$ according to its proposed direction.
    \end{itemize}
    \item For every connected component of $(V(G),E_a^\bad)$, deterministically compute an edge bipartition with degree discrepancy at most two at every vertex. Send its two parts to $E_{\LC(a)}$ and $E_{\RC(a)}$. This can be done deterministically by computing an Euler tour of the component after pairing odd-degree vertices.
    \item \textcolor{red}{Define $\bar V_a$ to be $\bar V_{a,m_{\dep(a)}}$. Set $V_{\LC(a)}^\star$ and $V_{\RC(a)}^\star$ to be $\bar V_a$}, and call $\mathcal A(\LC(a))$ and $\mathcal A(\RC(a))$.
\end{enumerate}

\Cref{tab:edge-coloring-static-transcript-bottom-up} summarizes the input and data that the modified static algorithm additionally keeps track of apart from \Cref{tab:edge-coloring-static-transcript-top-down}. For quantities indexed by a node $a$, write $d=\dep(a)$.

\begin{table}[htbp]
\centering
\begin{tabular}{ll}
\hline
Quantity & Meaning \\
\hline
$\routepath_{e,d}$ & The node at depth $d$ along its routing path \\
$\bar E_a$ & The ideal edge set $\{e\in E(G):\routepath_{e,d}=a\}$ \\
$\mathcal F$ & The complete set of seed failures supplied to the algorithm \\
\hline
$V_a^\star$ & The set of active vertices \\
$\bar V_{a,i}$ & Activated vertices after processing categories $1,\ldots,i$ \\
$\bar V_a$ & The final active set $\bar V_{a,m_d}$ \\
\hline
\end{tabular}
\caption{Additional input and data in the new static edge-splitting transcript}
\label{tab:edge-coloring-static-transcript-bottom-up}
\end{table}

\paragraph{Algorithm Correctness}

We start with some simple observations and then prove the algorithm's degree guarantee.

The first observation shows the monotonicity of active sets and is straightforward from the algorithm description.

\begin{observation}\label{obs:edge-coloring-const-active-set-monotonicity}
    If $x \in \bar V_{a,i}$ for some internal node $a$ and $i \in [0,m_{\dep(a)}]$, then $x \in \bar V_{a,j}$ for every $j \in [i,m_{\dep(a)}]$ and $x \in V_b^\star$ for every strict descendant $b$ of $a$.
\end{observation}

The second observation says that the process of activation is local.

\begin{observation}\label{obs:edge-coloring-const-bar-V-to-V-star}
    For an internal node $a$ and $i \in [0,m_{\dep(a)}]$, every vertex in $\bar V_{a,i}$ is within distance $2i$ of $\bar V_{a,0}$ in $E_a$.
\end{observation}
\begin{proof}
    We argue by induction on $i$. The base case $i = 0$ is trivial. For the induction step, $x \in \bar V_{a,i} \backslash \bar V_{a,i-1}$ only when it is incident to an edge moved to $E_{a,i}^\bad$. The failure $(y,a,i)$ responsible for the movement satisfies $y \in \bar V_{a,i-1}$, and the distance between $x$ and $y$ in $E_a$ is at most two. This completes the induction.
\end{proof}

The next observation shows that a passive vertex's adjacency list follows the ideal execution.

\begin{observation}\label{obs:edge-coloring-const-passive-incidence}
    For every node $a$ and $w \notin V_a^\star$, the sets of edges incident to $w$ are the same in $E_a$ and $\bar E_a$.
\end{observation}
\begin{proof}
    We argue by induction on $\dep(a)$. The base case $a = r_T$ is trivial. For the induction step, by \Cref{obs:edge-coloring-const-active-set-monotonicity}, $V_a^\star = \bar V_{\P(a)} \supseteq V_{\P(a)}^\star$. As a result, the induction hypothesis implies that the sets of edges incident to $w$ are the same in $E_{\P(a)}$ and $\bar E_{\P(a)}$. Since $w \notin \bar V_{\P(a)}$, none of the edges incident to $w$ in $E_{\P(a)}$ enters the bad set, and every edge follows its proposed direction. Then the statement follows since $\bar E_a$ contains the set of edges in $\bar E_{\P(a)}$ that propose to enter $a$.
\end{proof}

In the next lemma, we show that every activated vertex is in the cluster of some seed failure in $\mathcal{F}$.

\begin{lemma}\label{lem:edge-coloring-const-active-to-cluster}
    For every internal node $a$ and $w \in \bar V_a$, there is $(w^\star,a^\star) \in \mathcal{F}$ such that $a^\star$ is an ancestor of $a$, the distance between $w^\star$ and $w$ in $\bar E_{a^\star}$ is at most $\sum_{j=\dep(a^\star)}^{\dep(a)} 2m_j \le L_{\dep(a^\star)}$, and $(w,a)$ is in the cluster of $(w^\star,a^\star)$.
\end{lemma}
\begin{proof}
    We find $(w^\star,a^\star)$ by an iterative process. Each step in the process records an ancestor $c$ of $a$, a vertex $y \in \bar V_{c,0}$, and a path $P$ from $y$ to $w$ in $E_c$ of length at most $\sum_{j=\dep(c)}^{\dep(a)} 2m_j$. Initially, by \Cref{obs:edge-coloring-const-bar-V-to-V-star}, there is a vertex $y \in \bar V_{a,0}$ within distance at most $2m_{\dep(a)}$ from $w$. We set $c=a$ and let $P$ be the shortest path from $y$ to $w$ in $E_a$. In each step, we do the following checks.
\begin{enumerate}
    \item If $(y,c)\notin\mathcal F$, then $c\ne r_T$ and $y\in\bar V_{\P(c)}$. Apply \Cref{obs:edge-coloring-const-bar-V-to-V-star} to find a path of length at most $2m_{\dep(c)-1}$ from some $z\in \bar V_{\P(c),0}$ to $y$, add it to the beginning of $P$, eliminate loops when necessary, and replace $(y,c)$ with $(z,\P(c))$.
    \item If $(y,c)\in\mathcal F$ but an edge $uv$ of $P$ is absent from $\bar E_c$, then that edge departed from its routing path at a strict ancestor of $c$, and $u\in\bar V_{\P(c)}$ according to \Cref{obs:edge-coloring-const-active-set-monotonicity}. Replace the prefix ending at $u$ by a path of length at most $2m_{\dep(c)-1}$ from some $z\in \bar V_{\P(c),0}$ to $u$, and replace $(y,c)$ with $(z,\P(c))$.
    \item For the remaining case, $P\subseteq\bar E_c$. If a vertex of $P$ other than $w$ has degree greater than $U_{\dep(c)}$ in $\bar E_c$, let $z$ be the first such vertex. The prefix ending at $z$ witnesses that $(z,c)$ lies in the cluster of $(y,c)$. Properties (II) and (III) of completeness give $(z,\P(c))\in\mathcal F$. Retain the suffix from $z$ to $w$ and replace $(y,c)$ by $(z,\P(c))$.
    \item If none of these cases applies, $P$ witnesses that $(w,a)$ lies in the cluster of $(y,c)\in\mathcal F$, and the procedure stops.
\end{enumerate}

It is easy to see that every nonterminal step in this process moves $c$ to a strict ancestor and preserves the length invariant. The process enters step 4 and terminates when $c = r_T$. It cannot enter the first two branches because $E_{r_T}=\bar E_{r_T}$, and the third case does not apply because all vertices have degree at most $U_0$. Thus, the process terminates after finitely many steps and finds $(w^\star,a^\star) \in \mathcal{F}$ satisfying the statement.
\end{proof}
\begin{corollary}\label{cor:edge-coloring-const-activated-then-not-seed}
    For every internal node $a$ and $w \in \bar V_a \backslash \bar V_{a,0}$, $(w,a)$ is not a seed failure.
\end{corollary}
\begin{proof}
    From \Cref{lem:edge-coloring-const-active-to-cluster}, there is $(w^\star,a^\star) \in \mathcal{F}$ whose cluster contains $(w,a)$. Since $w \notin \bar V_{a,0}$, $(w,a) \notin \mathcal{F}$, so $(w,a)$ is not a seed failure by property (II) of completeness.
\end{proof}

The previous claims suffice to prove the degree guarantee for active vertices.

\begin{lemma}\label{lem:edge-coloring-const-static-degree-bound}
If $\mathcal F$ is complete, then for a node $a$ of depth $d$ and $w\in V_a^\star$, $\deg_{E_a}(w)\le U_d$.
\end{lemma}
\begin{proof}
We prove the statement by induction on $d$. For the base case $a = r_T$, every vertex in $E_a$ has degree at most $U_0$. For the induction step, denote $p = \P(a)$ and fix $w \in V_a^\star$. We claim, by a case analysis, that $\deg_{E_p}(w) \le U_{d-1}$:
\begin{itemize}
    \item When $w \in V_{p}^\star$, the induction hypothesis gives that $\deg_{E_{p}}(w) \le U_{d-1}$;
    \item When $w \in \bar V_{p,0} \backslash V_p^\star$, we know that $(w,p)$ is a seed failure. If $p = r_T$, then $\deg_{E_{p}}(w) \le U_0$; if $p \ne r_T$, $(w,\P(p)) \notin \mathcal{F}$ since otherwise $w \in \bar V_{\P(p),0} \subseteq \bar V_{\P(p)} = V_p^\star$ by \Cref{obs:edge-coloring-const-active-set-monotonicity}. This means, according to property (III) of completeness, $\deg_{\bar E_p}(w) \le U_{d-1}$. Finally, since $w \notin V_p^\star$, \Cref{obs:edge-coloring-const-passive-incidence} gives that $\deg_{E_p}(w) = \deg_{\bar E_p}(w) \le U_{d-1}$.
    \item Finally, when $w \in \bar V_p \backslash \bar V_{p,0}$, $(w,p)$ is not a seed failure by \Cref{cor:edge-coloring-const-activated-then-not-seed}, and $\deg_{E_p}(w) = \deg_{\bar E_p}(w) \le U_{d-1}$ according to \Cref{obs:edge-coloring-const-passive-incidence}.
\end{itemize}

As $E_{\LC(p)}$ and $E_{\RC(p)}$ partition $E_p$, to show that $\deg_{E_a}(w) \le U_d$, it suffices to show that the degree discrepancy of the bipartition generated in $\mathcal{A}(p)$ for $w$ is bounded by $2U_d - U_{d-1} = 10q_{d-1}U_{d-1}$. The post-shattering part generates discrepancy at most two for every vertex, and we consider pre-shattering next.

Write $U=U_{d-1}$, $q=q_{d-1}$, and $m=m_{d-1}$. If $w\in\bar V_{p,0}$, then following the same argument as in \cref{lem:edge-coloring-log-delta-top-down-degree}, the pre-shattering discrepancy is at most $3Uq$.

Now suppose that $w\notin\bar V_{p,0}$, and let $i^\star$ be the category in which $w$ becomes active. Before this category, no incident edge has become bad, and every incident edge follows its proposed direction. Since $(w,p)$ is not a seed failure and $w\notin V_p^\star$, \Cref{obs:edge-coloring-const-passive-incidence} bounds the total discrepancy of these earlier categories by $Uq$. Category $i^\star$ contributes at most its degree, $2U/m\le2Uq$, to the discrepancy. Starting with category $i^\star+1$, $w$ is tested, and \cref{lem:edge-coloring-log-delta-top-down-degree} bounds the remaining contribution by $3Uq$. Hence the pre-shattering discrepancy is at most $6Uq$, and the total discrepancy is at most $6Uq+2\le10Uq$, as required.
\end{proof}

For passive vertices at the leaves, \Cref{obs:edge-coloring-const-passive-incidence} and completeness guarantee (IV) imply that their final degrees are at most $U_R$. This, along with \Cref{lem:edge-coloring-const-static-degree-bound}, gives the degree guarantee of the algorithm.

\begin{corollary}\label{cor:edge-coloring-const-static-correctness}
    If $\mathcal{F}$ is complete, then every graph in $\mathcal{P}_R$ has maximum degree at most $U_R$.
\end{corollary}
\begin{proof}
    Fix a leaf $a$ and vertex $w$. From \Cref{lem:edge-coloring-const-static-degree-bound}, when $w \in V_a^\star$, $\deg_{E_a}(w) \le U_R$. When $w\notin V_a^\star$, we have $\deg_{E_a}(w) = \deg_{\bar E_a}(w)$ by \Cref{obs:edge-coloring-const-passive-incidence}. From property (IV) of completeness, if $\deg_{\bar E_a}(w) > U_R$, then $(w,\P(a))\in \mathcal{F}$ and $w \in V_a^\star$, a contradiction. This proves the claim.
\end{proof}

\subsection{The Dynamic Algorithm}

The dynamic algorithm maintains the transcript of the static algorithm above in two stages. The \emph{discovery stage} maintains a complete set $\mathcal F$, and the \emph{repair stage} updates the transcript using this set. During the dynamic algorithm, the random directions $\dir_{e,0},\dots,\dir_{e,R-1}$ and categories $\mathrm{cat}_{e,0},\dots,\mathrm{cat}_{e,R-1}$ remain unchanged throughout the lifetime of an edge $e$.

In the rest of this section, let us fix an update, denoted by $e=(u,v)$. We use $\bar E_b^\old$ and $\bar E_b^\new$ to denote the sets of edges whose routing paths pass through $b$ before and after this update, respectively. Further, define $\bar E_b^\cup=\bar E_b^\old\cup\bar E_b^\new$. The clusters of a pair in the old and new instances, called its old and new clusters, are defined by \Cref{def:edge-coloring-cluster} using $\bar E_b^\old$ and $\bar E_b^\new$ as the underlying graphs.

\subsubsection{Algorithm Description}

\paragraph{Failure Discovery Stage}

In the first stage, we explicitly maintain vertex degrees and adjacency lists of $\bar E_a$ only for leaves. For an internal node $b$, $\bar E_b$ is maintained implicitly as the union of $\bar E_a$ over $a \in \mathcal{L}_T(b)$. We also explicitly maintain the seed failure set $\mathcal F$. For an insertion, we sample $\dir_{e,0},\dir_{e,1},\dots,\dir_{e,R-1}$, while the random categories are sampled lazily. These choices remain unchanged throughout the lifetime of $e$. For either an insertion or a deletion, we then apply the update to $\bar E_{\routepath_{e,R}}$, thereby obtaining the new leaf edge sets.

We then call a recursive procedure $\Search$ on pairs in $V(G) \times T$ to update $\mathcal{F}$. We generate an \emph{initial} $\Search$ call for $(u,\routepath_{e,R})$, $(v,\routepath_{e,R})$, and every pair in $\mathcal{F}$ whose old cluster contains either of these two pairs. Every $\Search$ call generated recursively is called \emph{noninitial}.

$\Search(w,b)$ performs the following steps.
\begin{enumerate}
    \item If $(w,b)$ is no longer a seed failure after the update, remove it from $\mathcal F$ if present, and return.
    \item Add $(w,b)$ to $\mathcal F$ if absent.
    \item If $b\ne r_T$ and $\deg_{\bar E_b^\new}(w)>U_{\dep(b)}$, call $\Search(w,\P(b))$.
    \item Call $\Search(x,c)$ for every pair $(x,c)$ in the new cluster of $(w,b)$.
\end{enumerate}

Repeated calls to $\Search$ with the same parameter are redundant, so we use an additional bit for each pair to guarantee that every pair is visited only once throughout the stage. We use $\mathcal{F}^\old$ to denote the set $\mathcal{F}$ before the discovery stage and $\mathcal{F}^\new$ to denote the set $\mathcal{F}$ after all $\Search$ calls have finished.

\begin{lemma}\label{lem:discover-stage-invariants}
After every update, if $\mathcal F^\old$ is complete for the old instance, then $\mathcal F^\new$ is complete for the new instance.
\end{lemma}
\begin{proof}
We show the properties one by one. 

\begin{itemize}
    \item For (I), the status of $(w,b)$ as a seed failure can change only when $w\in\{u,v\}$ and $b$ lies on the routing path of $e$. Any such pair in $\mathcal F^\old$ has an old cluster containing $(w,\routepath_{e,R})$, so it generates an initial call and gets removed from $\mathcal{F}^\old$ if it is no longer a seed failure. All newly added members are seed failures by the algorithm description, so (I) holds.
    \item For (II), notice that the old and new cluster definitions differ. We use the following observation: For $(w,b) \in V(G) \times T$, if neither $(u,\routepath_{e,R})$ nor $(v,\routepath_{e,R})$ belongs to the old cluster of $(w,b)$, then the old and new clusters of $(w,b)$ are the same, and the sets of seed failures in this cluster are also the same. To see this, if $b$ is not on the routing path of $e$, then $e$ does not affect $\bar E_b$ and the statement is immediate. Otherwise, the path that witnesses the gain or loss of the cluster must use $e$ or pass through an endpoint whose degree changes in the old instance. The prefix then witnesses that $(u,\routepath_{e,R})$ or $(v,\routepath_{e,R})$ belongs to the old cluster of $(w,b)$, contradicting the assumption. The same argument also shows that the cluster of $(w,b)$ cannot contain a pair $(x,c)$ whose adjacency information changes in $\bar E_c$, so the set of seed failures remains unchanged.

    Now fix $(w,b)\in\mathcal F^\new$. If $\Search(w,b)$ is called, step 4 visits all seed failures in its new cluster and adds them to $\mathcal F^\new$. If $\Search(w,b)$ is not called, it is in $\mathcal{F}^\old$, and its cluster and the set of seed failures in that cluster remain unchanged. Since $\mathcal{F}^\old$ is complete for the old instance, (II) also holds in this case.
    \item For (III), write $d=\dep(b)$ and first observe that when $b \ne r_T$ and $\deg_{\bar E_b}(w) > U_{\dep(b)}$, $(w,\P(b))$ is a seed failure, since otherwise summing its category discrepancies would give that $\deg_{\bar E_b}(w) \le U_{d-1}(1+q_{d-1})/2 < U_d$, contradicting the degree overflow. As a result, if $\Search(w,b)$ is called, step 3 will call $\Search(w,\P(b))$ and add $(w,\P(b))$ to $\mathcal{F}^\new$. Otherwise, $(w,b) \in \mathcal{F}^\old$ and $\deg_{\bar E_b^\old}(w) = \deg_{\bar E_b^\new}(w)$, so $(w,\P(b)) \in \mathcal{F}^\old$ since $\mathcal{F}^\old$ is complete, and finally $(w,\P(b)) \in \mathcal{F}^\new$ since it is still a seed failure.
    \item Finally, if the event $\deg_{\bar E_c^\new}(x) > U_R$ is caused by the new edge $e$, then $x \in \{u,v\}$ and $c = \routepath_{e,R}$, and step 3 of $\Search(x,c)$ adds $(x,\P(c))$ to $\mathcal{F}^\new$ by property (III). Otherwise, $(x,\P(c))\in\mathcal F^\old$. The persistent overflow at $(x,c)$ implies that $(x,\P(c))$ is still a seed failure, so it remains in $\mathcal F^\new$.
\end{itemize}
\end{proof}

\paragraph{A Naive Failure Repair Stage}

As in \Cref{sec:dynamic-lll}, define $\mathcal{T}^\old$ as the static algorithm transcript of the old instance with seed failure set $\mathcal{F}^\old$, and $\mathcal{T}^\new$ as the static algorithm transcript of the new instance using $\mathcal{F}^\new$. The two transcripts use the same randomness for each edge. For a transcript item $X$ and $\alpha \in \{\old,\new\}$, we use $X^\alpha$ to refer to the value of $X$ in $\mathcal{T}^\alpha$. We assume that the algorithm knows $\mathcal{T}^\old$ before the update and constructs $\mathcal{T}^\new$ by updating $\mathcal{T}^\old$.

We use the following definition to group the transcript data.
\begin{definition}
    For a transcript $\mathcal{T}^\alpha$, an internal node $a$, an integer $i \in [0,m_{\dep(a)}]$, and a vertex $w$, we define the \emph{local information} $\sigma_{a,i}^\alpha(w)$ to be the following:
    \begin{itemize}
        \item The set of edges incident to $w$ in $E_a^\alpha$;
        \item The set of edges incident to $w$ in $E_{a,j}^{\bad,\alpha}$ for $1 \le j \le i$;
        \item Whether $w$ belongs to $\bar V_{a,j}^\alpha$ for $0 \le j \le i$.
    \end{itemize}
\end{definition}

For each internal node $a$ of depth $d$ and $i \in [0,m_d]$, we maintain a dirty set $D_{a,i}$. It is intended to contain all vertices $w$ with $\sigma_{a,i}^\old(w) \ne \sigma_{a,i}^\new(w)$, and we say $D_{a,i}$ is \emph{good} if it indeed contains all such vertices. For now, assume that all dirty sets are good.

The repair process for $a \ne r_T$ starts after the process for $\P(a)$ completes and produces a set $D_{\P(a)}^\star$. We initialize $D_{a,0}$ to be $D_{\P(a)}^\star$ when $a \ne r_T$, and $D_{r_T,0}$ is initialized to contain the two endpoints of the updated edge $e$. Additionally, for every vertex $w$ with $(w,a) \in \mathcal{F}^\old \oplus \mathcal{F}^\new$, we add $w$ to $D_{a,0}$. For every vertex in $D_{a,0}$, we compute $\sigma^\new_{a,0}$ based on $\mathcal{F}^\new$, $E_a^\new$ and $V_a^{\star,\new} = \bar V_{\P(a)}^\new$ when $a \ne r_T$, which have already been constructed during the discovery stage and the repair process for $\P(a)$.

If $D_{a,0}$ is empty, we can skip the process for repairing $\bar{\mathcal{A}}(a)$. Otherwise, categories are processed one by one to update the local information. Since the pre-shattering process for each category is local, we mainly focus on vertices around $D_{a,i}$ for recomputation. Specifically, define
\[
  J_a = (V, E_a^\old \cup E_a^\new).
\]
From the algorithm description, $\sigma_{a,i}^\old(x)$ depends only on $\sigma_{a,i-1}^\old(y)$ for $y$ inside the two-hop neighborhood of $x$ in $J_a$. The same holds for $\sigma_{a,i}^\new(x)$. As a result, if $x$ does not lie in the two-hop neighborhood of $D_{a,i-1}$ in $J_a$, $\sigma_{a,i}^\old(x) = \sigma_{a,i}^\new(x)$. So it suffices to compute $\sigma_{a,i}^\new(x)$ for every $x$ within the two-hop neighborhood of $D_{a,i-1}$ in $J_a$ by simulating step 4(b) in $\bar {\mathcal{A}}(a)$.

After the local information update, $E_a^{\bad,\new}$ is constructed. For each connected component $C$ in this graph, if $C$ does not intersect $D_{a,m_d}$, then the same component also appears in $\mathcal{T}^\old$, in which case we can reuse the old bipartition. For all other connected components, we recompute and store a new bipartition.

Let $D_a^\star$ be the set of vertices whose connected component in $E_a^{\bad,\new}$ touches $D_{a,m_d}$. Notice that $D_{a,m_d} \subseteq D_a^\star$. At the end of the repair process, $D_a^\star$ is passed to both children for initialization, and the repair process continues until we reach a leaf.

\paragraph{Failure Repair Stage, with a Clever Initialization}

The previous algorithm cannot avoid the $O(\log \Delta)$ overhead in work, since the edge update itself can affect the local information of $R = O(\log \Delta)$ nodes in the recursion tree even if it does not cause any change in the execution. To address this, we notice that \Cref{obs:edge-coloring-const-passive-incidence} describes the structure of $\sigma_{a,i}(x)$ when $x \notin \bar V_{a,i}$.

\begin{observation}\label{obs:edge-coloring-const-passive-vertex-local-info}
    For $\alpha \in \{\old, \new\}$, an internal node $a$ of depth $d$, an integer $i \in [0,m_d]$, and a vertex $x \notin \bar V_{a,i}^\alpha$, $\sigma_{a,i}^\alpha(x)$ stores the following: its adjacency list in $E_a^\alpha$ is the same as in $\bar E_a^\alpha = \cup_{b \in \mathcal{L}_T(a)} \bar E_b^\alpha$; none of its incident edges enter $E_{a,j}^{\bad,\alpha}$ for $1 \le j \le i$; and $x \notin \bar V_{a,j}^\alpha$ for $0 \le j \le i$.
\end{observation}
\begin{proof}
    By the algorithm description, $\bar V_{a,j-1}^\alpha \subseteq \bar V_{a,j}^\alpha$ for every $1 \le j \le i$. Thus, $x \notin \bar V_{a,j}^\alpha$ for $0 \le j \le i$, and $x \notin V_a^{\star, \alpha}$. By \Cref{obs:edge-coloring-const-passive-incidence}, its adjacency list in $E_a^\alpha$ is the same as that in $\bar E_a^\alpha$. Finally, since $\bar V_{a,j}^\alpha$ includes $x$ whenever an incident edge of $x$ is added to $E_{a,j}^{\bad,\alpha}$, we conclude that $E_{a,j}^{\bad,\alpha}$ has no edge incident to $x$ for every $1 \le j \le i$.
\end{proof}

Since we have already maintained $\bar E_b^\new$ for every leaf $b$ during failure discovery, we only need to explicitly store $\sigma_{a,i}(x)$ when $x$ is activated.

We can see the benefit of this implicit representation when neither endpoint of $e$ is activated in either $\mathcal{T}^\old$ or $\mathcal{T}^\new$ and $\mathcal{F}^\old = \mathcal{F}^\new$. In this case, the repair stage requires no work, since the discovery stage has already updated $\bar E_{\routepath_{e,R}}$, and all local information affected by this update is represented implicitly through these leaf edge sets.

This motivates our new initialization rule. We compute $s_e$ to be the shallowest internal node on the routing path of the updated edge $e$ such that an endpoint of $e$ belongs to $\bar V_{s_e}^\old$. Define $s_e = \routepath_{e,R}$ when such a node does not exist. Instead of adding $u$ and $v$ to $D_{r_T,0}$, we will add $u$ and $v$ to $D_{a,0}$ for every node $a$ along the path from $s_e$ to $\routepath_{e,R}$.

Next, we prove the correctness of the algorithm. Recall that $D_{a,i}$ is good if it contains every vertex $w$ whose local information differs between the two transcripts. We first show that the goodness property propagates from top to bottom.

\begin{observation}\label{obs:edge-coloring-const-dirty-set-goodness-per-category}
For an internal node $a$ and $0\le i<m_{\dep(a)}$, if $D_{a,i}$ is good, then so is $D_{a,i+1}$.
\end{observation}
\begin{proof}
Consider a vertex $x$ outside the two-hop neighborhood of $D_{a,i}$ in $J_a$. Since $D_{a,i}$ is good, every vertex within distance at most two of $x$ has the same local information for category $i$ in both transcripts. From the algorithm description, only these vertices can affect the local information for category $i+1$. Thus every vertex whose local information for category $i+1$ changes lies in the two-hop neighborhood of $D_{a,i}$ in $J_a$. The repair process examines exactly this neighborhood, so $D_{a,i+1}$ is good.
\end{proof}

\begin{observation}\label{obs:edge-coloring-const-dirty-set-goodness-in-subtree}
For an internal node $a\ne r_T$ of depth $d$, if $D_{\P(a),m_{d-1}}$ is good, then so is $D_{a,0}$.
\end{observation}
\begin{proof}
Consider a vertex $x$ whose local information $\sigma_{a,0}(x)$ differs between the two transcripts. Suppose that an edge $f$ incident to $x$ enters or leaves $E_a$. If its presence in $E_{\P(a)}$ or its status in the parent's pre-shattering process changes, the goodness of $D_{\P(a),m_{d-1}}$ places $x$ in $D_{\P(a),m_{d-1}}\subseteq D_{a,0}$. Otherwise, $f$ is bad in both transcripts and its post-shattering assignment changes. The deterministic bipartition for each component then implies that its component has changed. Its new component contains an endpoint of a changed bad edge, so $x\in D_{\P(a)}^\star\subseteq D_{a,0}$.

The remaining possibility is a change in membership in $\bar V_{a,0}=\bar V_{\P(a)}\cup\{w:(w,a)\in\mathcal F\}$. A change in the first set puts $x$ in $D_{\P(a),m_{d-1}}$, and a change in the second set adds $x$ directly to $D_{a,0}$. Thus $D_{a,0}$ is good.
\end{proof}

We next show that a good dirty set produces a correct repair.

\begin{lemma}\label{lem:edge-coloring-const-bottom-up-correctness-single-node}
Let $b$ be an internal node of depth $d$. Suppose that, before the repair at $b$, the algorithm has constructed $E_b^\new$ and also $\bar V_{\P(b)}^\new$ when $b \ne r_T$, and that $D_{b,0}$ is good. Then the repair at $b$ constructs $E_{\LC(b)}^\new$, $E_{\RC(b)}^\new$, $\bar V_b^\new$, and all local information at $b$ in $\mathcal T^\new$.
\end{lemma}
\begin{proof}
We use induction on the category $i$ to prove that $\sigma_{b,i}^\new(x)$ is constructed for every vertex $x$ during the repair process. The base case $i=0$ is implied by the fact that $D_{b,0}$ is good and the algorithm has access to $\mathcal{F}^\new$, $E_b^\new$, and $\bar V_{\P(b)}^\new$. For the induction step, by \Cref{obs:edge-coloring-const-dirty-set-goodness-per-category}, $D_{b,i}$ is good. For every vertex outside $D_{b,i}$, $\sigma_{b,i}^\old(x) = \sigma_{b,i}^\new(x)$; for vertices inside $D_{b,i}$, $\sigma_{b,i}^\new(x)$ is computed from category-$(i-1)$ local information given by the induction hypothesis.

As $\sigma_{b,m_d}^\new(x)$ tracks whether $x$ belongs to $\bar V_b^\new$, $\bar V_b^\new$ is also constructed, as is $E_b^{\bad,\new}$. The fixed randomness determines the child of $b$ to which each edge in $E_b^\new \backslash E_b^{\bad,\new}$ is sent. Additionally, the post-shattering process recomputes every component touching $D_{b,m_d}$ and reuses the bipartition for old components. Since the post-shattering process is deterministic, the resulting bipartition is the same as that generated in $\mathcal{T}^\new$. Thus, the algorithm constructs $E_{\LC(b)}^\new$ and $E_{\RC(b)}^\new$.
\end{proof}

Together, the preceding results would prove the correctness of the algorithm if $D_{r_T,0}$ were good. This is the case when we add $u$ and $v$ to $D_{r_T,0}$, but it does not hold for the new initialization. Define $\bar D_{a,i}$ to be the set of vertices $x \notin D_{a,i}$ such that $\sigma_{a,i}^\old(x) \ne \sigma_{a,i}^\new(x)$. We prove a stronger invariant for the algorithm with the new initialization, consisting of three items for every internal node $a$ of depth $d$ and $i \in [0,m_d]$:

\begin{itemize}
    \item[1.] For every vertex $x$, $\sigma_{a,i}^\new(x)$ is constructed.
    \item[2.] If $x \in \bar D_{a,i}$, then $a$ is a strict ancestor of $s_e$, $x$ is one of the endpoints of $e$, and $x \notin \bar V_{a,i}^\old \cup \bar V_{a,i}^\new$.
    \item[3.] $E_{\LC(a)}^\new$, $E_{\RC(a)}^\new$ and $\bar V_a^\new$ are constructed.
\end{itemize}

We first show that invariants 1 and 2 for $(a,i)$ imply those for $(a,i+1)$.

\begin{lemma}\label{lem:edge-coloring-const-correctness-same-node-1}
    For an internal node $a$ of depth $d$, if invariants 1 and 2 for $(a,i)$ hold for $i < m_d$, then every vertex $x$ that does not belong to either $\bar D_{a,i}$ or the two-hop neighborhood of $D_{a,i}$ in $J_a$ satisfies $\sigma_{a,i+1}^\old(x) = \sigma_{a,i+1}^\new(x)$.
\end{lemma}
\begin{proof}
    Since $x \notin D_{a,i} \cup \bar D_{a,i}$, $\sigma_{a,i}^\old(x) = \sigma_{a,i}^\new(x)$ by invariant 1. As a result, when $\sigma_{a,i+1}^\old(x) \ne \sigma_{a,i+1}^\new(x)$, either $x$ is incident to an edge in $E_{a,i+1}^{\bad,\old} \oplus E_{a,i+1}^{\bad,\new}$, or $x \in \bar V_{a,i+1}^\old \oplus \bar V_{a,i+1}^\new$.

    For the case $x \in \bar V_{a,i+1}^\old \oplus \bar V_{a,i+1}^\new$, either $x \in \bar V_{a,i}^\old \oplus \bar V_{a,i}^\new$, where $x \in D_{a,i}$ by invariants 1 and 2, a contradiction, or some $(x,y) \in E_{a,i+1}^{\bad,\old} \oplus E_{a,i+1}^{\bad,\new}$, which we will refute next.

    If $(x,y) \in E_{a,i+1}^{\bad,\old} \oplus E_{a,i+1}^{\bad,\new}$, a failure $(z,a,i+1)$ is responsible for moving it into $E_{a,i+1}^\bad$ in the corresponding transcript. $z \notin \bar D_{a,i}$ since $\bar D_{a,i} \cap (\bar V_{a,i}^\old \cup \bar V_{a,i}^\new) = \varnothing$ by invariant 2, and $z \notin D_{a,i}$ since $z$ and $x$ are within two hops. These imply that $\sigma_{a,i}^\old(z) = \sigma_{a,i}^\new(z)$, and $(z,a,i+1)$ is a failure in both transcripts. However, this failure adds $(x,y)$ to $E_{a,i+1}^\bad$ in only one transcript, so some $w \in \{x,y,z\}$ satisfies $\sigma_{a,i}^\old(w) \ne \sigma_{a,i}^\new(w)$. $w$ cannot belong to $D_{a,i}$ by the choice of $x$, so $w \in \bar D_{a,i}$ and $w$ must be $y$ since the other two vertices do not belong to $\bar D_{a,i}$.
    
    By invariant 2, $a$ is a strict ancestor of $s_e$, $y$ is an endpoint of $e$ and $y \notin \bar V_{a,i}^\old \cup \bar V_{a,i}^\new$. By \Cref{obs:edge-coloring-const-passive-vertex-local-info}, the only difference between $\sigma_{a,i}^\old(y)$ and $\sigma_{a,i}^\new(y)$ is the edge $e$. Either $e$ connects $y$ and $z$, meaning that $z$, as an endpoint of $e$, belongs to $\bar V_{a,i+1}^\old$; or $e$ does not connect $y$ and $z$, in which case $(y,z) \in E_a^\old$ and $y$, as an endpoint of $e$, belongs to $\bar V_{a,i+1}^\old$. In sum, some endpoint of $e$ belongs to $\bar V_{a,i+1}^\old \subseteq \bar V_a^\old$, and $a$ cannot be a strict ancestor of $s_e$, a contradiction.
\end{proof}

\begin{lemma}\label{lem:edge-coloring-const-correctness-same-node}
    For a node $a$ of depth $d$, invariants 1 and 2 for $(a,i)$ imply the same invariants for $(a,i+1)$ when $i \in [0,m_d-1]$.
\end{lemma}
\begin{proof}
    \Cref{lem:edge-coloring-const-correctness-same-node-1} immediately implies that $\bar D_{a,i+1}$ only contains endpoints of $e$. Now we show that for every $x \in \bar D_{a,i+1}$, $x \notin \bar V_{a,i+1}^\old \cup \bar V_{a,i+1}^\new$. Together with invariant 2 for $(a,i)$, this proves invariant 2 for $(a,i+1)$.

    The case $x \in \bar V_{a,i+1}^\old$ contradicts the choice of $s_e$. We may also assume that $x$ is outside the two-hop neighborhood of $D_{a,i}$ in $J_a$ since otherwise $x \in D_{a,i+1}$. As a consequence, the only case we need to refute is $x \in \bar V_{a,i+1}^\new \backslash \bar V_{a,i+1}^\old$. We can find a failure $(z,a,i+1)$ in $\mathcal{T}^\new$ that moves an edge $(x,y)$ to $E_{a,i+1}^{\bad,\new}$. As $z$ and $x$ are within two hops in $J_a$, $\sigma_{a,i}^\old(z) = \sigma_{a,i}^\new(z)$ and $(z,a,i+1)$ is also a failure in $\mathcal{T}^\old$. Following the same argument as \Cref{lem:edge-coloring-const-correctness-same-node-1}, we can show that some endpoint of $e$ belongs to $\bar V_{a,i+1}^\old$, contradicting the choice of $s_e$.

    Finally, from \Cref{lem:edge-coloring-const-correctness-same-node-1}, it remains to consider $x \in \bar D_{a,i+1}$ for invariant 1. Since $x \notin \bar V_{a,i+1}^\old \cup \bar V_{a,i+1}^\new$, its local information is represented implicitly in both transcripts, while the leaf edge sets are already updated during failure discovery. This shows that $\sigma_{a,i+1}(x)$ is updated to agree with $\mathcal{T}^\new$.
\end{proof}

Next, we show that invariants 1 and 2 for $(a,m_{\dep(a)})$ imply invariant 3 for $a$ and invariants 1 and 2 for its children.

\begin{lemma}\label{lem:edge-coloring-const-correctness-across-nodes}
    For an internal node $a$ of depth $d$, invariants 1 and 2 for $(a,m_d)$ imply invariant 3 for $a$ and invariants 1 and 2 for $(\LC(a),0)$ and $(\RC(a),0)$ when $\LC(a)$ and $\RC(a)$ are internal.
\end{lemma}
\begin{proof}
    First consider the case where $\bar D_{a,m_d} = \varnothing$. Following the post-shattering argument in \Cref{lem:edge-coloring-const-bottom-up-correctness-single-node}, $E_{\LC(a)}^\new$, $E_{\RC(a)}^\new$, and $\bar V_a^\new$ are constructed, giving invariant 3. Further, $\bar D_{\LC(a),0} = \bar D_{\RC(a),0} = \varnothing$ from \Cref{obs:edge-coloring-const-dirty-set-goodness-in-subtree}, and the initialization on the children computes new local information, establishing invariants 1 and 2 for both children.

    Now suppose $\bar D_{a,m_d}$ is not empty. As no vertex in $\bar D_{a,m_d}$ belongs to $\bar V_a^\old \cup \bar V_a^\new$, these vertices are isolated in both $E_a^{\bad,\old}$ and $E_a^{\bad,\new}$, and the post-shattering bipartition does not change when we add them to $D_{a,m_d}$. This means $E_{\LC(a)}^\new$ and $E_{\RC(a)}^\new$ are constructed. Additionally, $\bar V_a^\new$ can be generated from the local information given by invariant 1 for $(a,m_d)$. This shows invariant 3 for $a$.

    Now, fix an internal child $c$ of $a$. We know that adding $\bar D_{a,m_d}$ to $D_a^\star$ will make $D_{c,0}$ good since every vertex in $\bar D_{a,m_d}$ is isolated. Thus, compared with the execution using a good dirty set, initialization at node $c$ with dirty set $D_{c,0}$ ignores only vertices in $\bar D_{a,m_d}$. Consequently, $\bar D_{c,0} \subseteq \bar D_{a,m_d}$, and every vertex in $\bar D_{c,0}$ is an endpoint of $e$ by invariant 2 at $(a,m_d)$.

Now we show that $\bar D_{c,0}$ is empty when $c$ is not a strict ancestor of $s_e$. Fix $x\in\bar D_{c,0}$. By invariant 2 at $(a,m_d)$, $x \in \{u,v\}$ and $x \notin \bar V_a^\old \cup \bar V_a^\new$. If $x \in \bar V_{c,0}^\old \oplus \bar V_{c,0}^\new$, then $(x,c) \in \mathcal{F}^\old \oplus \mathcal{F}^\new$ and $x \in D_{c,0}$, a contradiction. Since $\sigma_{c,0}^\old(x) \ne \sigma_{c,0}^\new(x)$, the adjacency information at $x$ must be different between $E_c^\old$ and $E_c^\new$.

By \Cref{obs:edge-coloring-const-passive-incidence}, the sets of edges incident to $x$ are the same in $E_c^\old$ and $\bar E_c^\old$, as well as in $E_c^\new$ and $\bar E_c^\new$. If $c$ does not belong to the routing path of $e$, then the edges incident to $x$ are the same in $\bar E_c^\old$ and $\bar E_c^\new$, and $x \notin \bar D_{c,0}$. If $c$ belongs to the routing path of $e$ but is not a strict ancestor of $s_e$, both endpoints of $e$ are added to $D_{c,0}$. Thus, $\bar D_{c,0}$ is empty whenever $c$ is not a strict ancestor of $s_e$.

Finally, assume that $x \in \bar D_{c,0}$, so $c$ is a strict ancestor of $s_e$. Since $x$ cannot belong to $\bar V_{c,0}^\old \oplus \bar V_{c,0}^\new$, it either belongs to $\bar V_{c,0}^\old \cap \bar V_{c,0}^\new$ or does not belong to $\bar V_{c,0}^\old \cup \bar V_{c,0}^\new$. The first case cannot happen, since then $s_e$ would be an ancestor of $c$. This proves invariant 2 for $(c,0)$.

We finally show invariant 1 for $(c,0)$. By invariant 3 at $a$, $E_c^\new$ and $\bar V_a^\new$ are constructed, so each vertex $x \in D_{c,0}$ gets $\sigma_{c,0}^\new(x)$. Every vertex $x$ in $\bar D_{c,0}$ does not belong to $\bar V_{c,0}^\old \cup \bar V_{c,0}^\new$, so $\sigma_{c,0}$ is maintained implicitly in both transcripts, and it is updated to agree with $\mathcal{T}^\new$ during the discovery stage. Other vertices do not change their local information. This concludes invariant 1.

\end{proof}

We now combine these results to prove correctness.

\begin{lemma}\label{lem:edge-coloring-const-correctness}
After the failure repair process for an update $e$, all invariants hold for every internal node $a$ and $i \in [0,m_{\dep(a)}]$. Specifically, the algorithm constructs $\mathcal{T}^\new$.
\end{lemma}
\begin{proof}
We induct on $(\dep(a),i)$ to prove all invariants. The step from $(a,i)$ to $(a,i+1)$ is given in \Cref{lem:edge-coloring-const-correctness-same-node}, while the step from $(a,m_{\dep(a)})$ to $(\LC(a),0)$ and $(\RC(a),0)$ is given in \Cref{lem:edge-coloring-const-correctness-across-nodes}. It remains to consider the base case $a=r_T$ and $i=0$.

The local information for vertices whose membership in $\bar V_{r_T,0}$ changes is computed explicitly. Every other vertex $x$ with $\sigma_{r_T,0}^\old(x) \ne \sigma_{r_T,0}^\new(x)$ must be an endpoint of $e$, meaning that $\bar D_{r_T,0}$ only contains endpoints of $e$. If $x \notin D_{r_T,0}$, then $s_e \ne r_T$, meaning that $x \notin \bar V_{r_T,0}^\old$ and $\sigma_{r_T,0}(x)$ is represented implicitly. We have applied the update $e$ to $\bar E_{\routepath_{e,R}}$ in the failure discovery stage, so $\sigma_{r_T,0}(x)$ is updated correspondingly to agree with $\mathcal{T}^\new$, giving invariant 1. Finally, $x \notin \bar V_{r_T,0}^\new$ since otherwise $(x,r_T) \in \mathcal{F}^\old \oplus \mathcal{F}^\new$ and $x \in D_{r_T,0}$. This gives invariant 2.
\end{proof}

\subsubsection{Tools for Analyzing Expected Work}

We first introduce the objects needed to analyze the expected work in both stages.

\begin{definition}[Critical failure]\label{def:critical-failure}
For an update, a pair $(w,a) \in V(G) \times T$ is a \emph{critical failure} if there is $\alpha \in \{\old, \new\}$ such that $(w,a)$ is a seed failure in $\mathcal{T}^\alpha$ and either $a=r_T$ or $(w,\P(a))$ is not a seed failure in $\mathcal{T}^\alpha$.
\end{definition}
Critical failures can be seen as the starting points where one of $\mathcal{T}^\old$ and $\mathcal{T}^\new$ deviates from the ideal execution. If $(w,a)$ is a seed failure at a leaf, its parent is also a seed failure by the discrepancy argument in the proof of \Cref{lem:discover-stage-invariants}. Hence every critical failure is at an internal node.

We also define $\tau_d=U_dq_d^3=U_d/\log^6 U_d$ for $d\in[0,R]$. Since $U_R$ is large enough, one can show that the sequence $\tau_0,\dots,\tau_R$ decreases geometrically.

\begin{lemma}\label{lem:critical-failure-tail}
Fix an update in the dynamic algorithm. For every vertex $w$ and every internal node $a$ of depth $d$,
\[
    \Pr[(w,a)\text{ is a critical failure}]
    \le\exp(-\Omega(\tau_d)).
\]
The bound remains valid after conditioning on all randomness at levels smaller than $d$, on the level-$d$ randomness of edges not incident to $w$ in $\bar E_a^\cup = \bar E_a^\old \cup \bar E_a^\new$, and on the randomness of the edge $e$ being updated.
\end{lemma}
\begin{proof}
We prove the conditional statement; the unconditional one follows from the law of total probability.

Fix $\alpha \in \{\old, \new\}$. When $d>0$, whether $(w,\P(a))$ is a seed failure in $\mathcal{T}^\alpha$ is determined by the fixed randomness above $a$. If it is, $(w,a)$ cannot be a critical failure in $\mathcal{T}^\alpha$; otherwise, $\deg_{\bar E_a^\alpha}(w) \le U_{d-1}(1+q_{d-1})/2\le U_d$. For $a = r_T$, we also have $\deg_{\bar E_a^\alpha}(w) \le U_d$.

Now, fix $i \in [m_d]$. Let $X_i$ count the category-$i$ edges incident to $w$ in $\bar E_a^\alpha$. It is a sum of at most $U_d$ independent Bernoulli variables, each with success probability $1/m_d$, except that the contribution of the updated edge, if present, is fixed. Since $U_d/m_d$ is sufficiently large, a Chernoff bound gives
\[
    \Pr\!\left[X_i>2U_d/m_d\right] = \exp(-\Omega(U_dq_d)) = \exp(-\Omega(\tau_d)).
\]

Now condition on the category choices under the event $X_i\le2U_d/m_d$. The discrepancy $Y_i$ between $\LC$ and $\RC$ directions around $w$ for category-$i$ edges is a sum of at most $2U_d/m_d$ independent uniform signs, apart from a possible fixed $\pm1$ from the updated edge. Hoeffding's inequality then gives
\[
    \Pr\!\left[|Y_i|>U_dq_d/m_d\mid X_i \le 2U_d/m_d\right]
    = \exp\!\left(-\Omega\!\left(\frac{(U_dq_d/m_d)^2}{U_d/m_d}\right)\right)
    =\exp(-\Omega(\tau_d)).
\]
A union bound over the $m_d$ categories and $\alpha\in\{\old,\new\}$ proves the bound. The events do not depend on the randomness at level $d$ for edges not incident to $w$ in $\bar E_a^\cup$, so conditioning on this randomness has no effect.
\end{proof}

For $d<R$, put $\rho_d=2L_d+1$. For a critical failure $(w,a)$, let its \emph{affected region} $\mathcal R_{w,a}$ be the induced subgraph of the $\rho_{\dep(a)}$-hop neighborhood of $w$ in $\bar E_a^\cup$. Next, we define the most important auxiliary graph for the analysis.

\begin{definition}[$\hat G$]
The graph $\hat G=(\hat V,\hat E)$ has vertex set
\[
    \hat V=\{\langle x,a\rangle:x\in V(\mathcal R_{w,a})\text{ for some critical failure }(w,a)\}.
\]
For each affected region $\mathcal R_{w,a}$, there is a \emph{horizontal} edge in $\hat E$ between each pair of vertices $\langle x,a\rangle,\langle y,a\rangle$ for which $x$ and $y$ belong to $V(\mathcal R_{w,a})$. There is also a \emph{vertical} edge in $\hat E$ between $\langle x,a\rangle$ and $\langle x,b\rangle$ whenever $a\ne b$ are comparable and both vertices belong to $\hat V$.
\end{definition}

Notice that we use $\langle x,a \rangle$ for vertices in $\hat V$ to distinguish them from critical failures $(x,a)$. For $\langle w,a\rangle \in V(G) \times T$, let $\hat K(\langle w,a\rangle)$ be its component in $\hat G$ if $\langle w,a\rangle\in\hat V$, and let it be empty otherwise. The following lemma will be central in our expected work analysis.

\begin{lemma}\label{lem:hat-G-connected-component-size-bound}
For every $w\in V(G)$, $a\in T\setminus\mathcal L_T$, and $0\le d<R$,
\[
    \mathbb E\!\left[\left|\{\langle x,b\rangle\in\hat K(\langle w,a\rangle):\dep(b)=d\}\right|\right]
    \le\exp\!\left(-\Omega\!\left(\tau_d + \tau_{\dep(a)}\right)\right).
\]
The same bound holds after conditioning on all random choices of the updated edge.
\end{lemma}

We prove \Cref{lem:hat-G-connected-component-size-bound} in \Cref{sec:witness}. By a direct summation and the fact that $\tau_i$ is geometrically decreasing, the following corollary is immediate.

\begin{corollary}\label{cor:edge-coloring-const-component-size-bound-sum}
    For an update $e = (u,v)$, fix the random choices of $e$. Let $N_r$ count the number of vertices $\langle w,b\rangle$ in $\hat V$ connected to $\langle u,\routepath_{e,i}\rangle$ for some $0\le i<R$ such that $\dep(b) = r$. Then
\[
    \mathbb E[N_r]
    \le\sum_{i=0}^{R-1}\exp(-\Omega(\tau_r+\tau_i))
    =\exp(-\Omega(\tau_r)).
\]
\end{corollary}

\subsubsection{Implementation Details and Analysis}

We now describe the implementation of both stages. Before the algorithm processes edge updates, we compute $R$, $U_i$ and $q_i$ for $i\in[0,R]$, and $m_i$ for $i\in[0,R-1]$ in $O(\log \Delta)$ time.

\paragraph{Implementing the Discovery Stage}

For each edge update $e = (u,v)$, we generate $\dir_{e,0},\dots,\dir_{e,R-1}$ if it is an insertion, and insert or delete $e$ in $\bar E_{\routepath_{e,R}}$. Mapping $\LC$ to zero and $\RC$ to one, the sequence $\dir_{e,0},\dots,\dir_{e,R-1}$ corresponds to a uniform integer $B_e$ in $[0,2^R-1]$. We can therefore define a mapping $\phi$ from $[0,2^R-1]$ to $\mathcal{L}_T$ such that $\phi(B_e) = \routepath_{e,R}$ for every possible $B_e$. Since $R = O(\log \Delta) = O(\log n)$, we can generate the random directions and identify $\routepath_{e,R}$ in $O(1)$ time.

To apply $e$ in $\bar E_{\routepath_{e,R}}$ in constant time, we maintain the adjacency information with a hash table $\mathsf{Adj}_{a,x}$ of size $U_R$ for every leaf $a$ and every vertex $x$, using chaining to resolve collisions. The algorithm maintains an independent random function $f$ from $V$ to $[\Delta/2^R]$, each entry sampled lazily. Since $\Delta/2^R\le\Delta\le n-1$, each value of $f$ fits in an $O(\log n)$-bit word and can be sampled and accessed in $O(1)$ time. Each time we apply $e$, we insert or delete $v$ in $\mathsf{Adj}_{a,u,f(v)}$, and insert or delete $u$ in $\mathsf{Adj}_{a,v,f(u)}$. The length of $\mathsf{Adj}_{a,u,f(v)}$ is constant in expectation: indeed, an edge $(x,y) \in E(G)$ belongs to $\mathsf{Adj}_{a,u,f(v)}$ only when $x = u$, $B_{(x,y)} = B_e$ and $f(y) = f(v)$. The first condition ensures that there are at most $\Delta$ edges other than $e$ that can enter $\mathsf{Adj}_{a,u,f(v)}$, and the remaining two conditions hold with probability $1/2^R \times 2^R/\Delta = 1/\Delta$ for every such edge.

For every vertex $w$ and leaf $a$, we store a linked list of the pairs in $\mathcal F$ whose clusters contain $(w,a)$. This allows us to enumerate the linked lists at $(u,\routepath_{e,R})$ and $(v,\routepath_{e,R})$ for initial $\Search$ calls. According to the proof of \Cref{lem:discover-stage-invariants}, these lists also include all pairs in $\mathcal{F}$ whose cluster or status as a seed failure can change after the update.

During a $\Search(w,b)$ call, we need to check whether $(w,b)$ is a seed failure. We first introduce a subroutine that returns the adjacency list of $w$ in $\bar E_b$ when $\deg_{\bar E_b}(w) \le U_{\dep(b)}$ and returns \emph{fail} when $\deg_{\bar E_b}(w) > U_{\dep(b)}$. Since $\bar E_b = \cup_{a \in \mathcal{L}_T(b)} \bar E_a$, we iterate through $a \in \mathcal{L}_T(b)$ and the hash table $\mathsf{Adj}_{a,w}$ to construct the adjacency list. Once the adjacency list is longer than $U_{\dep(b)}$, we return fail; if the list always has length at most $U_{\dep(b)}$, we return the final list. The process iterates through $|\mathcal{L}_T(b)| \Delta/2^R = \Delta/2^{\dep(b)} = O(U_{\dep(b)})$ buckets. Since extracting each item takes $O(1)$ time, the total work performed by this process is $O(U_{\dep(b)})$.

With the process above, we can check whether $(w,b)$ is a seed failure in $\tilde O(U_{\dep(b)})$ time by gathering the adjacency list of $w$ in $\bar E_b$. If the process fails, $(w,b)$ is immediately a seed failure; otherwise, we perform the remaining checks in $\tilde O(U_{\dep(b)})$ time. The process also allows the algorithm to identify all pairs in the cluster of $(w,b)$ and update the linked lists by a straightforward BFS within $\exp(O(\log^4 U_{\dep(b)}))$ time.

\paragraph{Cost Analysis for the Discovery Stage}

We have argued that maintaining $\bar E_a$ takes $O(1)$ time; it remains to consider the $\Search$ calls. Say a $\Search(w,b)$ call \emph{succeeds} if $(w,b)$ is a seed failure in $\mathcal T^\new$, and \emph{fails} otherwise.

For every seed failure $(w,b)$ in either $\mathcal{T}^\old$ or $\mathcal{T}^\new$, we can always find a critical failure $(w,c)$ such that $c$ is an ancestor of $b$. Take one of maximum depth and call it the \emph{corresponding critical failure} of $(w,b)$.

\begin{lemma}\label{lem:bottom-up-changed-failures-to-hat-G}
For every successful call $\Search(w,b)$, the vertex $\langle w,c\rangle$ associated with its corresponding critical failure $(w,c)$ is connected in $\hat G$ to $\langle u,\routepath_{e,i}\rangle$ for some $0\le i<R$. The same conclusion holds for every failed initial call $\Search(w,b)$ with $(w,b) \in \mathcal F^\old$.
\end{lemma}
\begin{proof}

First, notice that $\langle w,c \rangle \in \hat G$ for every critical failure $(w,c)$. We induct on the forest of $\Search$ calls to prove the statement, where the parent of a noninitial $\Search(w,b)$ call is the first successful $\Search(x,c)$ call that generates it. For the base case, consider a successful initial $\Search(w,b)$ call. Then $(w,b)$ is a seed failure in $\mathcal{T}^\new$, and the old cluster of $(w,b)$ contains either $(u,\routepath_{e,R})$ or $(v,\routepath_{e,R})$. This means that the distance between $u$ and $w$ in $\bar E_b^\cup$ is at most $L_{\dep(b)}+1$. Its corresponding critical failure $(w,c)$ satisfies $u \in \mathcal R_{w,c}$ and $L_{\dep(b)}\le L_{\dep(c)}$. Finally, $b$ is an ancestor of $\routepath_{e,R}$ and $c$ is an ancestor of $b$, so $c = \routepath_{e,\dep(c)}$, and $\langle w,c\rangle$ has the connectivity required by the lemma. The statement for a failed initial call $\Search(w,b)$ with $(w,b) \in \mathcal{F}^\old$ is the same as above, except that $(w,b)$ is a seed failure in $\mathcal{T}^\old$ rather than $\mathcal{T}^\new$.

For the induction step, suppose a successful noninitial $\Search(w,b)$ call is generated by $\Search(x,s)$. The induction hypothesis shows that the vertex $\langle x,s'\rangle$ associated with the corresponding critical failure $(x,s')$ of $(x,s)$ is connected to $\langle u,\routepath_{e,i}\rangle$ in $\hat G$ for some $i \in [0,R-1]$. By the algorithm description, either $w=x$ and $b = \P(s)$, or $(w,b)$ is in the new cluster of $(x,s)$. In both cases, the distance between $w$ and $x$ in $\bar E_s^\cup \subseteq \bar E_{s'}^\cup$ is at most $L_{\dep(s)}\le \rho_{\dep(s')}$, meaning that $\langle x,s'\rangle$ is connected to $\langle w,s'\rangle$ by horizontal edges in $\hat G$. Let $(w,c)$ be the corresponding critical failure of $(w,b)$. Then $c$ is comparable to $s'$ since $b$ is, and $\langle w,s'\rangle$ and $\langle w,c\rangle$ are connected by a vertical edge. This finishes the induction.
\end{proof}

\begin{lemma}\label{lem:bottom-up-exploration-cost}
The discovery stage has $O(1)$ expected worst-case update time.
\end{lemma}
\begin{proof}
Fix the random choices for the updated edge. Let $S_d$ count the number of successful $\Search(w,b)$ calls with $\dep(b) = d$. By \Cref{lem:bottom-up-changed-failures-to-hat-G}, each is associated with a vertex $\langle w,c\rangle$ counted by $N_r$ for some $r\le\min(d,R-1)$. A fixed vertex $\langle w,c\rangle$ counted by $N_r$ contributes at most $2^{d-r}$ to $S_d$. Therefore, by \Cref{cor:edge-coloring-const-component-size-bound-sum},
\[
    \mathbb E[S_d]
    \le\sum_{r=0}^{\min(d,R-1)}2^{d-r}
       \exp(-\Omega(\tau_r))
    =\exp(-\Omega(\tau_d)).
\]

Charge the work on every failed noninitial call to the successful call that generates it. A successful $\Search(w,b)$ call at depth $d$ generates at most $\exp(O(\log^4 U_d))$ $\Search(x,s)$ calls with $U_{\dep(s)}=O(U_d)$, so finding these pairs, filtering out seed failures among them, and maintaining the linked lists take $\exp(O(\log^4 U_d))$ work in total. Consequently, the total expected work for successful calls is at most
\[
    \sum_{d=0}^R
       \exp(O(\log^4 U_d)-\Omega(\tau_d))
    =O(1)
\]
since $\tau_d \gg \log^4 U_d$, the sequence $\tau_d$ decreases geometrically, and $\tau_R$ is sufficiently large. For the failed initial $\Search(w,b)$ calls, the cases in which $(w,b) = (u,\routepath_{e,R})$ or $(w,b) = (v,\routepath_{e,R})$ take $O(1)$ time since we can keep track of vertex degrees in $\bar E_{\routepath_{e,R}}$ along the way. The other cases follow the same analysis as above due to the second part of \Cref{lem:bottom-up-changed-failures-to-hat-G}.
\end{proof}

\paragraph{Cost Analysis for the Repair Stage} We have already described the implementation, so we turn to the cost analysis. We first relate the dirty sets to connected components in $\hat G$.

\begin{lemma}\label{lem:active-vertices-to-affected-regions}
For every internal node $a$ of depth $d$ and every $x\in \left(\cup_{i \in [0,m_d]}D_{a,i}\right) \cup D_a^\star$, there is a critical failure $(w,b)$ such that $b$ is an ancestor of $a$, $\langle w,b\rangle$ is connected in $\hat G$ to $\langle u,\routepath_{e,d'}\rangle$ for some $d'\in[0,R-1]$, and $\dist_{\bar E_b^\cup}(w,x) \le 1+L_{\dep(b)} + \sum_{j=\dep(b)}^{\dep(a)} 2m_j \le \rho_{\dep(b)}$. Consequently, $\langle x,b\rangle$ is connected to $\langle u,\routepath_{e,d'}\rangle$ in $\hat G$.
\end{lemma}
\begin{proof}
    We induct on $\dep(a)$. First consider $D_{a,0}$. A vertex $x$ is included for one of three reasons:
\begin{itemize}
    \item $(x,a) \in \mathcal{F}^\old \oplus \mathcal{F}^\new$. From \Cref{lem:bottom-up-changed-failures-to-hat-G}, the vertex $\langle x,b\rangle$ associated with its corresponding critical failure $(x,b)$ is connected to $\langle u,\routepath_{e,d'}\rangle$ in $\hat G$ for some $d' \in [0,R-1]$, and $b$ is an ancestor of $a$, so it satisfies the claim.
    \item $a \ne r_T$ and $x \in D_{\P(a)}^\star$. From the induction hypothesis we get the desired critical failure $(w,b)$ with distance bound $1+L_{\dep(b)} + \sum_{j=\dep(b)}^{\dep(a)-1} 2m_j$.
    \item $x \in \{u,v\}$ and $a$ is along the path from $s_e$ to $\routepath_{e,R}$. Either $u$ or $v$ belongs to $\bar V_{s_e}^\old$, and from \Cref{lem:edge-coloring-const-active-to-cluster}, we know that there is a seed failure $(w^\star,a^\star)$ in $\mathcal{T}^\old$ whose cluster contains $(u,a)$ or $(v,a)$. As $(u,v)$ is the updated edge and $a^\star$ is on its routing path, the distance between $w^\star$ and $x$ in $\bar E_{a^\star}^\cup$ is at most $L_{\dep(a^\star)}+1$. Let $(w^\star,b)$ be the corresponding critical failure of $(w^\star,a^\star)$. Then $\langle w^\star,b\rangle$ is connected to $\langle u,b\rangle$ in $\hat G$, so it satisfies the claim.
\end{itemize}

The argument above gives a critical failure $(w,b)$ satisfying all the conditions and the tighter distance bound $\dist_{\bar E_b^\cup}(w,x) \le 1+L_{\dep(b)} + \sum_{j=\dep(b)}^{\dep(a)-1} 2m_j$. It proves the base case for the following induction statement on $i$: For every $x \in D_{a,i}$, there is a critical failure $(w,b)$ satisfying all conditions in the lemma statement and a tighter distance bound
\[
    \dist_{\bar E_b^\cup}(w,x) \le 1+2i+L_{\dep(b)} + \sum_{j=\dep(b)}^{\dep(a)-1} 2m_j \le 1+L_{\dep(b)} + \sum_{j = \dep(b)}^{\dep(a)} 2m_j.
\]

For the induction step, $D_{a,i}$ is within the two-hop neighborhood of $D_{a,i-1}$ for every $i \in [1,m_{\dep(a)}]$ in $J_a$. Notice that the underlying edge set of $J_a$ is $E_a^\old \cup E_a^\new$, rather than $\bar E_a^\cup$. For vertices in $D_{a,i-1}$, apply the induction hypothesis directly. For other vertices $x$, take a shortest path $(x,y,z)$ to $D_{a,i-1}$ in $J_a$, and set $y=z$ if the path has length one. If it lies entirely in $\bar E_a^\cup$, then the induction step holds trivially.

If not, then either $(x,y)$ or $(y,z)$ belongs to $(E_a^\old \cup E_a^\new) \backslash \bar E_a^\cup$, so that edge deviates from its routing path. Let $a'$ be the highest ancestor of $a$ such that either $(x,y)$ or $(y,z)$ belongs to $(E_{a'}^\old \cup E_{a'}^\new) \backslash \bar E_{a'}^\cup$. Then some $p \in \{x,y,z\}$ belongs to $\bar V_{\P(a')}^\old \cup \bar V_{\P(a')}^\new$. By \Cref{lem:edge-coloring-const-active-to-cluster}, there is a seed failure $(w,a'')$ in either transcript whose cluster contains $(p,\P(a'))$. Take its corresponding critical failure $(w,b)$.

We know that $b$ is an ancestor of $a$, and the distance between $w$ and $p$ in $\bar E_b^\cup$ is at most $L_{\dep(b)}$. Additionally, $(x,y)$ and $(y,z)$ belong to $\bar E_b^\cup$ by the choice of $a'$, so $x,y,z$ are in the affected region of $(w,b)$, and the distance between $w$ and $x$ in $\bar E_b^\cup$ is at most $L_{\dep(b)}+2 \le 1+2i+L_{\dep(b)}$. Finally, we show that $\langle w,b\rangle$ is connected to $\langle u,\routepath_{e,d'}\rangle$ in $\hat G$ for some $d' \in [0,R-1]$. The vertex $\langle w,b\rangle$ is connected to $\langle z,b\rangle$ by a horizontal edge, whereas the induction hypothesis gives an ancestor $b'$ of $a$ such that $\langle z,b'\rangle$ is connected to $\langle u,\routepath_{e,d'}\rangle$. Since $b$ and $b'$ are comparable, there is a vertical edge between $\langle z,b\rangle$ and $\langle z,b'\rangle$.

It remains to consider $D_a^\star$. Fix $x \in D_a^\star$ and a shortest path $x_0,x_1,\ldots,x_t$ in $E_a^{\bad,\new}$ from $D_{a,m_{\dep(a)}}$ to $x$. The case $t=0$ is trivial; hence, consider $t \ge 1$. From the argument above, there is some ancestor $b_0'$ of $a$ such that $\langle x_0,b_0'\rangle$ is connected to $\langle u,\routepath_{e,d'}\rangle$ for some $d' \in [0,R-1]$.

For each $j\ge1$, let $a_j$ be the highest ancestor of $a$ such that the edge $(x_{j-1},x_j)$ enters $E_{a_j}^{\bad,\new}$. Such an $a_j$ exists since $a$ is a candidate. We know that $x_j \in \bar V_{a_j}^\new$, so we can apply \Cref{lem:edge-coloring-const-active-to-cluster} to find a seed failure $(w_j,b_j) \in \mathcal{F}^\new$ whose new cluster contains $(x_j,a_j)$ and consider its corresponding critical failure $(w_j,b_j')$. By the choice of $a_j$, $(x_{j-1},x_j)$ belongs to $\bar E_{b_j'}^\cup$, so the distance between $w_j$ and either endpoint of $(x_{j-1},x_j)$ is at most $L_{\dep(b_j')} + 1$, meaning that $\langle x_{j-1},b_j'\rangle$ and $\langle x_j,b_j'\rangle$ are connected by horizontal edges in $\hat G$. At the same time, $\langle x_{j-1},b_j'\rangle$ is connected to $\langle x_{j-1},b_{j-1}'\rangle$ by a vertical edge since they are both ancestors of $a$. Consequently, $(w_t, b_t')$ is a critical failure such that $\langle w_t,b_t'\rangle$ is connected to $\langle u,\routepath_{e,d'}\rangle$, and the distance between $w_t$ and $x$ in $\bar E_{b_t'}^\cup$ is at most $L_{\dep(b_t')}+1$. This proves the claim.
\end{proof}

The following corollary is immediate from \Cref{lem:active-vertices-to-affected-regions,cor:edge-coloring-const-component-size-bound-sum}.

\begin{corollary}\label{cor:dirty-vertex-bound-per-level}
    Define $Z_d = \sum_{\dep(a) = d} |D_a^\star \cup (\cup_{i \in [0,m_d]} D_{a,i})|$. Then $\mathbb{E}[Z_d] = \exp(-\Omega(\tau_d))$. The statement also holds conditional on the randomness of the updated edge.
\end{corollary}
\begin{proof}
    From \Cref{lem:active-vertices-to-affected-regions}, every pair $(x,a)$ with $\dep(a) = d$ and $x \in D_a^\star \cup (\cup_{i \in [0,m_d]} D_{a,i})$ corresponds to a vertex $\langle x,b \rangle$ in $\hat G$ connected to $\langle u,\routepath_{e,d'}\rangle$ for some ancestor $b$ of $a$. Such a vertex $\langle x,b \rangle$ counts toward $N_{\dep(b)}$ in \Cref{cor:edge-coloring-const-component-size-bound-sum}. Each vertex $\langle x,b\rangle$ with $\dep(b) = d'$ contributes at most $2^{d-d'}$ pairs to $Z_d$, so the expected value of $Z_d$ is bounded by
    \[
        \sum_{d'=0}^d 2^{d-d'} \exp(-\Omega(\tau_{d'})) = \exp(-\Omega(\tau_d)).
    \]
\end{proof}

\begin{lemma}\label{lem:edge-coloring-const-repair-stage-run-time}
    The expected work in the repair stage is $O(1)$.
\end{lemma}
\begin{proof}
    We split the main cost in the repair stage into the pre-shattering and post-shattering parts.
    
    We first notice that the maximum degree of $E_a^\alpha$ is $2^{R-\dep(a)}U_R=O(U_{\dep(a)})$ for $\alpha \in \{\old, \new\}$. This follows from $E_a^\alpha = \cup_{b \in \mathcal{L}_T(a)} E_b^\alpha$ and \Cref{cor:edge-coloring-const-static-correctness}. Thus, the maximum degree of $J_a$ is $O(U_{\dep(a)})$.
    
    In the pre-shattering part, for every node $a$ of depth $d$, $i \in [0,m_d]$, and $x \in D_{a,i}$, $\sigma_{a,i}^\new(x)$ is computed. If $x \notin \bar V_{a,i}^\new$, then $\sigma_{a,i}^\new(x)$ is maintained implicitly, and we spend $O(1)$ work to mark it as implicit; otherwise, we collect the set of vertices and local information in its two-hop neighborhood in $\poly(U_{\dep(a)})$ time to construct $\sigma_{a,i}^\new(x)$. 

    We also need to take into account work to compute $s_e$. Since $x \in \bar V_{a}^\old$ implies $x \in \bar V_b^\old$ for internal nodes $b$ inside the subtree of $a$, we can find $s_e$ by tracing upward along the routing path. Testing whether $x \in \bar V_a^\old$ can be done by checking $\sigma_{a,m_{\dep(a)}}^\old(x)$ in $O(1)$ time, so the total time to find $s_e$ is $O(1+R - \dep(s_e)) = O(U_{\dep(s_e)})$.

    To skip repairs on nodes $a$ with empty $D_{a,0}$, we maintain an array of lists $Q_0,Q_1,\dots,Q_{R-1}$ in which $Q_i$ contains the set of nodes $a$ with $D_{a,0} \ne \varnothing$ and $\dep(a) = i$. We also maintain the smallest index $k$ such that $Q_k \ne \varnothing$. Each pair $(x,a) \in \mathcal{F}^\old \oplus \mathcal{F}^\new$ adds $a$ to $Q_{\dep(a)}$, and the two endpoints of $e$ add $s_e$ to $Q_{\dep(s_e)}$. Finally, we go through $Q_k,Q_{k+1},\dots,Q_{R-1}$ for the repair. This incurs an additional cost of $O(R-k) = O(U_k)$, dominated by the previous $\poly(U_k)$ cost.

    In the post-shattering part, each vertex in $D_a^\star$ participates in computing the new post-shattering bipartition. The total work is the number of vertices, $|D_a^\star|$, plus the number of edges in the affected components. For $x \in D_a^\star$, if the number of edges incident to $x$ in $E_a^{\bad, \new}$ is nonzero, then $x \in \bar V_a^\new$ and the degree of $x$ in $E_a^{\bad, \new}$ is $O(U_d)$. In sum, the total work in the post-shattering part is $O(U_d|D_a^\star|)$.

    Combining the two parts, we know that each vertex counted in $Z_d$, defined in \Cref{cor:dirty-vertex-bound-per-level}, generates $\poly(U_d)$ work. From \Cref{cor:dirty-vertex-bound-per-level}, $\mathbb{E}[Z_d] = \exp(-\Omega(\tau_d))$, so the expected total work is
    \[
        O(1) + \sum_{d=0}^{R-1} \exp(O(\log U_d) - \Omega(\tau_d)) = O(1)
    \]
    since $\tau_d \gg \log U_d$, the sequence $\tau_i$ decreases geometrically, and $\tau_R$ is large enough.
\end{proof}

We are ready for the final theorem.

\begin{theorem}\label{thm:edge-coloring-const}
There is a randomized algorithm that maintains a proper $(1+O(\epsilon))\Delta$ edge coloring for a dynamic graph with maximum degree $\Delta\ge\Delta_\epsilon$. Against an oblivious adversary, each edge insertion or deletion takes $\poly(1/\epsilon)$ expected worst-case time.
\end{theorem}
\begin{proof}

If $\Delta \le \exp(\Delta_\epsilon)$, apply \Cref{lem:edge-coloring-small-deg} directly. Otherwise, use the dynamic algorithm in this section to maintain the partition $\mathcal{P}_R = \{E_a \mid a \in \mathcal{L}_T\}$, and apply \Cref{lem:edge-coloring-small-deg} to each subgraph with disjoint palettes. This application uses the facts that $U_R \in [\exp(\Delta_\epsilon) / 2, \exp(\Delta_\epsilon)]$ and every subgraph in $\mathcal{P}_R$ always has maximum degree at most $U_R$ according to \Cref{lem:discover-stage-invariants,lem:edge-coloring-const-correctness,cor:edge-coloring-const-static-correctness}. This results in a coloring of size $(1+O(\epsilon))U_R 2^R = (1+O(\epsilon)) \Delta$ as $U_R = (1+O(\epsilon))\Delta/2^R$.

By \Cref{lem:bottom-up-exploration-cost,lem:edge-coloring-const-repair-stage-run-time}, maintaining $\mathcal{P}_R$ takes $O(1)$ expected work and produces $O(1)$ expected edge updates, so the total expected cost of the \Cref{lem:edge-coloring-small-deg} instances is $\poly(1/\epsilon)$. This means that the whole algorithm has $\poly(1/\epsilon)$ expected work.
\end{proof}

%% file: witness.tex
\section{Proof of the Component Size Upper Bound}\label{sec:witness}

In this section, we prove \Cref{lem:hat-G-connected-component-size-bound}. Recall the definition of $\hat G(\hat V,\hat E)$: it has vertex set
\[
    \hat V=\{\langle x,a\rangle:x\in V(\mathcal R_{w,a})\text{ for some critical failure }(w,a)\}.
\]
For each affected region $\mathcal R_{w,a}$, there is a \emph{horizontal} edge in $\hat E$ between each pair of vertices $\langle x,a\rangle,\langle y,a\rangle$ for which $x$ and $y$ belong to $V(\mathcal R_{w,a})$. There is also a \emph{vertical} edge in $\hat E$ between $\langle x,a\rangle$ and $\langle x,b\rangle$ whenever $a\ne b$ are comparable and both vertices belong to $\hat V$.

Fix the updated edge $e = (u,v)$ and condition on all random choices for $e$. In the rest of this section, for a node $c$ and $d\in[0,R]$, define $A_d(c)$ to be $c$ when $\dep(c) \le d$ and to be the depth-$d$ ancestor of $c$ when $\dep(c) > d$.

\subsection{A Witness for Each Vertex in the Connected Component}

Let $\mathcal H$ be a graph whose vertices are the critical failures in the instance. Two distinct critical failures $(x,c)$ and $(y,b)$ are adjacent in $\mathcal H$ if $c$ and $b$ are comparable and
\[
    V(\mathcal R_{x,c})\cap V(\mathcal R_{y,b})\ne\varnothing.
\]

\begin{lemma}\label{lem:hat-component-region-chain}
If $\langle x,a\rangle$ and $\langle y,b\rangle$ are in the same connected component of $\hat G$, then there is a sequence of critical failures
\[
    (z_1,c_1),(z_2,c_2),\ldots,(z_\ell,c_\ell)
\]
such that $c_1=a$, $c_\ell=b$, $x\in V(\mathcal R_{z_1,c_1})$, $y\in V(\mathcal R_{z_\ell,c_\ell})$, and neighboring failures are adjacent in $\mathcal H$.
\end{lemma}
\begin{proof}
If $\langle x,a\rangle=\langle y,b\rangle$, choose any critical failure $(z,a)$ whose affected region contains $x$. Otherwise, choose a path $P = (\langle p_1,c_1\rangle,\langle p_2,c_2\rangle,\dots,\langle p_\ell,c_\ell\rangle)$ in $\hat G$ from $\langle x,a\rangle$ to $\langle y,b\rangle$. Associate a critical failure $(z_i,c_i)$ with each $1 \le i \le \ell$. When $i \ne \ell$ and $c_i = c_{i+1}$, $(z_i,c_i)$ is a critical failure that introduces the horizontal edge between $\langle p_i,c_i\rangle$ and $\langle p_{i+1},c_{i+1}\rangle$ in $\hat G$; otherwise, it is a critical failure that introduces $\langle p_i,c_i\rangle$.

We claim that the sequence $(z_1,c_1),\dots,(z_\ell,c_\ell)$ satisfies the statement. From the definition, $c_1 = a$, $c_\ell = b$, $x \in V(\mathcal R_{z_1,c_1})$, and $y \in V(\mathcal R_{z_\ell,c_\ell})$. For every $1 \le i < \ell$, if $c_i = c_{i+1}$, then $p_{i+1} \in V(\mathcal R_{z_i,c_i}) \cap V(\mathcal R_{z_{i+1},c_{i+1}})$ and the two failures are adjacent in $\mathcal{H}$; otherwise, $c_i$ and $c_{i+1}$ are comparable, and $p_i \in V(\mathcal R_{z_i,c_i}) \cap V(\mathcal R_{z_{i+1},c_{i+1}})$. Consecutive equal failures can be suppressed, and all remaining neighboring failures are adjacent in $\mathcal H$, giving the required sequence.
\end{proof}

We now define a witness certifying that $\langle y,b\rangle$ lies in the same component of $\hat G$ as $\langle x,a\rangle$.

\begin{definition}[Witness]\label{def:witness}
For $x,y\in V(G)$ and $a,b\in T\setminus\mathcal{L}_T$, a \emph{witness} of length $\ell\ge1$ from $\langle x,a\rangle$ to $\langle y,b\rangle$ consists of pairs $(z_i,c_i)\in V(G)\times (T\setminus\mathcal{L}_T)$ for $i\in[\ell]$, vertices $w_0,\ldots,w_\ell\in V(G)$, and two simple paths $P_i^L,P_i^R$ for every $i\in[\ell]$. Write $d_i=\dep(c_i)$, $T_i=P_i^L\cup P_i^R$, and $F_i=E(T_i)$. The following properties hold for every $i \in [\ell]$:
\begin{itemize}
    \item[1.] $c_1=a$, $c_\ell=b$, $w_0=x$, $w_\ell=y$, $c_i$ and $c_{i+1}$ are comparable when $i < \ell$.
    \item[2.] $(z_i,c_i)$ is a critical failure.
    \item[3.] $P_i^L$ is a $z_i$--$w_{i-1}$ path and $P_i^R$ is a $z_i$--$w_i$ path in $\bar E_{c_i}^\cup$, each of length at most $\rho_{d_i}$.
    \item[4.] $T_i$ is a tree with at most one vertex of degree three and all other vertices of degree at most two.
\end{itemize}
The following properties hold for every $i,j\in[\ell]$ with $|i-j|>1$:
\begin{itemize}
    \item[5.] If $c_i$ and $c_j$ are comparable, then $z_j\notin V(T_i)$.
    \item[6.] If $c_i=c_j$, then $\dist_{\bar E_{c_i}^\cup}(z_i,z_j)>1$.
    \item[7.] $F_i\cap F_j=\varnothing$.
\end{itemize}
\end{definition}

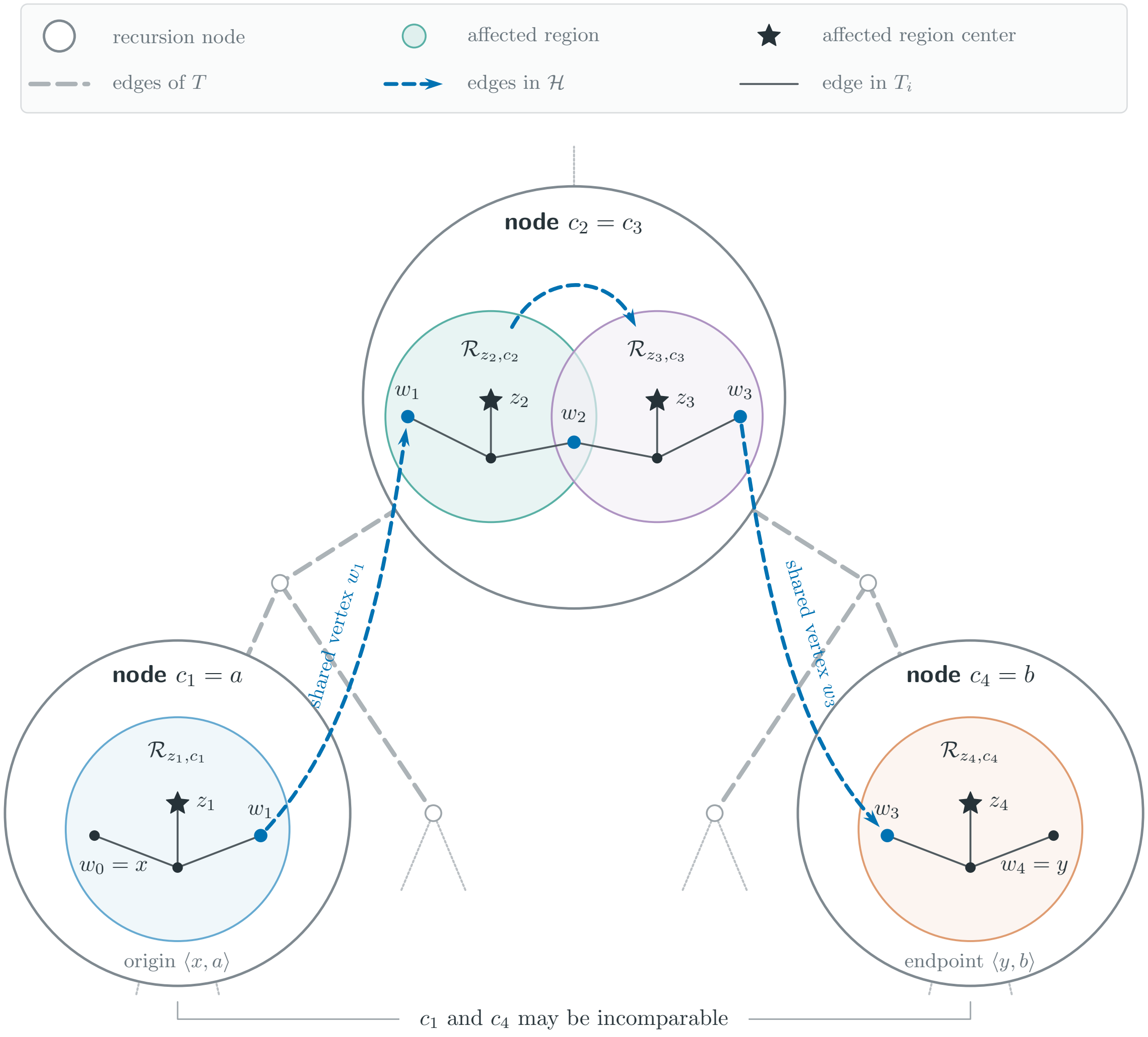
\begin{figure}[htbp]
    \centering
    \resizebox{0.95\linewidth}{!}{%
        \input{figures/witness-definition.tex}%
    }
    \caption{A possible witness of length four from $\langle x,a\rangle$ to $\langle y,b\rangle$}
    \label{fig:witness-definition}
\end{figure}

\Cref{fig:witness-definition} provides an illustration of a possible witness satisfying properties one through four.

Next, we show that a witness necessarily exists whenever a pair of vertices lies in the same connected component.
\begin{lemma}\label{lem:vertex-to-witness}
If $\langle x,a\rangle$ and $\langle y,b\rangle$ are in the same connected component of $\hat G$, then there is a witness from $\langle x,a\rangle$ to $\langle y,b\rangle$.
\end{lemma}
\begin{proof}
By \Cref{lem:hat-component-region-chain}, there is a path in $\mathcal H$ from a critical failure $(z_1,a)$ whose affected region contains $x$ to a critical failure $(z_\ell,b)$ whose affected region contains $y$. Choosing a shortest path $(z_1,c_1),\ldots,(z_\ell,c_\ell)$ and setting $w_0=x$ and $w_\ell=y$ gives properties 1 and 2.

For each $i<\ell$, choose any $w_i\in V(\mathcal R_{z_i,c_i})\cap V(\mathcal R_{z_{i+1},c_{i+1}})$. For each $i$, fix a shortest-path tree of $\mathcal R_{z_i,c_i}$ rooted at $z_i$, and let $P_i^L$ and $P_i^R$ be its paths to $w_{i-1}$ and $w_i$. Because $\mathcal R_{z_i,c_i}$ has radius $\rho_{d_i}$, these paths satisfy property 3. Their union is a tree in which only the branching point can have degree three, proving property 4.

Suppose $|i-j|>1$ and $c_i,c_j$ are comparable. If $V(\mathcal R_{z_i,c_i})$ and $V(\mathcal R_{z_j,c_j})$ intersect, then $(z_i,c_i)$ and $(z_j,c_j)$ are adjacent in $\mathcal H$, contradicting the minimality of the path in $\mathcal H$. Since $T_i\subseteq \mathcal R_{z_i,c_i}$ and $z_j\in V(\mathcal R_{z_j,c_j})$, this proves property 5. This also shows that $F_i \cap F_j = \varnothing$ when $c_i$ and $c_j$ are comparable; for noncomparable $c_i$ and $c_j$, $\bar E_{c_i}^\cup$ and $\bar E_{c_j}^\cup$ are disjoint. This gives property 7. Finally, if $c_i=c_j$ and the centers were equal or adjacent in $\bar E_{c_i}^\cup$, the affected regions would intersect; this proves property 6.
\end{proof}

From \Cref{lem:vertex-to-witness}, we know that to prove \Cref{lem:hat-G-connected-component-size-bound}, it suffices to bound the expected number of witnesses starting at $\langle w,a\rangle$.

\subsection{Level-by-Level Exposure for Counting}

We expose one level of randomness at a time to count the number of witnesses.

\begin{definition}[Partial witness]\label{def:hat-partial-witness}
Fix $\langle u,a\rangle$ and $\mathcal W=\bigl((z_i,c_i)_{i=1}^{\ell},(w_i)_{i=0}^{\ell},(P_i^L,P_i^R)_{i=1}^{\ell}\bigr)$ for $\ell \ge 1$. Write $d_i=\dep(c_i)$, $T_i=P_i^L\cup P_i^R$, and $F_i=E(T_i)$. For $d\in[0,R]$, the descriptor $\mathcal W$ is a $d$-\emph{witness} ending at $\langle w_\ell,c_\ell\rangle$ if the following properties hold for every $i \in [\ell]$:
\begin{itemize}
    \item[1.] $c_1=a$, $w_0=u$, $c_i$ and $c_{i+1}$ are comparable for $i < \ell$.
    \item[2.] If $d_i<d$, then $(z_i,c_i)$ is a critical failure.
    \item[3.] $P_i^L$ is a simple $z_i$--$w_{i-1}$ path and $P_i^R$ is a simple $z_i$--$w_i$ path in $\bar E_{A_d(c_i)}^\cup$, each of length at most $\rho_{d_i}$.
    \item[4.] $T_i$ is a tree with at most one vertex of degree three, and all its other vertices have degree at most two.
\end{itemize}
Additionally, the following properties hold for every $i, j \in [\ell]$ with $|i-j| > 1$:
\begin{itemize}
    \item[5.] If $c_i$ and $c_j$ are comparable, then $z_j\notin V(T_i)$.
    \item[6.] If $c_i=c_j$ and $d_i\le d$, then $\dist_{\bar E_{c_i}^\cup}(z_i,z_j)>1$.
    \item[7.] $F_i\cap F_j=\varnothing$.
\end{itemize}
\end{definition}

By comparing the definitions, a witness is an $R$-witness. The next two lemmas explain the usefulness of the partial witness definition.

\begin{lemma}\label{lem:witness-monotonicity}
If $\mathcal{W}$ is a $d$-witness for some $d\in[1,R]$, then it is also a $(d-1)$-witness.
\end{lemma}
\begin{proof}
Only properties 2, 3, and 6 depend on $d$. Properties 2 and 6 become weaker when $d$ decreases. For property 3, $A_{d-1}(c_i)$ is an ancestor of $A_d(c_i)$, so $\bar E_{A_d(c_i)}^\cup\subseteq\bar E_{A_{d-1}(c_i)}^\cup$. Hence property 3 also becomes weaker.
\end{proof}

\begin{lemma}\label{lem:witness-randomness-dependency}
Relative to the fixed randomness of the updated edge $e$, whether $\mathcal{W}$ is a $d$-witness depends only on the randomness of all other edges at levels before $d$. In particular, whether it is a $0$-witness does not depend on the randomness.
\end{lemma}
\begin{proof}
Only properties 2, 3, and 6 depend on the edge randomness. If $d_i<d$, whether $(z_i,c_i)$ is critical is determined by the seed failure tests at $(z_i,c_i)$ and at $(z_i,\P(c_i))$ when $c_i \ne r_T$, which depend on randomness at levels at most $d_i<d$. The graph $\bar E_{A_d(c_i)}^\cup$ in property 3 is determined by the random directions on each edge at levels smaller than $d$. Finally, when property 6 applies, $d_i\le d$, and $\bar E_{c_i}^\cup$ depends only on the direction randomness at levels smaller than $d_i\le d$. Thus all three properties depend only on levels smaller than $d$.
\end{proof}

For a fixed descriptor $\mathcal W$ and $d\in[0,R-1]$, define
\begin{gather*}
    I_d=\{i\in[\ell]:d_i=d\},\\
    I_d^+=\{i\in[\ell]:d_i>d\},\\
    M_d=\left|\left(\cup_{i\in I_d^+}F_i\right)\setminus\{e\}\right|.
\end{gather*}
We next bound the probability that a $d$-witness is also a $(d+1)$-witness.

\begin{lemma}\label{lem:hat-one-level-exposure}
For every descriptor $\mathcal W$ and every $d\in[0,R-1]$ such that $\mathcal{W}$ has nonzero probability of being a $d$-witness,
\[
    \Pr[\mathcal W\text{ is a $(d+1)$-witness}\mid \mathcal W\text{ is a $d$-witness}] \le \exp(-\Omega(\tau_d|I_d|))2^{-M_d}.
\]
\end{lemma}
\begin{proof}
By \Cref{lem:witness-randomness-dependency}, the event that $\mathcal W$ is a $d$-witness depends only on levels before $d$. We condition on any randomness from those levels under which it is a $d$-witness and prove a uniform bound over the conditioning.

For every $i\in I_d^+$, property 3 at level $d+1$ requires every edge in $F_i$ to be routed from $A_d(c_i)$ to $A_{d+1}(c_i)$. If there are conflicting requirements, or the fixed updated edge violates a requirement, the probability that they all hold is zero; otherwise, they impose exactly $M_d$ mutually independent requirements on the random directions. With probability $2^{-M_d}$, all requirements are fulfilled.

Partition $I_d$ by the parity of the index and let $S$ be a larger class, so $|S|\ge|I_d|/2$. For $i\in S$, let $\mathcal E_i$ be the event that $(z_i,c_i)$ is a critical failure. For distinct $i,j\in S$, we have $|i-j|>1$. As a result, property 6 ensures that the distance between $z_i$ and $z_j$ is greater than one when $c_i=c_j$, and their incident edges in $\bar E_{c_i}^\cup$ are disjoint. If $c_i\ne c_j$, $\bar E_{c_i}^\cup$ and $\bar E_{c_j}^\cup$ are always disjoint because two nodes are at the same depth. Therefore, the events $\{\mathcal E_i:i\in S\}$ depend on disjoint sets of level-$d$ random choices and are mutually independent. By \Cref{lem:critical-failure-tail},
\[
    \Pr\!\left[\bigcap_{i\in S}\mathcal E_i\right]
    \le\exp(-\Omega(\tau_d|S|))
    =\exp(-\Omega(\tau_d|I_d|)).
\]

It remains to account for dependencies between the routing requirements and the critical failures. We claim that for each $i\in S$, all but six routing requirements are independent of $\mathcal{E}_i$. To see this, if an edge in $F_j$ with $d_j > d$ is incident to $z_i$ in $\bar E_{c_i}^\cup$, then $c_j$ is a descendant of $c_i$, and property 5 implies $|j-i|\le1$. Since $d_i=d$, we have $j\ne i$, so this edge can only come from $F_{i-1} \cup F_{i+1}$. The claim follows because property 4 bounds the degree of $z_i$ in each of these two edge sets by three.

After discarding at most $6|S|$ routing requirements, the remaining requirements and the critical failure events are mutually independent. As a result, the probability that $\mathcal{W}$ becomes a $(d+1)$-witness is at most
\[
    \exp(-\Omega(\tau_d|I_d|))2^{-M_d+6|S|} = \exp(-\Omega(\tau_d|I_d|))2^{-M_d}.
\]
\end{proof}

For an edge $g\in\bigcup_iF_i$, define its \emph{witness depth} by $d_{\mathcal W}(g)=\max\{d_i:g\in F_i\}$.
Applying \Cref{lem:hat-one-level-exposure} successively gives the following corollary.

\begin{corollary}\label{cor:hat-complete-witness-probability}
There is an absolute constant $C_2>0$ such that, for every fixed $0$-witness $\mathcal W$,
\[
    \Pr[\mathcal W\text{ is a witness}]
    \le \left(\prod_{i=1}^{\ell}\exp(-C_2\tau_{d_i})\right) \left(
    \prod_{g\in(\bigcup_{i=1}^{\ell}F_i)\setminus\{e\}}2^{-d_{\mathcal W}(g)}\right).
\]
\end{corollary}
\begin{proof}
The statement is immediate when the probability is zero; otherwise, the probability that $\mathcal{W}$ is a $d$-witness is always nonzero for every $d \in [0,R]$ according to \Cref{lem:witness-monotonicity}. By \Cref{lem:witness-monotonicity,lem:witness-randomness-dependency}, the probability that $\mathcal{W}$ is a witness is the product of $\Pr[\mathcal{W}\text{ is an $i$-witness} \mid \mathcal{W}\text{ is an $(i-1)$-witness}]$ for every $1 \le i \le R$. Applying \Cref{lem:hat-one-level-exposure} bounds this probability by
\[
    \prod_{d=0}^{R-1}\exp(-C_2\tau_d|I_d|)2^{-M_d}= \left(\prod_{i=1}^{\ell}\exp(-C_2\tau_{d_i})\right)\left(\prod_{g\in(\bigcup_iF_i)\setminus\{e\}}2^{-d_{\mathcal W}(g)}\right).
\]
\end{proof}

\subsection{Counting Witnesses}

\begin{lemma}\label{lem:witness-count}
Fix $x \in V(G)$, $a\in T\setminus\mathcal{L}_T$, and the randomness of the updated edge, and write $d'=\dep(a)$. For every $d<R$, the expected number of $R$-witnesses from $\langle w,a\rangle$ to vertices $\langle x,b\rangle$ with $\dep(b)=d$ is $\exp(-\Omega(\tau_{d'}+\tau_d))$.
\end{lemma}
\begin{proof}
We enumerate all $0$-witnesses from $\langle w,a \rangle$ to $\langle x, b\rangle$ and sum the bound in \Cref{cor:hat-complete-witness-probability}. Fix the length $\ell$ of the witness and the depth sequence $d_1,\ldots,d_\ell$ with $d_1=d'$ and $d_\ell=d$. We have $c_1=a$ and $w_0=w$. Suppose $c_i$ and $w_{i-1}$ have been chosen. We enumerate $P_i^L$ from $w_{i-1}$ to $z_i$, then enumerate $P_i^R$ from $z_i$ to $w_i$, and compute the product in \Cref{cor:hat-complete-witness-probability} incrementally for every possible witness along the way.

We extend $P_i^L$ and $P_i^R$ by one edge at a time. By property 7, an edge in $P_i^L$ or $P_i^R$ does not belong to $\cup_{j < i-1} F_j$. As a result, the new edge is either the updated edge $e$, one of at most six edges in $F_{i-1}\cup F_i$ according to property 4, or an edge that has not appeared before. In the last case, we have at most $\Delta$ choices, and, denoting this edge by $g$, we need to charge $2^{-d_i}$ since $d_{\mathcal{W}}(g)$ grows from zero to $d_i$. When an edge reappears at a larger depth, we omit its additional routing penalty for an upper bound. Each extension therefore contributes a multiplicative $\Delta/2^{d_i}+7=O(U_{d_i})$ to the sum of probabilities. There are at most $(\rho_{d_i}+1)^2$ choices for the two path lengths and at most $2\rho_{d_i}$ extensions. Since $\rho_{d_i}=O(\log^3 U_{d_i})$, the weighted number of choices for $T_i$, $z_i$, and $w_i$ is at most
\[
    (\rho_{d_i}+1)^2O(U_{d_i})^{2\rho_{d_i}}
    =\exp(O(\log^4 U_{d_i})).
\]

For $i<\ell$, we additionally need to choose a node $c_{i+1}$ of depth $d_{i+1}$ comparable with $c_i$. If $d_{i+1}\le d_i$, the choice is unique. Otherwise, the number of choices is $2^{d_{i+1}-d_i}\le2^{R-d_i}\le\Delta / 2^{d_i}\le U_{d_i}$. Finally, accounting for the critical-failure probability $\exp(-\Omega(\tau_{d_i}))$ in the weight and multiplying over $1\le i\le\ell$, we conclude that the sum of probabilities over all $0$-witnesses with the fixed length and depth sequence is
\[
    \prod_{i=1}^{\ell}\exp\bigl(-\Omega(\tau_{d_i})+O(\log^4 U_{d_i})\bigr)
    \le\prod_{i=1}^{\ell}\exp(-\Omega(\tau_{d_i})).
\]
Summing over all possible depth sequences $d_2,d_3,\dots,d_{\ell-1}$, the total sum of probabilities for all $0$-witnesses of length $\ell$ is $\exp(-\Omega(\tau_{d'} + \tau_d)) C^{\ell - 2}$, where $C = \sum_{h=0}^{R-1} \exp(-\Omega(\tau_h)) \le 1/2$ when $\epsilon$ is small enough. As a consequence, summing over all lengths $\ell\ge2$ gives an expected number of witnesses at most $\exp(-\Omega(\tau_{d'} + \tau_d))$.
When $\ell=1$, necessarily $d'=d$, and its contribution is at most $\exp(-\Omega(\tau_d))$. This proves the claim.
\end{proof}

\begin{proof}[Proof of \Cref{lem:hat-G-connected-component-size-bound}]
Fix $\langle w,a\rangle$ and $d\in[0,R-1]$, and write $d'=\dep(a)$. For every $\langle x,b\rangle\in\hat K(\langle w,a\rangle)$ with $\dep(b)=d$, \Cref{lem:vertex-to-witness} supplies an $R$-witness from $\langle w,a\rangle$ to $\langle x,b\rangle$. Witnesses ending at distinct vertices are distinct; thus, the number of depth-$d$ vertices in the component is at most the number of $R$-witnesses ending at depth $d$. Taking expectations and applying \Cref{lem:witness-count} proves the conditional bound. Averaging over the choices fixed for $e$ proves the unconditional bound.
\end{proof}

%% file: figures/witness-definition.tex
\begingroup

\definecolor{witnessBlue}{HTML}{0072B2}
\definecolor{witnessOrange}{HTML}{C95A12}
\definecolor{witnessTeal}{HTML}{008575}
\definecolor{witnessPurple}{HTML}{8B65A8}
\definecolor{witnessInk}{HTML}{263238}
\definecolor{witnessGray}{HTML}{68747C}

\begin{tikzpicture}[
    x=1cm,y=1cm,
    font=\fontsize{10}{12}\selectfont,
    text=witnessInk,
    line cap=round,line join=round,
    tree edge/.style={draw=witnessGray!55,line width=2pt,
        dash pattern=on 8pt off 4pt},
    continuation/.style={draw=witnessGray!45,line width=.9pt,densely dotted},
    tree node/.style={circle,draw=witnessGray!65,fill=white,
        line width=.8pt,inner sep=2.5pt},
    node boundary/.style={draw=witnessGray!85,fill=white,line width=1.1pt},
    node label/.style={font=\sffamily\bfseries\fontsize{11}{13}\selectfont},
    vertex/.style={circle,fill=witnessInk,inner sep=1.65pt},
    joining vertex/.style={circle,fill=witnessBlue,inner sep=2.1pt},
    failure/.style={star,star points=5,star point ratio=2.15,
        draw=witnessInk,fill=witnessInk,minimum size=3.7mm,inner sep=0pt},
    local path/.style={draw=witnessInk!80,line width=.85pt},
    chain/.style={draw=witnessBlue,line width=1.7pt,
        dash pattern=on 5pt off 3pt,->,shorten <=4pt,shorten >=5pt},
    note/.style={font=\fontsize{9}{11}\selectfont,text=witnessGray},
    >={Stealth[length=2.8mm,width=1.8mm]}
]


\coordinate (ancestor) at (9,12);
\coordinate (leftparent) at (4.4,9.1);
\coordinate (rightparent) at (13.6,9.1);
\coordinate (c1) at (2.8,5.5);
\coordinate (leftother) at (6.8,5.5);
\coordinate (rightother) at (11.2,5.5);
\coordinate (c4) at (15.2,5.5);
\draw[tree edge] (c1)--(leftparent)--(ancestor)--(rightparent)--(c4);
\draw[tree edge] (leftparent)--(leftother);
\draw[tree edge] (rightparent)--(rightother);
\draw[continuation] (9,15.3)--(9,15.92);
\foreach \xx in {2.8,15.2} {
    \draw[continuation] (\xx,5.5)--(\xx-.65,2.65);
    \draw[continuation] (\xx,5.5)--(\xx+.65,2.65);
}
\foreach \xx in {6.8,11.2} {
    \draw[continuation] (\xx,5.5)--(\xx-.50,4.30);
    \draw[continuation] (\xx,5.5)--(\xx+.50,4.30);
}
\foreach \cc in {leftparent,rightparent,leftother,rightother} {
    \node[tree node] at (\cc) {};
}

\path[node boundary] (ancestor) circle [radius=3.3cm];
\path[node boundary] (c1) circle [radius=2.7cm];
\path[node boundary] (c4) circle [radius=2.7cm];
\node[node label] at (9,14.73) {node $c_2=c_3$};
\node[node label] at (2.8,7.65) {node $c_1=a$};
\node[node label] at (15.2,7.65) {node $c_4=b$};

\path[fill=witnessGray!3,draw=witnessGray!25,line width=.6pt,rounded corners=3pt]
    (.35,16.45) rectangle (17.65,18.15);
\path[node boundary] (.95,17.65) circle [radius=.24cm];
\node[note,anchor=west] at (1.65,17.65) {recursion node};
\path[fill=witnessTeal!12,draw=witnessTeal!65,line width=.8pt]
    (6.50,17.65) circle [radius=.18cm];
\node[note,anchor=west] at (7.20,17.65) {affected region};
\node[failure] at (12.05,17.65) {};
\node[note,anchor=west] at (12.75,17.65) {affected region center};
\draw[tree edge] (.50,16.90)--(1.40,16.90);
\node[note,anchor=west] at (1.65,16.90) {edges of $T$};
\draw[chain,shorten <=0pt,shorten >=0pt] (6.05,16.90)--(6.95,16.90);
\node[note,anchor=west] at (7.20,16.90) {edges in $\mathcal{H}$};
\draw[local path] (11.60,16.90)--(12.50,16.90);
\node[note,anchor=west] at (12.75,16.90) {edge in $T_i$};

\path[fill=witnessTeal!9,draw=witnessTeal!65,line width=.85pt]
    (7.7,11.7) circle [radius=1.65cm];
\path[fill=witnessPurple!10,fill opacity=.65,
      draw=witnessPurple!70,line width=.85pt]
    (10.3,11.7) circle [radius=1.65cm];
\node at (7.7,12.72) {$\mathcal R_{z_2,c_2}$};
\node at (10.3,12.72) {$\mathcal R_{z_3,c_3}$};

\path[fill=witnessBlue!6,draw=witnessBlue!60,line width=.85pt]
    (2.8,5.25) circle [radius=1.75cm];
\path[fill=witnessOrange!6,draw=witnessOrange!60,line width=.85pt]
    (15.2,5.25) circle [radius=1.75cm];
\node at (2.8,6.45) {$\mathcal R_{z_1,c_1}$};
\node at (15.2,6.45) {$\mathcal R_{z_4,c_4}$};

\coordinate (z1) at (2.8,5.65);
\coordinate (fork1) at (2.8,4.65);
\coordinate (x) at (1.50,5.15);
\coordinate (w1lower) at (4.10,5.15);
\draw[local path] (x)--(fork1)--(w1lower);
\draw[local path] (z1)--(fork1);

\coordinate (z2) at (7.7,11.95);
\coordinate (fork2) at (7.7,11.05);
\coordinate (w1upper) at (6.40,11.70);
\coordinate (w2) at (9,11.30);
\draw[local path] (w1upper)--(fork2)--(w2);
\draw[local path] (z2)--(fork2);

\coordinate (z3) at (10.3,11.95);
\coordinate (fork3) at (10.3,11.05);
\coordinate (w3upper) at (11.60,11.70);
\draw[local path] (w2)--(fork3)--(w3upper);
\draw[local path] (z3)--(fork3);

\coordinate (z4) at (15.2,5.65);
\coordinate (fork4) at (15.2,4.65);
\coordinate (w3lower) at (13.90,5.15);
\coordinate (y) at (16.50,5.15);
\draw[local path] (w3lower)--(fork4)--(y);
\draw[local path] (z4)--(fork4);

\draw[chain] (w1lower)
    .. controls (5.55,6.75) and (6.00,8.90) .. (w1upper)
    node[pos=.53,sloped,above=5pt,text=witnessBlue,
         font=\fontsize{9}{11}\selectfont] {shared vertex $w_1$};
\draw[chain] (w3upper)
    .. controls (12.00,8.90) and (12.45,6.75) .. (w3lower)
    node[pos=.47,sloped,above=5pt,text=witnessBlue,
         font=\fontsize{9}{11}\selectfont] {shared vertex $w_3$};

\draw[chain,shorten <=0pt,shorten >=0pt]
    (8.03,13.10) .. controls (8.48,13.95) and (9.52,13.95) .. (9.97,13.10);

\foreach \i in {1,2,3,4} {
    \node[failure] at (z\i) {};
    \node[anchor=west,xshift=5pt,inner sep=3pt] at (z\i) {$z_{\i}$};
    \node[vertex] at (fork\i) {};
}
\foreach \cc in {w1lower,w1upper,w2,w3upper,w3lower} {
    \node[joining vertex] at (\cc) {};
}
\foreach \cc in {x,y} {\node[vertex] at (\cc) {};}
\node at (1.80,4.65) {$w_0=x$};
\node at (16.20,4.65) {$w_4=y$};
\node[above=4pt] at (w1lower) {$w_1$};
\node[above=4pt] at (w1upper) {$w_1$};
\node[above=5pt] at (w2) {$w_2$};
\node[above=4pt] at (w3upper) {$w_3$};
\node[above=4pt] at (w3lower) {$w_3$};
\node[note] at (2.8,3.17) {origin $\langle x,a\rangle$};
\node[note] at (15.2,3.17) {endpoint $\langle y,b\rangle$};


\draw[witnessGray!65,line width=.65pt]
    (2.8,2.55)--(2.8,2.28)--(15.2,2.28)--(15.2,2.55);
\node[fill=white,inner xsep=9pt,inner ysep=2pt] at (9,2.28)
    {$c_1$ and $c_4$ may be incomparable};


\end{tikzpicture}%
\endgroup